\documentclass[12pt]{elsarticle}
\usepackage{amssymb}
\usepackage{txfonts}
\usepackage{mathbbold}
\usepackage{epsfig,bm,dcolumn}
\usepackage{graphicx}
\usepackage{color}
\usepackage{overpic}

\journal{Annals of Physics (N. Y.)}
\begin{document}

\begin{frontmatter}
\title{Superfluidity and collective modes in Rashba spin-orbit coupled Fermi gases}
\author{\normalsize{Lianyi He$^{a}$\footnote{E-mail address: lianyi@th.physik.uni-frankfurt.de}
and Xu-Guang Huang$^{b}$\footnote{E-mail address: xhuang@th.physik.uni-frankfurt.de}}}
\address{$^a$ Frankfurt Institute for Advanced Studies and Institute for Theoretical Physics, J. W. Goethe University,
60438 Frankfurt am Main, Germany\\
$^b$ Center for Exploration of Energy and Matter and Physics Department, Indiana University, Bloomington, IN 47408, USA}

\date{\today}

\begin{abstract}
We present a theoretical study of the superfluidity and the corresponding collective modes in two-component atomic Fermi gases with $s$-wave attraction
and synthetic Rashba spin-orbit coupling. The general effective action for the collective modes is derived from the functional path integral formalism.
By tuning the spin-orbit coupling from weak to strong, the system undergoes a crossover from an ordinary BCS/BEC superfluid to a Bose-Einstein condensate
of rashbons. We show that the properties of the superfluid density and the Anderson-Bogoliubov mode manifest this crossover. At large spin-orbit coupling,
the superfluid density and the sound velocity become independent of the strength of the $s$-wave attraction. The two-body interaction among the rashbons
is also determined. When a Zeeman field is turned on, the system undergoes quantum phase transitions to some exotic superfluid phases which are
topologically nontrivial. For the two-dimensional system, the nonanalyticities of the thermodynamic functions and the sound velocity across the phase
transition are related to the bulk gapless fermionic excitation which causes infrared singularities. The superfluid density and the sound velocity behave
nonmonotonically: they are suppressed by the Zeeman field in the normal superfluid phase, but get enhanced in the topological superfluid phase. The
three-dimensional system is also studied.
\end{abstract}

\begin{keyword}
        Fermi superfluidity \sep
        BCS-BEC crossover\sep
        Rashba spin-orbit coupling
%% keywords here, in the form: keyword \sep keyword
%% MSC codes here, in the form: \MSC code \sep code
%% or \MSC[2008] code \sep code (2000 is the default)
\end{keyword}
\end{frontmatter}

%%%%%%%%%%%%%%%%%%%%%%%%%%%%%%%%%%%%%%%%%%%%%%%%%%%%%%%%%%%%%%%%%%%%%%%%%%%%%%%%%%%%%%%%%%%%%%%%%%%%%%%%%%%%%%
\section{Introduction}
%%%%%%%%%%%%%%%%%%%%%%%%%%%%%%%%%%%%%%%%%%%%%%%%%%%%%%%%%%%%%%%%%%%%%%%%%%%%%%%%%%%%%%%%%%%%%%%%%%%%%%%%%%%%%%
It is generally believed that, by tuning the strength of the attractive interaction in a many-fermion system, we can realize a smooth crossover from
the Bardeen--Cooper--Schrieffer (BCS) superfluidity at weak attraction to Bose--Einstein condensation (BEC) of difermion molecules at strong
attraction~\cite{Eagles,Leggett,BCSBEC1,BCSBEC2,BCSBEC3,BCSBEC4,BCSBEC5,BCSBEC6,BCSBEC7,BCSBEC8}. One typical example is the dilute Fermi gas in three
dimensions with short-range attractive interaction, where the effective range $r_0$ of the interaction is much smaller than the inter-particle distance
characterized by $k_{\rm F}^{-1}$ where $k_{\rm F}$ is the Fermi momentum in the absence of interaction. The attraction strength is characterized by
a dimensionless parameter $1/(k_{\rm F}a_s)$ where $a_s$ is the $s$-wave scattering length of the short-range interaction. The BCS-BEC crossover has
been confirmed in the experiments of ultracold fermionic atoms~\cite{BCSBECexp1,BCSBECexp2,BCSBECexp3}, where the $s$-wave scattering length and hence
the parameter $1/(k_{\rm F}a_s)$ was tuned by means of the Feshbach resonance.

On the other hand, the effect of a nonzero Zeeman field $h$ has been a longstanding problem of fermionic superconductivity/superfluidity for several
decades~\cite{LOFFreview}. It is generally believed that the superfluidity is completely destroyed when the Zeeman field becomes large enough. The
well-known theoretical result for $s$-wave weak-coupling superconductors is that, at a critical Zeeman field $h_{\rm CC}=\Delta_0/\sqrt{2}$
(called Chandrasekhar-Clogston limit) where $\Delta_0$ is the zero temperature gap at $h=0$, a first-order phase transition from the BCS state to the
normal state occurs~\cite{CClimit,Sarma}. Further theoretical studies showed that the inhomogeneous Fulde-Ferrell-Larkin-Ovchinnikov state~\cite{FFLO}
may survive in a narrow window between $h_{\rm CC}$ and $h_{\rm FFLO}\approx0.754\Delta_0$. The Zeeman field effects in the BCS-BEC crossover have been
experimentally studied by using cold fermionic atoms~\cite{Imexp,Imth}. The atom numbers of the two lowest hyperfine states of $^6$Li are adjusted to
create a population imbalance which simulates effectively the Zeeman field $h$. The experimental results show that the fermionic superfluidity in the
BCS-BEC crossover regime is also completely destroyed when the Zeeman field is large enough.

The recent experimental breakthroughs in generating synthetic non-Abelian gauge field and synthetic spin-orbit coupling
\cite{SOC,SOCF01,SOCF02,SOCrmp,SOC01,SOC02,SOC03,3DSOC} have opened up the way to study the spin-orbit coupling effects as well as the combined
spin-orbit coupling and Zeeman field effects on the BCS-BEC crossover~\cite{SOC-BCSBEC,SOC-Hu,SOC-Yu,SOC-Gong,SOC-Iskin,SOC-He,SOC-Han,SOC-Yi,SOC-other}.
For solid state systems, it was shown that the topologically nontrivial superconducting phase appears in spin-orbit coupled systems if the Zeeman field
is large enough~\cite{TSC01,TSC02,TSC03,TSC04,TSC05,TSC06,TSC07,TSC08,TSC09}. For neutral atoms, the spin-orbit coupling can be generated through a
synthetic non-Abelian gauge potential ${\bf A}$~\cite{SOCrmp}. The well-known Rashba spin-orbit coupling for spin-1/2 fermions can be generated via a
2D synthetic vector potential~\cite{SOC02,SOC03}
\begin{equation}
{\bf A}=-\lambda\hbar\mbox{\boldmath{$\sigma$}}_\bot=-\lambda\hbar(\sigma_x{\bf e}_x+\sigma_y{\bf e}_y), \label{EQ001}
\end{equation}
where ${\bf a}_\bot=a_x{\bf e}_x+a_y{\bf e}_y$ for any vector ${\bf a}$. The single-particle Hamiltonian for a fermion moving in the synthetic gauge
field is given by ${\cal H}_{0} = \left(\hat{\bf p}-{\bf A}\right)^2/(2m)$ where $\hat{\bf p}=-i\hbar\nabla$ is the momentum operator. In this paper
we use the natural units $\hbar=k_{\rm B}=m=1$ for convenience.

For the 2D synthetic vector potential ${\bf A}$ given in (\ref{EQ001}), the single-particle Hamiltonian can be reduced to
\begin{equation}
{\cal H}_{0} =\frac{\hat{\bf p}^2}{2}+\lambda\mbox{\boldmath{$\sigma$}}_\bot\cdot\hat{\bf p}_\bot,
\end{equation}
where an irrelevant constant $\lambda^2/2$ has been omitted. The spin-dependent term $\lambda\mbox{\boldmath{$\sigma$}}_\bot\cdot\hat{\bf p}_\bot$ can be
mapped to the standard Rashba spin-orbit coupling $\lambda(\sigma_x\hat{p}_y-\sigma_y\hat{p}_x)$ by a spin rotation $\sigma_x\rightarrow\sigma_y$ and
$\sigma_y\rightarrow-\sigma_x$. The gauge field strength $\lambda$ characterizes the strength of the spin-orbit coupling, which can be tuned from weak
to strong in cold atom experiments. Since the final physical results depend only on $\lambda^2$, we set $\lambda>0$ in this paper without loss of
generality. For many-fermion systems, the spin-orbit coupling strength can be characterized by the dimensionless ratio $\lambda/k_{\rm F}$. While for
solid state systems this ratio is very small, it can reach the order $O(1)$ in cold atom systems~\cite{SOCF01,SOCF02}. Therefore, cold fermionic atoms
provide the way to study the fermionic superfluidity in the presence of a strong spin-orbit coupling.

Motivated by the experimental progress of realizing spin-orbit coupled atomic Fermi gases~\cite{SOCF01,SOCF02}, the fermionic superfluidity with
spin-orbit coupling has been extensively studied~\cite{SOC-BCSBEC,SOC-Hu,SOC-Yu,SOC-Gong,SOC-Iskin,SOC-He,SOC-Han,SOC-Yi,SOC-Zhou,SOC-other}. It was
shown that, in the presence of the Rashba spin-orbit coupling, the two-body bound state exists even for $a_s<0$ where the bound state does not exist
for $\lambda=0$~\cite{3Dbound}. With increased $\lambda$, the binding energy is generally enhanced. The bound state at $\lambda\neq0$ possesses a
nontrivial effective mass which is generally larger than twice of the fermion mass~\cite{SOC-Hu,SOC-Yu,SOC-He}. Such a novel bound state is referred
to as rashbon in the studies~\cite{rashbon}. For many-fermion systems, it has been proposed that a spin-orbit coupled Fermi gas can undergo a smooth
crossover from the ordinary BCS/BEC superfluidity to the Bose-Einstein condensation of rashbons if $\lambda/k_{\rm F}$ is tuned from small to large
values~\cite{SOC-BCSBEC,SOC-Hu,SOC-Yu,SOC-Gong,SOC-Iskin,SOC-He,SOC-Han}. On the other hand, if a Zeeman field $h$ is turned on, some topologically
nontrivial superfluid phases emerge~\cite{SOC-Gong,SOC-Han,SOC-Yi}.

In this paper, we study the bulk superfluid properties and the collective modes in Rashba spin-orbit coupled Fermi superfluids. We mainly consider two
aspects: (1) the bulk superfluid properties and the collective modes from weak to strong spin-orbit coupling at zero Zeeman field, which manifest the
crossover from ordinary Fermi superfluidity to the Bose-Einstein condensation of rashbons, and (2) the quantum phase transitions from the normal superfluid
phase to topologically nontrivial superfluid phases in the presence of nonzero Zeeman field and their effects on the bulk superfluid properties and the
collective modes.

The paper is organized as follows. In Sec. \ref{s2}, we derive the general effective action for the superfluid ground state and the collective modes
with arbitrary spin-orbit coupling and Zeeman field by using the functional path integral method. In Sec. \ref{s3} and Sec. \ref{s4}, we study the
systems with zero Zeeman field in three and two spatial dimensions, respectively. The systems with nonzero Zeeman fields are studied in Sec. \ref{s5}.
We summarize in Sec. \ref{s6}.

%%%%%%%%%%%%%%%%%%%%%%%%%%%%%%%%%%%%%%%%%%%%%%%%%%%%%%%%%%%%%%%%%%%%%%%%%%%%%%%%%%%%%%%%%%%%%%%%%%%%%%%%%%%%%%
\section{General formalism}\label{s2}
%%%%%%%%%%%%%%%%%%%%%%%%%%%%%%%%%%%%%%%%%%%%%%%%%%%%%%%%%%%%%%%%%%%%%%%%%%%%%%%%%%%%%%%%%%%%%%%%%%%%%%%%%%%%%%
We consider a homogeneous spin-1/2 Fermi gas with a short-range $s$-wave attractive interaction in the spin-singlet channel. For cold atom experiments,
the attractive strength can be tuned from weak to strong~\cite{BCSBEC6}. In the dilute limit where the effective range $r_0$ is much smaller than the
characteristic length scales of the system, that is, $r_0\ll k_{\rm F}^{-1}, a_s, \lambda^{-1}$, the attractive interaction can be modeled by a contact
one~\cite{Cui}. The many-body Hamiltonian of the system can be written as
\begin{eqnarray}
H =H_0+H_Z+H_{\rm int},
\end{eqnarray}
where
\begin{eqnarray}
&&H_0=\int d^3 {\bf r}\psi^{\dagger}({\bf r})\left(\frac{\hat{\bf p}^2}{2}
+\lambda\mbox{\boldmath{$\sigma$}}_\bot\cdot\hat{\bf p}_\bot-\mu\right)\psi({\bf r}),\nonumber\\
&&H_Z=-h\int d^3 {\bf r}\psi^{\dagger}({\bf r})\sigma_z \psi({\bf r}),\nonumber\\
&&H_{\rm int}=-U\int d^3 {\bf r}^{\phantom{\dag}}\psi^\dagger_{\uparrow}({\bf r})\psi^\dagger_{\downarrow}({\bf r})
\psi^{\phantom{\dag}}_{\downarrow}({\bf r})\psi^{\phantom{\dag}}_{\uparrow}({\bf r}).
\end{eqnarray}
Here $\psi({\bf r})= [\psi_\uparrow({\bf r}), \psi_\downarrow({\bf r})]^{\rm T}$ represents the two-component fermion fields, $\mu$ is the chemical
potential, and $h$ is the Zeeman magnetic field. We set $h>0$ in this paper without loss of generality. The contact coupling $U>0$ denotes the
attractive $s$-wave interaction between unlike spins.

In the functional path integral formalism, the partition function of the system at finite temperature $T$ is
\begin{eqnarray}
{\cal Z} = \int \mathcal{ D} \psi
\mathcal{D}\bar{\psi}\exp\left\{-{\cal S}[\psi,\bar{\psi}]\right\},
\end{eqnarray}
where
\begin{eqnarray}
{\cal S}[\psi,\bar{\psi}]=\int_0^{1/T} d\tau\int d^3{\bf r}
\bar{\psi}\partial_\tau \psi+\int_0^{1/T} d\tau H(\psi,\bar{\psi}).
\end{eqnarray}
Here $H(\psi,\bar{\psi})$ is obtained by replacing the field operators $\psi^\dagger$ and $\psi$ with the Grassmann variables $\bar{\psi}$ and $\psi$,
respectively. To decouple the interaction term we introduce the auxiliary complex pairing field $\Phi(x) = -U\psi_\downarrow(x)\psi_\uparrow(x)$
~$[x=(\tau,{\bf r})]$ and apply the Hubbard-Stratonovich transformation.  Using the Nambu-Gor'kov representation
\begin{eqnarray}
\Psi(x)=\left(\begin{array}{cc} \psi(x)\\ i\sigma_y\bar{\psi}^{\rm T}(x)\end{array}\right),\ \ \
\bar{\Psi}(x)=\left(\ \bar{\psi}(x)\ \ -\psi^{\rm T}(x)i\sigma_y\ \right),
\end{eqnarray}
we express the partition function as
\begin{eqnarray}
{\cal Z}=\int {\cal D}\Psi{\cal D}\bar{\Psi}{\cal D}\Phi {\cal D}\Phi^{\dagger}
\exp\Big\{-\tilde{{\cal S}}[\Psi,\bar{\Psi},\Phi,\Phi^\dagger]\Big\},
\end{eqnarray}
where
\begin{eqnarray}
\tilde{{\cal S}}[\Psi,\bar{\Psi},\Phi,\Phi^\dagger]=\frac{1}{U}\int dx|\Phi(x)|^2
-\frac{1}{2}\int dx\int dx^\prime\bar{\Psi}(x){\bf G}^{-1}(x,x^\prime)\Psi(x^\prime).
\end{eqnarray}
The inverse single-particle Green's function ${\bf G}^{-1}(x,x^\prime)$ is given by
\begin{eqnarray}
{\bf G}^{-1}(x,x^\prime)=\left(\begin{array}{cc}{\bf G}_+^{-1}(x)&\Phi(x)\\
\Phi^\dagger(x)& {\bf G}_-^{-1}(x)\end{array}\right)\delta(x-x^\prime),
\end{eqnarray}
where
\begin{eqnarray}
{\bf G}_\pm^{-1}(x)=-\partial_{\tau}+h\sigma_z \mp(\hat{\bf p}^2/2+\lambda\mbox{\boldmath{$\sigma$}}_\bot\cdot\hat{\bf p}_\bot-\mu).
\end{eqnarray}
Integrating out the fermion fields, we obtain
\begin{eqnarray}
{\cal Z}=\int{\cal D} \Phi {\cal D} \Phi^{\dagger}
\exp\Big\{-{\cal S}_{\rm{eff}}[\Phi, \Phi^{\dagger}]\Big\},
\end{eqnarray}
where the effective action reads
\begin{eqnarray}
{\cal S}_{\rm{eff}}[\Phi, \Phi^{\dagger}] = \frac{1}{U}\int dx|\Phi(x)|^{2}
-\frac{1}{2}\mbox{Trln} [{\bf G}^{-1}(x,x^\prime)].
\end{eqnarray}

The effective action ${\cal S}_{\rm{eff}}[\Phi, \Phi^{\dagger}]$ cannot be evaluated precisely. In this work, we consider mainly the zero temperature
case. Therefore, we follow the conventional approach to the BCS-BEC crossover problem, that is, we first consider the superfluid ground state which
corresponds to the saddle point of the effective action, and then study the Gaussian fluctuations around the saddle point. The Gaussian fluctuations
correspond to the collective modes, including the gapless Goldstone mode and the massive Higgs mode. In ordinary fermionic superfluids, only the
Goldstone mode or the so-called Anderson-Bogoliubov mode remains at low energy whereas the Higgs mode is pushed up to the two-particle continuum.
Therefore, the Higgs mode usually appears as a broad resonance at the large characteristic energy scale of the system.

In the superfluid ground state, the pairing field $\Phi(x)$ acquires a nonzero expectation value $\langle\Phi(x)\rangle=\Delta$, which serves as the
order parameter of the superfluidity. Due to the U$(1)$ symmetry, we can set $\Delta$ to be real without loss of generality. Then we decompose the
pairing field as $\Phi(x)=\Delta+\phi(x)$, where $\phi(x)$ is the fluctuation around the mean field. The effective action
${\cal S}_{\rm{eff}}[\Phi,\Phi^\dagger]$ can be expanded in powers of the fluctuation $\phi(x)$, that is,
\begin{eqnarray}
{\cal S}_{\rm{eff}}[\Phi,\Phi^\dagger]={\cal S}_{\rm{eff}}^{(0)}(\Delta)+{\cal S}_{\rm{eff}}^{(2)}[\phi,\phi^\dagger]+\cdots,
\label{sexpansion}
\end{eqnarray}
where ${\cal S}_{\rm{eff}}^{(0)}(\Delta)\equiv{\cal S}_{\rm eff}[\Delta,\Delta]$ is the saddle-point or mean-field effective action with $\Delta$
determined by the saddle point condition $\partial{\cal S}_{\rm{eff}}^{(0)}/\partial\Delta=0$. Note that under the saddle point condition the linear
terms in $\phi$ and $\phi^\dag$ in Eq.~(\ref{sexpansion}) vanish automatically.

%%%%%%%%%%%%%%%%%%%%%%%%%%%%%%%%%%%%%%%%%%%%%%%%%%%%%%%%%%%%%%%%%%%%%%%%%%%%%%%%%%%%%%%%%%%%%%%%%%%%%%%%%%%%%%
\subsection{Saddle point: mean-field approximation }
%%%%%%%%%%%%%%%%%%%%%%%%%%%%%%%%%%%%%%%%%%%%%%%%%%%%%%%%%%%%%%%%%%%%%%%%%%%%%%%%%%%%%%%%%%%%%%%%%%%%%%%%%%%%%%
The mean-field effective action or the thermodynamic potential $\Omega$ can be expressed as
\begin{eqnarray}
\Omega=\frac{{\cal S}_{\rm{eff}}^{(0)}(\Delta)}{\beta V}=\frac{\Delta^2}{U}
-\frac{1}{2}\sum_{K}{\rm ln}{\rm det}[{\cal G}^{-1}(K)],
\end{eqnarray}
where the inverse fermion Green's function reads
\begin{eqnarray}
{\cal  G}^{-1}(K)=\left(\begin{array}{cc}{\cal G}_+^{-1}(K) &\Delta\\ \Delta& {\cal G}_-^{-1}(K)\end{array}\right)
\end{eqnarray}
and ${\cal G}_\pm^{-1}(K)$ is given by
\begin{eqnarray}
{\cal G}_\pm^{-1}(K)=i\omega_n+h\sigma_z\mp(\xi_{\bf k}+\lambda\mbox{\boldmath{$\sigma$}}_\bot\cdot{\bf k}_\bot).
\end{eqnarray}
The dispersion $\xi_{\bf k}$ is defined as $\xi_{\bf k}=\epsilon_{\bf k}-\mu$ with $\epsilon_{\bf k}={\bf k}^2/2$. In this paper $K=(i\omega_n,{\bf k})$
denotes the energy and the momentum of fermions with $\omega_n=(2n+1)\pi T$ ($n$ integer) being the fermion Matsubara frequency. We use the notation
$\sum_{K}=T\sum_n\sum_{\bf k}$ with $\sum_{\bf k}=\int d^3{\bf k}/(2\pi)^3$ for the 3D system.

The determinant of the inverse fermion propagator, ${\rm{det}}[{\cal G}^{-1}(K)]$, can be evaluated as
\begin{eqnarray}
\rm{det}[{\cal G}^{-1}(K)]&=&\prod_{s=\pm}\left[(i\omega_n+sh)^2-E_{\bf k}^2-\lambda^2{\bf k}_\bot^2\right]
-4\lambda^2{\bf k}_\bot^2(\xi_{\bf k}^2-h^2)\nonumber\\
&=&\left[(i\omega_n)^2-(E_{\bf k}^+)^2\right]\left[(i\omega_n)^2-(E_{\bf k}^-)^2\right],
\end{eqnarray}
where $E_{\bf k}=\sqrt{\xi_{\bf k}^2+\Delta^2}$. The quasiparticle excitation spectra $E_{\bf k}^\pm$ are given by
\begin{eqnarray}
E_{\bf k}^\pm=\sqrt{E_{\bf k}^2+\eta_{\bf k}^2\pm2\zeta_{\bf k}}.
\end{eqnarray}
The quantities $\eta_{\bf k}$ and $\zeta_{\bf k}$ are defined as $\eta_{\bf k}=\sqrt{\lambda^2{\bf k}_\bot^2+h^2}$ and
$\zeta_{\bf k}=\sqrt{\xi_{\bf k}^2\eta_{\bf k}^2+h^2\Delta^2}$. Completing the Matsubara frequency sum we obtain the explicit form of the mean-field
effective action
\begin{eqnarray}
\Omega=\frac{\Delta^2}{U}-\sum_{\bf k}\sum_{\alpha=\pm}
\left[\frac{E_{\bf k}^\alpha-\xi_{\bf k}^\alpha}{2}+T\ln\left(1+e^{-E_{\bf k}^\alpha/T}\right)\right],
\end{eqnarray}
where $\xi_{\bf k}^\pm=\xi_{\bf k}\pm\eta_{\bf k}$. Here the term $\sum_{\alpha}\sum_{\bf k}\xi_{\bf k}^\alpha/2=\sum_{\bf k}\xi_{\bf k}$ is added to
recover the correct limit for $\Delta\rightarrow0$. The integral over the fermion momentum ${\bf k}$ is divergent and the contact coupling $U$ needs
to be regularized. For a short-range interaction potential with its $s$-wave scattering length $a_s$, it is natural to regularize $U$ by the two-body
T-matrix in the absence of SOC. We have
\begin{eqnarray}
\frac{1}{U}=-\frac{1}{4\pi a_s}+\sum_{\bf k}\frac{1}{2\epsilon_{\bf k}}.
\end{eqnarray}

The superfluid order parameter $\Delta$ should satisfy the saddle-point condition $\partial\Omega/\partial\Delta=0$, or the so-called gap equation
\begin{eqnarray}
\frac{1}{U}=\sum_{\bf k}\sum_{\alpha=\pm}\left(1+\alpha\frac{h^2}{\zeta_{\bf k}}\right)
\frac{1-2f(E_{\bf k}^\alpha)}{4E_{\bf k}^\alpha},
\end{eqnarray}
where $f(E)=1/(e^{E/T}+1)$ is the Fermi-Dirac distribution. Meanwhile, if the total fermion density $n$ is imposed, the chemical potential $\mu$ should
be determined by the number equation $-\partial\Omega/\partial\mu=n$, that is,
\begin{eqnarray}
n=\sum_{\bf k}\sum_{\alpha=\pm}\left[\frac{1}{2}-\left(1+\alpha\frac{\eta_{\bf k}^2}{\zeta_{\bf k}}\right)\xi_{\bf k}
\frac{1-2f(E_{\bf k}^\alpha)}{2E_{\bf k}^\alpha}\right].
\end{eqnarray}
In general, $\Delta$ and $\mu$ are obtained by solving the gap and number equations simultaneously. As a convention, we define the Fermi momentum
$k_{\rm F}$ through the noninteracting form $n=k_{\rm F}^3/(3\pi^2)$, and the Fermi energy is given by $\epsilon_{\rm F}=k_{\rm F}^2/2$.

In the Nambu-Gor'kov space, the fermion propagator ${\cal G}(K)$ takes the form
\begin{eqnarray}
{\cal G}(K)=\left(\begin{array}{cc}{\cal G}_{11}(K)&{\cal G}_{12}(K)\\ {\cal G}_{21}(K)& {\cal G}_{22}(K)\end{array}\right).
\end{eqnarray}
The matrix elements can be evaluated as
\begin{eqnarray}
&&{\cal G}_{11}(K)={\cal G}_-^{-1}(K)\frac{L_+(K)L_-(K)-\Delta^2}{[(i\omega_n)^2-(E_{\bf k}^+)^2][(i\omega_n)^2-(E_{\bf k}^-)^2]},\nonumber\\
&&{\cal G}_{22}(K)={\cal G}_+^{-1}(K)\frac{L_-(K)L_+(K)-\Delta^2}{[(i\omega_n)^2-(E_{\bf k}^+)^2][(i\omega_n)^2-(E_{\bf k}^-)^2]},\nonumber\\
&&{\cal G}_{12}(K)=-\Delta\frac{L_-(K)L_+(K)-\Delta^2}{[(i\omega_n)^2-(E_{\bf k}^+)^2][(i\omega_n)^2-(E_{\bf k}^-)^2]},\nonumber\\
&&{\cal G}_{21}(K)=-\Delta\frac{L_+(K)L_-(K)-\Delta^2}{[(i\omega_n)^2-(E_{\bf k}^+)^2][(i\omega_n)^2-(E_{\bf k}^-)^2]},
\end{eqnarray}
where $L_\pm(K)$ are given by
\begin{eqnarray}
L_\pm(K)=i\omega_n-h\sigma_z\pm(\xi_{\bf k}-\lambda\mbox{\boldmath{$\sigma$}}_\bot\cdot{\bf k}_\bot).
\end{eqnarray}
To evaluate the collective mode propagator, we also express the fermion propagator in an alternative form by using the following projectors
\begin{eqnarray}
{\cal P}_{\bf k}^\pm(h)=\frac{1}{2}\left(1\pm\frac{\lambda\mbox{\boldmath{$\sigma$}}_\bot\cdot{\bf k}_\bot+h\sigma_z}{\eta_{\bf k}}\right)
\end{eqnarray}
which possess the following properties
\begin{eqnarray}
{\cal P}_{\bf k}^+(h)+{\cal P}_{\bf k}^-(h)=1,\ \ \
{\cal P}_{\bf k}^\alpha(h){\cal P}_{\bf k}^\beta(h)=\delta_{\alpha\beta}{\cal P}_{\bf k}^\alpha(h).
\end{eqnarray}
With the help of these projectors, the fermion propagator can be expressed as
\begin{eqnarray}
&&{\cal G}_{11}(K)=\sum_{\alpha=\pm}(i\omega_n+\xi_{\bf k}^\alpha)
\frac{[(i\omega_n)^2-(\xi_{\bf k}^{-\alpha})^2]{\cal P}_{\bf k}^\alpha(-h)-\Delta^2{\cal P}_{\bf k}^\alpha(h)}
{[(i\omega_n)^2-(E_{\bf k}^+)^2][(i\omega_n)^2-(E_{\bf k}^-)^2]},\nonumber\\
&&{\cal G}_{22}(K)=\sum_{\alpha=\pm}(i\omega_n-\xi_{\bf k}^\alpha)
\frac{[(i\omega_n)^2-(\xi_{\bf k}^{-\alpha})^2]{\cal P}_{\bf k}^\alpha(h)-\Delta^2{\cal P}_{\bf k}^\alpha(-h)}
{[(i\omega_n)^2-(E_{\bf k}^+)^2][(i\omega_n)^2-(E_{\bf k}^-)^2]},\nonumber\\
&&{\cal G}_{12}(K)=-\Delta\sum_{\alpha=\pm}\frac{[(i\omega_n)^2-(\xi_{\bf k}^{-\alpha})^2-\Delta^2]{\cal P}_{\bf k}^\alpha(h)
-2h(i\omega_n+\xi_{\bf k}^{-\alpha})\sigma_z{\cal P}_{\bf k}^\alpha(h)}
{[(i\omega_n)^2-(E_{\bf k}^+)^2][(i\omega_n)^2-(E_{\bf k}^-)^2]},\nonumber\\
&&=-\Delta\sum_{\alpha=\pm}\frac{[(i\omega_n)^2-(\xi_{\bf k}^{-\alpha})^2-\Delta^2]{\cal P}_{\bf k}^\alpha(-h)
-2h(i\omega_n-\xi_{\bf k}^{-\alpha}){\cal P}_{\bf k}^\alpha(-h)\sigma_z}
{[(i\omega_n)^2-(E_{\bf k}^+)^2][(i\omega_n)^2-(E_{\bf k}^-)^2]},\nonumber\\
&&{\cal G}_{21}(K)=-\Delta\sum_{\alpha=\pm}\frac{[(i\omega_n)^2-(\xi_{\bf k}^{-\alpha})^2-\Delta^2]{\cal P}_{\bf k}^\alpha(h)
-2h(i\omega_n+\xi_{\bf k}^{-\alpha}){\cal P}_{\bf k}^\alpha(h)\sigma_z}
{[(i\omega_n)^2-(E_{\bf k}^+)^2][(i\omega_n)^2-(E_{\bf k}^-)^2]}\nonumber\\
&&=-\Delta\sum_{\alpha=\pm}\frac{[(i\omega_n)^2-(\xi_{\bf k}^{-\alpha})^2-\Delta^2]{\cal P}_{\bf k}^\alpha(-h)
-2h(i\omega_n-\xi_{\bf k}^{-\alpha})\sigma_z{\cal P}_{\bf k}^\alpha(-h)}
{[(i\omega_n)^2-(E_{\bf k}^+)^2][(i\omega_n)^2-(E_{\bf k}^-)^2]}.
\end{eqnarray}

We note that the anomalous Green's function ${\cal G}_{12}(K)$ is not diagonal in the spin space for $\lambda\neq0$. Therefore, the spin-orbit
coupling generates spin-triplet pairing even though the order parameter $\Delta$ has the $s$-wave symmetry. According to the Green's function
relation, the spin-singlet and spin-triplet pairing amplitudes can be read from the diagonal and off-diagonal components of ${\cal G}_{12}(K)$.
We obtain
\begin{eqnarray}
\langle\psi_\uparrow({\bf k})\psi_\downarrow({\bf k})\rangle
=-\langle\psi_\downarrow({\bf k})\psi_\uparrow({\bf k})\rangle
=\Delta\sum_{\alpha=\pm}\left(1+\alpha\frac{h^2}{\zeta_{\bf k}}\right)\frac{1-2f(E_{\bf k}^\alpha)}{4E_{\bf k}^\alpha}
\end{eqnarray}
for the spin-singlet pairing amplitudes and
\begin{eqnarray}
\langle\psi_\uparrow({\bf k})\psi_\uparrow({\bf k})\rangle&=&-\lambda(k_x-ik_y)\Delta\frac{\xi_{\bf k}+h}{\zeta_{\bf k}}
\sum_{\alpha=\pm}\alpha\frac{1-2f(E_{\bf k}^\alpha)}{4E_{\bf k}^\alpha},\nonumber\\
\langle\psi_\downarrow({\bf k})\psi_\downarrow({\bf k})\rangle&=&\lambda(k_x+ik_y)\Delta\frac{\xi_{\bf k}-h}{\zeta_{\bf k}}
\sum_{\alpha=\pm}\alpha\frac{1-2f(E_{\bf k}^\alpha)}{4E_{\bf k}^\alpha}
\end{eqnarray}
for the spin-triplet pairing amplitudes. In the mean-field theory, the condensation density $n_0$ is half of the summation of all pairing amplitudes
squared~\cite{FC,FCFM}, that is,
\begin{eqnarray}
n_0=\frac{1}{2}\sum_{\bf k}\sum_{\sigma,\sigma^\prime=\uparrow,\downarrow}|\langle\psi_\sigma({\bf k})\psi_{\sigma^\prime}({\bf k})\rangle|^2.
\end{eqnarray}

It is also useful to reexpress the mean-field theory in the helicity representation~\cite{TSC08}. The helicity basis $(\psi_+,\psi_-)^{\rm T}$ is
related to basis $(\psi_\uparrow,\psi_\downarrow)^{\rm T}$ by a ${\bf k}$-dependent SU$(2)$ transformation. In the helicity basis, $H_0+H_Z$ is
diagonal, that is
\begin{eqnarray}
\label{helicity1}
H_0+H_Z=\sum_{\bf k}\left[\xi_{\bf k}^+ \psi^\dagger_+({\bf k})\psi^{\phantom{\dag}}_+({\bf k})
+\xi_{\bf k}^-\psi_-^\dagger({\bf k})\psi^{\phantom{\dag}}_-({\bf k})\right].
\end{eqnarray}
Therefore, the spin-orbit coupled Fermi gas can be viewed as a two-band system. The Zeeman field provides a band gap $2h$ at ${\bf k}=0$. In the
presence of attraction, the mean-field approximation for $H_{\rm int}$ reads
\begin{eqnarray}
\label{helicity2}
H_{\rm int}\simeq\frac{1}{2}\sum_{\alpha,\beta=\pm}\sum_{\bf k}
\left[\Delta_{\alpha\beta}({\bf k})\psi_\alpha^\dagger({\bf k})\psi^\dagger_\beta({\bf k})+h.c.\right].
\end{eqnarray}
The new momentum-dependent pair potentials $\Delta_{\alpha\beta}({\bf k})$ read
\begin{eqnarray}
\label{helicity3}
\Delta_{+-}({\bf k})=-\Delta_{-+}({\bf k})=-\Delta_{\rm s}({\bf k}),\ \ \ \Delta_{++}({\bf k})=\Delta^*_{--}({\bf k})=-\Delta_{\rm t}({\bf k}),
\end{eqnarray}
where the interband and intraband pair potentials are given by
\begin{eqnarray}
\label{helicity4}
\Delta_{\rm s}({\bf k})=\frac{h}{\eta_{\bf k}}\Delta,\ \ \ \ \ \ \ \Delta_{\rm t}({\bf k})=\frac{\lambda(k_x-ik_y)}{\eta_{\bf k}}\Delta.
\end{eqnarray}
Using these new pair potentials, the quasiparticle spectra $E_{\bf k}^\pm$ can be expressed as
\begin{eqnarray}
\label{helicity5}
E_{\bf k}^\pm=\sqrt{\left(\sqrt{\xi_{\bf k}^2+|\Delta_{\rm s}({\bf k})|^2}\pm\eta_{\bf k}\right)^2+|\Delta_{\rm t}({\bf k})|^2}.
\end{eqnarray}
The above expressions in the helicity basis will help us understand some results in Sec. \ref{s5}.

%%%%%%%%%%%%%%%%%%%%%%%%%%%%%%%%%%%%%%%%%%%%%%%%%%%%%%%%%%%%%%%%%%%%%%%%%%%%%%%%%%%%%%%%%%%%%%%%%%%%%%%%%%%%%%
\subsection{Gaussian fluctuation: collective excitations }
%%%%%%%%%%%%%%%%%%%%%%%%%%%%%%%%%%%%%%%%%%%%%%%%%%%%%%%%%%%%%%%%%%%%%%%%%%%%%%%%%%%%%%%%%%%%%%%%%%%%%%%%%%%%%%
Then we consider the fluctuations around the mean field. The linear terms which are of order $O(\phi)$ vanish precisely once the saddle-point
condition $\Delta=\Delta_0$ is imposed. The quadratic terms, corresponding to the Gaussian fluctuations, can be evaluated as
\begin{eqnarray}
\frac{{\cal S}_{\rm{eff}}^{(2)}[\phi,\phi^\dag]}{\beta V}=\sum_Q\Bigg\{\frac{|\phi(Q)|^2}{U}
+\frac{1}{4}\sum_K{\rm{Tr}}\left[{\cal G}(K+Q)\Sigma(Q){\cal G}(K)\Sigma(-Q)\right]\Bigg\},
\end{eqnarray}
where $\Sigma(Q)$ is defined as
\begin{eqnarray}
\Sigma(Q)=\left(\begin{array}{cc}0&\phi(Q)\\ \phi^\dagger(-Q)& 0\end{array}\right).
\end{eqnarray}
In this paper $Q=(i\nu_n,{\bf q})$ denotes the energy and momentum of bosons with $\nu_n=2n\pi T$ being the boson Matsubara frequency.

After taking the trace in the Nambu-Gor'kov space, we find that ${\cal S}_{\rm{eff}}^{(2)}$ can be written in a bilinear form
\begin{eqnarray}
\frac{{\cal S}_{\rm{eff}}^{(2)}[\phi,\phi^\dag]}{\beta V}=\frac{1}{2}\sum_Q
\left(\begin{array}{cc} \phi^\dagger(Q) & \phi(-Q)\end{array}\right)
{\bf M}(Q)\left(\begin{array}{cc} \phi(Q)\\ \phi^\dagger(-Q)\end{array}\right),
\end{eqnarray}
where the inverse boson propagator ${\bf M}(Q)$ is a $2\times2$ matrix,
\begin{eqnarray}
{\bf M}(Q)=\left(\begin{array}{cc}{\bf M}_{11}(Q)&{\bf M}_{12}(Q)\\ {\bf M}_{21}(Q)& {\bf M}_{22}(Q)\end{array}\right).
\end{eqnarray}
The matrix elements of ${\bf M}(Q)$ can be expressed in terms of the fermion propagator ${\cal G}(K)$. We have
\begin{eqnarray}
&&{\bf M}_{11}(Q)=\frac{1}{U}+\frac{1}{2}\sum_K{\rm{Tr}}\left[{\cal G}_{11}(K+Q){\cal G}_{22}(K)\right],\nonumber\\
&&{\bf M}_{22}(Q)=\frac{1}{U}+\frac{1}{2}\sum_K{\rm{Tr}}\left[{\cal G}_{22}(K+Q){\cal G}_{11}(K)\right],\nonumber\\
&&{\bf M}_{12}(Q)=\frac{1}{2}\sum_K{\rm{Tr}}\left[{\cal G}_{12}(K+Q){\cal G}_{12}(K)\right],\nonumber\\
&&{\bf M}_{21}(Q)=\frac{1}{2}\sum_K{\rm{Tr}}\left[{\cal G}_{21}(K+Q){\cal G}_{21}(K)\right].
\end{eqnarray}
The explicit forms of these functions are evaluated in Appendix A. It is straightforward to show that ${\bf M}_{11}(i\nu_n,{\bf q})
={\bf M}_{22}(-i\nu_n,{\bf q})$. However, we have ${\bf M}_{12}(i\nu_n,{\bf q})\neq{\bf M}_{21}(i\nu_n,{\bf q})$ if the spin-orbit coupling $\lambda$
and the Zeeman field $h$ are both nonzero. Taking the analytical continuation $i\nu_n\rightarrow\omega+i0^+$, the dispersion $\omega({\bf q})$ of the
collective mode is determined by the equation
\begin{equation}
\det{{\bf M}[\omega({\bf q}), {\bf q}]}=0.
\end{equation}

We can decompose ${\bf M}_{11}(\omega,{\bf q})$ as ${\bf M}_{11}(\omega,{\bf q})={\bf M}_{11}^+(\omega,{\bf q})+{\bf M}_{11}^-(\omega,{\bf q})$, where
${\bf M}_{11}^+(\omega,{\bf q})$ and ${\bf M}_{11}^-(\omega,{\bf q})$ are even and odd functions of $\omega$, respectively. Their explicit
forms read
\begin{eqnarray}
{\bf M}_{11}^+(\omega,{\bf q})&=&\frac{1}{U}+\frac{1}{2}\sum_{\alpha,\beta=\pm}\sum_{\bf k}{\cal W}_1^{\alpha\beta}({\bf k},{\bf q})
\left[\frac{1}{\omega-E_{{\bf k}+{\bf p}}^\alpha-E_{{\bf k}-{\bf p}}^\beta}
-\frac{1}{\omega+E_{{\bf k}+{\bf p}}^\alpha+E_{{\bf k}-{\bf p}}^\beta}\right]\nonumber\\
&\times&\left[1-f(E_{{\bf k}+{\bf p}}^\alpha)-f(E_{{\bf k}-{\bf p}}^\beta)\right]\nonumber\\
&+&\frac{1}{2}\sum_{\alpha,\beta=\pm}\sum_{\bf k}{\cal U}_1^{\alpha\beta}({\bf k},{\bf q})
\left[\frac{1}{\omega+E_{{\bf k}+{\bf p}}^\alpha-E_{{\bf k}-{\bf p}}^\beta}
-\frac{1}{\omega-E_{{\bf k}+{\bf p}}^\alpha+E_{{\bf k}-{\bf p}}^\beta}\right]\nonumber\\
&\times&\left[f(E_{{\bf k}+{\bf p}}^\alpha)-f(E_{{\bf k}-{\bf p}}^\beta)\right]
\end{eqnarray}
and
\begin{eqnarray}
{\bf M}_{11}^-(\omega,{\bf q})&=&\frac{1}{2}\sum_{\alpha,\beta=\pm}\sum_{\bf k}{\cal W}_2^{\alpha\beta}({\bf k},{\bf q})
\left[\frac{1}{\omega-E_{{\bf k}+{\bf p}}^\alpha-E_{{\bf k}-{\bf p}}^\beta}
+\frac{1}{\omega+E_{{\bf k}+{\bf p}}^\alpha+E_{{\bf k}-{\bf p}}^\beta}\right]\nonumber\\
&\times&\left[1-f(E_{{\bf k}+{\bf p}}^\alpha)-f(E_{{\bf k}-{\bf p}}^\beta)\right]\nonumber\\
&+&\frac{1}{2}\sum_{\alpha,\beta=\pm}\sum_{\bf k}{\cal U}_2^{\alpha\beta}({\bf k},{\bf q})
\left[\frac{1}{\omega+E_{{\bf k}+{\bf p}}^\alpha-E_{{\bf k}-{\bf p}}^\beta}
+\frac{1}{\omega-E_{{\bf k}+{\bf p}}^\alpha+E_{{\bf k}-{\bf p}}^\beta}\right]\nonumber\\
&\times&\left[f(E_{{\bf k}+{\bf p}}^\alpha)-f(E_{{\bf k}-{\bf p}}^\beta)\right],
\end{eqnarray}
where ${\bf p}={\bf q}/2$ for convenience. On the other hand ${\bf M}_{12}(\omega,{\bf q})$ and ${\bf M}_{21}(\omega,{\bf q})$ are even functions
of $\omega$ and can be expressed as
\begin{eqnarray}
&&{\bf M}_{12}(\omega,{\bf q})={\bf M}_{12}^+(\omega,{\bf q})+i{\bf M}_{12}^-(\omega,{\bf q}),\nonumber\\
&&{\bf M}_{21}(\omega,{\bf q})={\bf M}_{12}^+(\omega,{\bf q})-i{\bf M}_{12}^-(\omega,{\bf q}).
\end{eqnarray}
Here ${\bf M}_{12}^+(\omega,{\bf q})$ and ${\bf M}_{12}^-(\omega,{\bf q})$ read
\begin{eqnarray}
{\bf M}_{12}^+(\omega,{\bf q})&=&-\frac{1}{2}\sum_{\alpha,\beta=\pm}\sum_{\bf k}{\cal W}_3^{\alpha\beta}({\bf k},{\bf q})
\left[\frac{1}{\omega-E_{{\bf k}+{\bf p}}^\alpha-E_{{\bf k}-{\bf p}}^\beta}
-\frac{1}{\omega+E_{{\bf k}+{\bf p}}^\alpha+E_{{\bf k}-{\bf p}}^\beta}\right]\nonumber\\
&\times&\left[1-f(E_{{\bf k}+{\bf p}}^\alpha)-f(E_{{\bf k}-{\bf p}}^\beta)\right]\nonumber\\
&+&\frac{1}{2}\sum_{\alpha,\beta=\pm}\sum_{\bf k}{\cal U}_3^{\alpha\beta}({\bf k},{\bf q})
\left[\frac{1}{\omega+E_{{\bf k}+{\bf p}}^\alpha-E_{{\bf k}-{\bf p}}^\beta}
-\frac{1}{\omega-E_{{\bf k}+{\bf p}}^\alpha+E_{{\bf k}-{\bf p}}^\beta}\right]\nonumber\\
&\times&\left[f(E_{{\bf k}+{\bf p}}^\alpha)-f(E_{{\bf k}-{\bf p}}^\beta)\right]
\end{eqnarray}
and
\begin{eqnarray}
{\bf M}_{12}^-(\omega,{\bf q})&=&-\frac{1}{2}\sum_{\alpha,\beta=\pm}\sum_{\bf k}{\cal W}_4^{\alpha\beta}({\bf k},{\bf q})
\left[\frac{1}{\omega-E_{{\bf k}+{\bf p}}^\alpha-E_{{\bf k}-{\bf p}}^\beta}
-\frac{1}{\omega+E_{{\bf k}+{\bf p}}^\alpha+E_{{\bf k}-{\bf p}}^\beta}\right]\nonumber\\
&\times&\left[1-f(E_{{\bf k}+{\bf p}}^\alpha)-f(E_{{\bf k}-{\bf p}}^\beta)\right]\nonumber\\
&+&\frac{1}{2}\sum_{\alpha,\beta=\pm}\sum_{\bf k}{\cal U}_4^{\alpha\beta}({\bf k},{\bf q})
\left[\frac{1}{\omega+E_{{\bf k}+{\bf p}}^\alpha-E_{{\bf k}-{\bf p}}^\beta}
-\frac{1}{\omega-E_{{\bf k}+{\bf p}}^\alpha+E_{{\bf k}-{\bf p}}^\beta}\right]\nonumber\\
&\times&\left[f(E_{{\bf k}+{\bf p}}^\alpha)-f(E_{{\bf k}-{\bf p}}^\beta)\right].
\end{eqnarray}
The explicit expressions of the functions ${\cal W}_i^{\alpha\beta}({\bf k},{\bf q})$ and ${\cal U}_i^{\alpha\beta}({\bf k},{\bf q})$
($i=1,2,3,4$) are presented in Appendix A. We note that ${\cal W}_4^{\alpha\beta}({\bf k},{\bf q})$ and
${\cal U}_4^{\alpha\beta}({\bf k},{\bf q})$ are odd functions of $h$, that is, they are proportional to $h\lambda^2$. However, the determinant
of the matrix ${\bf M}$ is an even function of $h$, as we expect.

To make the results more physically transparent, we decompose the complex fluctuation field $\phi(x)$ into its amplitude part $\rho(x)$ and
phase part $\theta(x)$, $\phi(x)=\rho(x)+i\Delta_0\theta(x)$. Converting to the variables $\rho(x)$ and $\theta(x)$, we obtain
\begin{equation}
\frac{{\cal S}_{\rm{eff}}^{(2)}[\rho,\theta]}{\beta V}=\frac{1}{2}\sum_Q\left(\begin{array}{cc} \rho(-Q)&\theta(-Q)\end{array}\right)
{\bf N}(Q)\left(\begin{array}{c} \rho(Q)\\ \theta(Q)\end{array}\right),
\end{equation}
where the matrix ${\bf N}(Q)$ reads
\begin{equation}
{\bf N}(Q)=2\left(\begin{array}{cc} {\bf M}_{11}^++{\bf M}_{12}^+&i\Delta_0({\bf M}_{11}^--i{\bf M}_{12}^-)\\
-i\Delta_0({\bf M}_{11}^-+i{\bf M}_{12}^-) & \Delta_0^2({\bf M}_{11}^+-{\bf M}_{12}^+)\end{array}\right).
\end{equation}
From the expressions of ${\bf M}_{11}^-$ and ${\bf M}_{12}^-$, we have ${\bf M}_{11}^-(0,{\bf q})=0$ and ${\bf M}_{12}^-(\omega, {\bf 0})=0$.
Therefore the amplitude and phase modes decouple completely at $\omega=0$ and ${\bf q}=0$. Furthermore, at the saddle point $\Delta=\Delta_0$
we have precisely
\begin{eqnarray}
{\bf M}_{11}^+(0,{\bf 0})={\bf M}_{12}^+(0,{\bf 0}).
\end{eqnarray}
Therefore the phase mode at ${\bf q}=0$ is gapless, that is, the Goldstone mode. For neutral fermionic superfluids, this mode is also called the
Anderson-Bogoliubov mode. Another collective mode or the so-called Higgs mode is massive. It is likely heavily damped since its mass gap is generally
larger than the two-particle continuum at ${\bf q}=0$.

We are interested in the low energy behaviors of these collective modes. For this purpose, we make a small ${\bf q}$ and $\omega$ expansion of
${\bf N}(Q)$. The spin-orbit coupling breaks the O$(3)$ rotational symmetry to a O$(2)$ circular rotational symmetry. Therefore, the expansion takes
the form
\begin{eqnarray}
2({\bf M}_{11}^++{\bf M}_{12}^+) &=& A+C_\bot{\bf q}_\bot^2+C_\|{\bf q}_\|^2-D\omega^2+\cdots,\nonumber\\
2\Delta_0^2({\bf M}_{11}^+-{\bf M}_{12}^+) &=& J_\bot{\bf q}_\bot^2+J_\|{\bf q}_\|^2-R\omega^2+\cdots,\nonumber\\
2\Delta_0{\bf M}_{11}^- &=& -B\omega+\cdots. \label{Expansion}
\end{eqnarray}
Here ${\bf a}_\|=a_z{\bf e}_z$ for any vector ${\bf a}$. Note that the term ${\bf M}_{12}^-(\omega,{\bf q})$ has no contribution to this expansion up
to the order $O(\omega^2, {\bf q}^2)$. We should emphasize that such an expansion is only possible at zero temperature, since the terms proportional
to $f(E_{\bf k+p}^\alpha)-f(E_{\bf k-p}^\alpha)$ have the Landau theory singularity for ${\bf q}$ and $\omega$ going to zero~\cite{BCSBEC3}. In the
remaining of this paper, we restrict our studies to the zero temperature case.

The parameter $A$ ($J_\bot$ and $J_\|$) characterizes the stability of the saddle point $\Delta=\Delta_0$ against the amplitude (phase) fluctuation.
First, it is easy to show that
\begin{equation}
A=\frac{\partial^2\Omega(\Delta)}{\partial\Delta^2}\bigg|_{\Delta=\Delta_0}.
\end{equation}
Therefore, the stability against the amplitude fluctuation requires $A>0$. Second, the superfluid phase stiffness $J_\bot$ ($J_\|$) is precisely
proportional to the superfluid density $n_s^\bot$ ($n_s^\|$). We have
\begin{equation}
J_\bot=\frac{n_s^\bot}{4m},\ \ \ \ \ J_\|=\frac{n_s^\|}{4m}.
\end{equation}
On the other hand, the superfluid density can be obtained from another equivalent definition \cite{NS01,NS02}. When the superfluid moves with a uniform
velocity $\mbox{\boldmath{$\upsilon$}}_s=\mbox{\boldmath{$\upsilon$}}_\bot{\bf e}_\bot+\mbox{\boldmath{$\upsilon$}}_\|{\bf e}_z$, the superfluid
order parameter transforms like $\Phi\rightarrow \Phi e^{2im\mbox{\boldmath{$\upsilon$}}_s\cdot {\bf r}}$ and
$\Phi^*\rightarrow \Phi^* e^{-2im\mbox{\boldmath{$\upsilon$}}_s\cdot {\bf r}}$ ($m=1$ in our units). The superfluid density $n_s$ is defined as the
response of the thermodynamic potential $\Omega$ to an infinitesimal velocity $\mbox{\boldmath{$\upsilon$}}_s$, i.e.,
\begin{eqnarray}
\Omega(\mbox{\boldmath{$\upsilon$}}_s)=\Omega({\bf 0})+\frac{1}{2}n_s^\bot\mbox{\boldmath{$\upsilon$}}_\bot^2
+\frac{1}{2}n_s^\|\mbox{\boldmath{$\upsilon$}}_\|^2+O(\mbox{\boldmath{$\upsilon$}}_s^4).
\end{eqnarray}
Therefore, the stability against the phase fluctuation requires $J_\bot>0$ and $J_\|>0$. Once the superfluid phase stiffness becomes negative, the
saddle-point state is unstable and some Fulde-Ferrell-Larkin-Ovchinnikov-like state with inhomogeneous phase and/or amplitude modulation will be
energetically favored~\cite{NSI01,NSI02,NSI03,NSI04,NSI05,NSI06}.

As long as the stability conditions are satisfied, the dispersion of the Goldstone mode at small momentum is given by
\begin{equation}
\omega({\bf q})=\sqrt{(c_s^\bot)^2{\bf q}_\bot^2+(c_s^\|)^2{\bf q}_\|^2},
\end{equation}
where the transverse and longitudinal sound velocities are given by
\begin{equation}
c_s^\bot=\sqrt{J_\bot\over B^2/A+R},\ \ \ \ c_s^\|=\sqrt{J_\|\over B^2/A+R}.
\end{equation}
The Higgs mode is massive and its mass gap $M_{\rm H}$ reads
\begin{eqnarray}
M_{\rm H}=\sqrt{\frac{B^2+AR}{DR}}.
\end{eqnarray}
We note that the expansion (\ref{Expansion}) is valid only for small frequency $\omega$, therefore this formula only gives the qualitative behavior of
the Higgs mode. In general, the Higgs mode is a resonance since its spectral density arises above the two-particle continuum
$E_{\rm c}({\bf q})=\min_{\bf k}\{E_{{\bf k}+{\bf q}/2}^-+E_{{\bf k}-{\bf q}/2}^-\}$.

Finally, we summarize the explicit expressions for the expansion parameters in Eq. (\ref{Expansion}). For details of the calculations, see Appendix B.
First, $A$ can be evaluated as
\begin{equation}
A=\frac{1}{2}\sum_{\alpha=\pm}\sum_{\bf k}\left[\frac{\Delta^2}{(E_{\bf k}^\alpha)^3}
\left(1+\alpha\frac{h^2}{\zeta_{\bf k}}\right)^2+\alpha\frac{h^4\Delta^2}{E_{\bf k}^\alpha\zeta_{\bf k}^3}
-2\frac{\Delta^2}{(E_{\bf k}^\alpha)^2}\left(1+\alpha\frac{h^2}{\zeta_{\bf k}}\right)^2\delta(E_{\bf k}^\alpha)\right].
\end{equation}
The expansion parameters in the frequency expansion read
\begin{eqnarray}
B&=&\Delta\sum_{\bf k}\frac{\xi_{\bf k}}{(E_{\bf k}^++E_{\bf k}^-)^2}\left[\left(\frac{1}{E_{\bf k}^+}+\frac{1}{E_{\bf k}^-}\right)
\frac{h^2E_{\bf k}^2}{\zeta_{\bf k}^2}+\left(\frac{1}{E_{\bf k}^+}-\frac{1}{E_{\bf k}^-}\right)\frac{h^2}{\zeta_{\bf k}}\right]\nonumber\\
&+&\frac{\Delta}{4}\sum_{\alpha\pm}\sum_{\bf k}\frac{\xi_{\bf k}}{(E_{\bf k}^\alpha)^3}
\left(1+\alpha\frac{\lambda^2{\bf k}_\bot^2}{\zeta_{\bf k}}-\frac{h^2E_{\bf k}^2}{\zeta_{\bf k}^2}\right),\nonumber\\
D&=&\frac{1}{8}\sum_{\alpha=\pm}\sum_{\bf k}\left[\frac{\xi_{\bf k}^2+\eta_{\bf k}^2+2\alpha\zeta_{\bf k}}{(E_{\bf k}^\alpha)^5}
\frac{\lambda^2{\bf k}_\bot^2\xi_{\bf k}^2}{\zeta_{\bf k}^2}+\frac{\Delta^2}{(E_{\bf k}^\alpha)^5}
\frac{\lambda^2{\bf k}_\bot^2h^2}{\zeta_{\bf k}^2}\right]\nonumber\\
&+&\sum_{\bf k}\frac{1}{(E_{\bf k}^++E_{\bf k}^-)^3}\frac{h^2\xi_{\bf k}^2}{\zeta_{\bf k}^2}
\left(1+\frac{E_{\bf k}^2-\eta_{\bf k}^2}{E_{\bf k}^+E_{\bf k}^-}\right),\nonumber\\
R&=&\Delta^2\sum_{\bf k}\frac{1}{(E_{\bf k}^++E_{\bf k}^-)^3}\frac{h^2E_{\bf k}^2}{\zeta_{\bf k}^2}
\left(1+\frac{E_{\bf k}^2-\eta_{\bf k}^2}{E_{\bf k}^+E_{\bf k}^-}
+\frac{2\lambda^2{\bf k}_\bot^2\Delta^2}{E_{\bf k}^+E_{\bf k}^-E_{\bf k}^2}\right)\nonumber\\
&+&\frac{\Delta^2}{8}\sum_{\alpha=\pm}\sum_{\bf k}\frac{1}{(E_{\bf k}^\alpha)^3}
\frac{\lambda^2{\bf k}_\bot^2\xi_{\bf k}^2}{\zeta_{\bf k}^2}.
\end{eqnarray}
The transverse phase stiffness $J_\bot$ and the longitudinal one $J_\|$ take the form
\begin{eqnarray}
J_\bot&=&\frac{1}{4m}\Bigg\{n-\sum_{\bf k}\sum_{\alpha=\pm}\frac{{\bf k}_\bot^2}{2}
\left(1+\alpha\frac{\lambda^2\xi_{\bf k}}{\zeta_{\bf k}}\right)^2\delta(E_{\bf k}^\alpha)\nonumber\\
&-&\sum_{\bf k}\sum_{\alpha=\pm}\frac{\lambda^2}{2E_{\bf k}^\alpha}\left[\alpha\left(1+\frac{h^2E_{\bf k}^2}{\zeta_{\bf k}^2}
+\frac{\lambda^2{\bf k}_\bot^2h^2\Delta^2}{\zeta_{\bf k}^2E_{\bf k}^2}\right)\frac{E_{\bf k}^2}{2\zeta_{\bf k}}
+\left(1-\frac{\lambda^2{\bf k}_\bot^2\xi_{\bf k}^2}{2\zeta_{\bf k}^2}\right)\right]\Bigg\},\nonumber\\
J_\|&=&\frac{1}{4m}\left[n-\sum_{\bf k}\sum_{\alpha=\pm}{\bf k}_\|^2\delta(E_{\bf k}^\alpha)\right].
\end{eqnarray}
The delta functions $\delta(E_{\bf k}^\alpha)$ in the expressions of $J_\bot$,$J_\|$ and $A$ come from the zero-temperature limit of the function
$(1/4T){\rm sech}^2(E_{\bf k}^\alpha/2T)=-\partial f(E_{\bf k}^\alpha)/\partial E_{\bf k}^\alpha$. The integrations over these delta functions vanish
precisely when the excitation spectra $E_{\bf k}^\alpha$ are fully gapped. However, at large enough Zeeman field $h$, the superfluid state may become
gapless where the lower excitation spectrum $E_{\bf k}^-$ has zeros. In this case, these terms may have finite contributions, depending on whether
these zeros correspond to gapless Fermi surfaces~\cite{NSI06}. For vanishing spin-orbit coupling, the expansion parameters $A,B,D,R$ reduce to the
known expressions in the previous studies~\cite{Gubankova,Lamacraft}. They are given by
\begin{eqnarray}
&&A=\sum_{\bf k}\frac{\Delta^2}{E_{\bf k}^2}\left[\frac{\Theta(E_{\bf k}-h)}{E_{\bf k}}-\delta(E_{\bf k}-h)\right],
\ \ \ \ B=\frac{\Delta}{2}\sum_{\bf k}\frac{\xi_{\bf k}}{E_{\bf k}^3}\Theta(E_{\bf k}-h),\nonumber\\
&&D=\frac{1}{4}\sum_{\bf k}\frac{\xi_{\bf k}^2}{E_{\bf k}^5}\Theta(E_{\bf k}-h),\ \ \ \ \ \ \ \ \ \ \ \ \ \ \ \ \ \ \ \ \ \ \ \ \
R=\frac{\Delta^2}{4}\sum_{\bf k}\frac{1}{E_{\bf k}^3}\Theta(E_{\bf k}-h).
\end{eqnarray}
The phase stiffness becomes isotropic for $\lambda=0$. It reads
\begin{eqnarray}
J_\bot=J_\|=\frac{1}{4m}\left[n-\sum_{\bf k}\frac{{\bf k}^2}{3}\delta(E_{\bf k}-h)\right].
\end{eqnarray}
Therefore, for $\lambda=0$ the zeros of $E_{\bf k}^-$ correspond to some gapless Fermi surfaces, which leads to large negative contributions to the
parameters $J_\bot,J_\|$ and $A$~\cite{NSI01,NSI02,NSI03,NSI04,NSI05,NSI06,NSI07,Gubankova,Lamacraft}.

%%%%%%%%%%%%%%%%%%%%%%%%%%%%%%%%%%%%%%%%%%%%%%%%%%%%%%%%%%%%%%%%%%%%%%%%%%%%%%%%%%%%%%%%%%%%%%%%%%%%%%%%%%%%%%
\subsection{Two-body problem}
%%%%%%%%%%%%%%%%%%%%%%%%%%%%%%%%%%%%%%%%%%%%%%%%%%%%%%%%%%%%%%%%%%%%%%%%%%%%%%%%%%%%%%%%%%%%%%%%%%%%%%%%%%%%%%
To understand the behavior for the collective modes in the BCS-BEC crossover, it is useful to compare it with the result of the two-body problem in
the absence of medium effect. In the functional path integral formalism, the two-body vertex function $\Gamma^{-1}(Q)$ can be obtained from its
coordinate representation defined as
\begin{eqnarray}
\Gamma^{-1}(x,x^\prime)= \frac{1}{\beta V}\frac{\delta^2{\cal S}_{\rm{eff}}[\Phi, \Phi^{\ast}]}
{\delta \Phi^{\ast}(x)\delta\Phi(x^\prime)}\bigg|_{\Phi=0}.
\end{eqnarray}

For $\Delta=0$, we have ${\cal G}_{12}={\cal G}_{21}=0$ and the
single-particle Green's function ${\cal G}(K)$ reduces to the form
\begin{eqnarray}
{\cal G}(K)=\left(\begin{array}{cc}{\cal G}_{+}(K)&0\\ 0& {\cal G}_{-}(K)\end{array}\right),
\end{eqnarray}
where the diagonal elements read
\begin{eqnarray}
{\cal G}_{+}(i\omega_n,{\bf k})=\sum_{\alpha=\pm}\frac{{\cal P}_{\bf k}^\alpha(-h)}{i\omega_n-\xi_{\bf k}^\alpha},\ \ \ \ \
{\cal G}_{-}(i\omega_n,{\bf k})=\sum_{\alpha=\pm}\frac{{\cal P}_{\bf k}^\alpha(h)}{i\omega_n+\xi_{\bf k}^\alpha}.
\end{eqnarray}
Therefore, $\Gamma^{-1}(Q)$ can be expressed as
\begin{eqnarray}
\Gamma^{-1}(Q)=\frac{1}{U}+\frac{1}{2}\sum_K{\rm{Tr}}\left[{\cal G}_{+}(K+Q){\cal G}_{-}(K)\right].
\end{eqnarray}
The explicit expression can be evaluated as
\begin{eqnarray}
\Gamma^{-1}(Q)=\frac{1}{U}-\frac{1}{2}\sum_{\alpha,\beta=\pm}\sum_{\bf k}
\frac{1-f(\xi_{{\bf k}+{\bf p}}^\alpha)-f(\xi_{{\bf k}-{\bf p}}^\beta)}
{\xi_{{\bf k}+{\bf p}}^\alpha+\xi_{{\bf k}-{\bf p}}^\beta-i\nu_n}{\cal T}_{\bf kq}^{\alpha\beta},
\end{eqnarray}
where
\begin{eqnarray}
{\cal T}_{\bf kq}^{\alpha\beta}=\frac{1}{2}\left[1+\alpha\beta\frac{\lambda^2({\bf k}_\bot^2-{\bf p}_\bot^2)-h^2}
{\eta_{{\bf k}+{\bf p}}\eta_{{\bf k}-{\bf p}}}\right].
\end{eqnarray}

To study the two-body problem in the absence of medium effect, we discard the Fermi-Dirac distribution functions. The energy-momentum dispersion
$\omega_{\bf q}$ of the pair excitation is defined as the solution $\omega+2\mu=\omega_{\bf q}$ of the equation
${\rm Re}\Gamma^{-1}(\omega+i\epsilon,{\bf q})=0$ . For $h=0$, after some manipulations, the two-body equation becomes
\begin{eqnarray}
\label{twobody}
\frac{1}{U}=\sum_{\bf k}\frac{{\cal E}_{\bf kq}}{{\cal E}_{\bf kq}^2-4\lambda^2{\bf k}_\bot^2
\left[1+\frac{\lambda^2{\bf q}_\bot^2\sin^2\varphi}{{\cal E}_{\bf kq}^2-\lambda^2{\bf q}_\bot^2}\right]},
\end{eqnarray}
where ${\cal E}_{\bf kq}={\bf k}^2+{\bf q}^2/4-\omega_{\bf q}$ and $\varphi$ is the angle between ${\bf k}_\bot$ and ${\bf q}_\bot$.

%%%%%%%%%%%%%%%%%%%%%%%%%%%%%%%%%%%%%%%%%%%%%%%%%%%%%%%%%%%%%%%%%%%%%%%%%%%%%%%%%%%%%%%%%%%%%%%%%%%%%%%%%%%%%%
\subsection{Two-dimensional case}
%%%%%%%%%%%%%%%%%%%%%%%%%%%%%%%%%%%%%%%%%%%%%%%%%%%%%%%%%%%%%%%%%%%%%%%%%%%%%%%%%%%%%%%%%%%%%%%%%%%%%%%%%%%%%%

The above formalism applies also to the case of two spatial dimensions. To apply the general formalism, we only need to freeze the longitudinal ($z$)
degree of freedom, that is, perform the following replacement:
\begin{eqnarray}
{\bf k}^2\rightarrow {\bf k}^2={\bf k}_\bot^2,\ \ \ \ \ \
\sum_{\bf k}\rightarrow \sum_{\bf k}=\int \frac{d^2{\bf k}}{(2\pi)^2}.
\end{eqnarray}
In contrast to the anisotropic 3D case, the superfluid ground state in the 2D case is isotropic. A quasi-2D cold atomic gas can be realized by arranging
a one-dimensional optical lattice along the axial ($z$) direction and a weak harmonic trapping potential in the radial ($x-y$) plane~\cite{2Dfermion},
such that atoms are strongly confined along the axial direction and form a series of pancake-shaped quasi-2D clouds. The strong anisotropy of the trapping
potentials, namely $\omega_{z}\gg \omega_\bot$ where $\omega_z$ ($\omega_\bot$) is the axial (radial) frequency, allows us to use an effective 2D
Hamiltonian to deal with the radial degrees of freedom.

For the 2D case, the two-body bound state exists for arbitrarily weak attraction. The coupling constant $U$ should be regularized in the following
way~\cite{BCSBEC4}
\begin{eqnarray}
\frac{1}{U}=\sum_{\bf k}\frac{1}{2\epsilon_{\bf k}+\epsilon_{\rm B}},
\end{eqnarray}
where $\epsilon_{\rm B}$ is the binding energy of the two-body bound state in the absence of spin-orbit coupling. For convenience, we define a 2D
scattering length $a_{2\rm D}$ by~\cite{2Dsl}
\begin{eqnarray}
\epsilon_{\rm B}=\frac{4e^{-2\gamma}}{ma_{2\rm D}^2},
\end{eqnarray}
where $\gamma\simeq0.577216$ is Euler's constant. For quasi-2D cold atoms confined by an axial trapping frequency $\omega_z$, the binding energy is
related to the the 3D $s$-wave scattering length $a_s$ by $\epsilon_{\rm B}=(C\hbar\omega_z/\pi)\exp(\sqrt{2\pi}l_z/a_s)$~\cite{Petrov}, where
$l_z=\sqrt{\hbar/\omega_z}$ and $C\simeq0.915$.

%%%%%%%%%%%%%%%%%%%%%%%%%%%%%%%%%%%%%%%%%%%%%%%%%%%%%%%%%%%%%%%%%%%%%%%%%%%%%%%%%%%%%%%%%%%%%%%%%%%%%%%%%%%%%%
\section{Results for zero Zeeman field: 3D case}\label{s3}
%%%%%%%%%%%%%%%%%%%%%%%%%%%%%%%%%%%%%%%%%%%%%%%%%%%%%%%%%%%%%%%%%%%%%%%%%%%%%%%%%%%%%%%%%%%%%%%%%%%%%%%%%%%%%%
In this section, we study the 3D system with zero Zeeman field ($h=0$). We focus on some bulk superfluid properties and the collective modes from
weak to strong spin-orbit coupling $\lambda$. Some results for the bound state and the BCS-BEC crossover have been addressed in the previous
studies~\cite{SOC-BCSBEC,SOC-Hu,SOC-Yu,3Dbound}. Here we present these results for the sake of completeness.

%%%%%%%%%%%%%%%%%%%%%%%%%%%%%%%%%%%%%%%%%%%%%%%%%%%%%%%%%%%%%%%%%%%%%%%%%%%%%%%%%%%%%%%%%%%%%%%%%%%%%%%%%%%%%%
\subsection{Bound state and BCS-BEC crossover}
%%%%%%%%%%%%%%%%%%%%%%%%%%%%%%%%%%%%%%%%%%%%%%%%%%%%%%%%%%%%%%%%%%%%%%%%%%%%%%%%%%%%%%%%%%%%%%%%%%%%%%%%%%%%%%
First, we show that there exists a two-body bound state in the presence of spin-orbit coupling even for negative values of $a_s$~\cite{3Dbound}.
For small ${\bf q}$, the dispersion $\omega_{\bf q}$ in Eq. (\ref{twobody}) can be written as $\omega_{\bf q}=-E_{\rm B}
+{\bf q}_\bot^2/(2M_{\rm B}^\bot)+{\bf q}_\|^2/(2M_{\rm B}^\|)$. From the imaginary part of the retarded Green's function
$\Gamma^{-1}(\omega+i\epsilon,{\bf q})$, we conclude that the bound state exists if the binding energy $E_{\rm B}>\lambda^2$. The binding energy
$E_{\rm B}$ is determined by
the equation
\begin{eqnarray}
-\frac{1}{4\pi a_s}=\sum_{\bf k}\left[\frac{{\bf k}^2+E_{\rm B}}{({\bf k}^2+E_{\rm B})^2-4\lambda^2{\bf k}_\bot^2}-\frac{1}{{\bf k}^2}\right].
\end{eqnarray}
Using the condition $E_{\rm B}>\lambda^2$ we find that the solution can be expressed as
\begin{eqnarray}
\frac{E_{\rm B}}{\lambda^2}=1+\eta(\kappa),
\end{eqnarray}
where $\kappa=1/(\lambda a_s)$. Completing the integral we obtain
\begin{eqnarray}
\sqrt{1+\eta(\kappa)}-\frac{1}{2}\ln\frac{\sqrt{1+\eta(\kappa)}+1}{\sqrt{1+\eta(\kappa)}-1}=\kappa.
\end{eqnarray}
Solving this equation, we find that a positive solution for $\eta$ always exists for arbitrary $\kappa$. The solution behaves as
$\eta(\kappa)\rightarrow e^{2\kappa}$ for $\kappa\rightarrow-\infty$ and $\eta(\kappa)\rightarrow\kappa^2-1$ for $\kappa\rightarrow+\infty$.
Therefore, the bound state can form in the presence of spin-orbit coupling even for negative values of $a_s$.

\begin{figure}[!htb]
\begin{center}
\includegraphics[width=10cm]{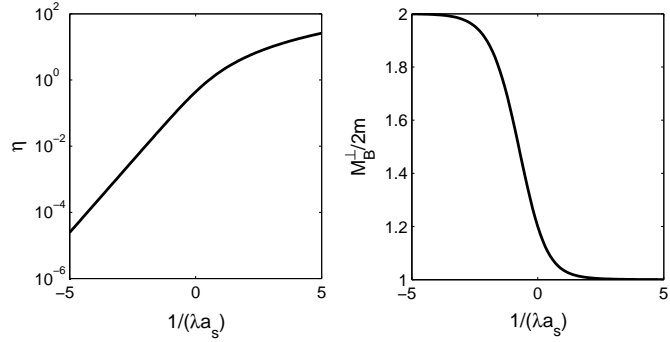}
\caption{The binding energy $E_{\rm B}$ (we show the dimensionless quantity
$\eta$) and the transverse effective mass $M_{\rm B}^\perp$ (divided by $2m$) as functions of $1/(\lambda a_s)$.
 \label{rashbon3D}}
\end{center}
\end{figure}

Meanwhile, expanding Eq. (\ref{twobody}) to the order $O({\bf q}^2)$, we obtain $M_{\rm B}^\|=2m$ for arbitrary $\kappa$ and the equation for
the transverse effective mass $M_{\rm B}^\bot$,
\begin{eqnarray}
\left(1-\frac{2m}{M_{\rm B}^\bot}\right)\sum_{\bf k}\frac{({\bf k}^2+E_{\rm B})^2+4\lambda^2{\bf k}_\bot^2}
{[({\bf k}^2+E_{\rm B})^2-4\lambda^2{\bf k}_\bot^2]^2}
=\sum_{\bf k}\frac{8\lambda^4{\bf k}_\bot^2}{({\bf k}^2+E_{\rm B})[({\bf k}^2+E_{\rm B})^2-4\lambda^2{\bf k}_\bot^2]^2}.
\end{eqnarray}
Completing the integral, we get
\begin{eqnarray}
\frac{2m}{M_{\rm B}^\bot}=1-\frac{1}{2}\left[\frac{\eta(\kappa)}{1+\eta(\kappa)}
\ln\frac{\eta(\kappa)}{1+\eta(\kappa)}+\frac{1}{1+\eta(\kappa)}\right].
\end{eqnarray}
The $\kappa$-dependence of the binding energy and the transverse effective mass can be numerically obtained. The results are shown in
Fig. \ref{rashbon3D}. We find that the transverse effective mass $M_{\rm B}^\bot$ is always larger than $2m$, and it approaches $4m$ for
$\kappa\rightarrow-\infty$. At unitary ($a_s\rightarrow\pm\infty$) or for large spin-orbit coupling ($\lambda\rightarrow\infty$), we have
$\kappa\rightarrow0$. In this case, the binding energy and the transverse effective mass read
\begin{eqnarray}
\eta(0)=0.439,\ \ \ \frac{M_{\rm B}^\bot}{2m}=1.20.
\end{eqnarray}
This novel bound state is referred to as rashbons in the studies~\cite{rashbon}.

%%%%%%%%%%%%%%%%%%%%%%%%%%%%%%%%%%%%%%%%%%%%%%%%%%%%%%%%%%%%%%%%%%%%%%%
%%%%%%%%%%%%%%%%%%%%%%%%%%%%%%%%%%%%%%%%%%%%%%%%%%%%%%%%%%%%%%%%%%%%%%%%

Then we turn to the superfluid state. For zero Zeeman field, the single particle excitation spectra reduce to
$E_{\bf k}^\pm=\sqrt{(\xi_{\bf k}\pm\lambda k_\bot)^2+\Delta^2}$. The fermion Green's function takes the following form
\begin{eqnarray}
&&{\cal G}_{11}(i\omega_n,{\bf k})=-{\cal G}_{22}(-i\omega_n,{\bf k})=\sum_{\alpha=\pm}
\frac{i\omega_n+\xi_{\bf k}^\alpha}{(i\omega_n)^2-(E_{\bf k}^\alpha)^2}{\cal P}_{\bf k}^\alpha(0),\nonumber\\
&&{\cal G}_{12}(i\omega_n,{\bf k})={\cal G}_{21}(i\omega_n,{\bf k})=\sum_{\alpha=\pm}
\frac{-\Delta}{(i\omega_n)^2-(E_{\bf k}^\alpha)^2}{\cal P}_{\bf k}^\alpha(0),
\end{eqnarray}
where $\xi_{\bf k}^\pm=\xi_{\bf k}\pm\lambda k_\bot$ and the projectors become
${\cal P}_{\bf k}^\pm(0)=\frac{1}{2}\left(1\pm\mbox{\boldmath{$\sigma$}}_\bot\cdot{\bf k}_\bot/k_\bot\right)$. Since the total density
$n=k_{\rm F}^3/(3\pi^2)$ is fixed, the system can be characterized by two dimensionless parameters  $1/(k_{\rm F}a_s)$ and $\lambda/k_{\rm F}$.
The order parameter $\Delta$ and the chemical potential $\mu$ can be determined in units of the Fermi energy $\epsilon_{\rm F}=k_{\rm F}^2/2$ from
the gap and number equations. For $h=0$, they become \begin{eqnarray}
-\frac{1}{4\pi a_s}=\sum_{\bf k}\left[\sum_{\alpha=\pm}\frac{1}{4\sqrt{(\xi_{\bf k}+\alpha\lambda k_\bot)^2+\Delta^2}}-\frac{1}{{\bf k}^2}\right]
\end{eqnarray}
and
\begin{eqnarray}
n=\frac{1}{2}\sum_{\alpha=\pm}\sum_{\bf k}\left[1-\frac{\xi_{\bf k}+\alpha\lambda k_\bot}
{\sqrt{(\xi_{\bf k}+\alpha\lambda k_\bot)^2+\Delta^2}}\right].
\end{eqnarray}
Applying the transformation $k_\bot\rightarrow k_\bot\pm\lambda$, the above equations can be written in another form,
\begin{eqnarray}
-\frac{1}{4\pi a_s}=\sum_{\bf k}\left(\frac{1}{2E_{\bf k}}-\frac{1}{{\bf k}^2}\right)+\frac{\lambda}{4\pi^2}
\int_0^\infty dk_z\int_0^\lambda \frac{dk_\bot}{\sqrt{(\xi_{\bf k}-\lambda^2/2)^2+\Delta^2}}
\end{eqnarray}
and
\begin{eqnarray}
n=\sum_{\bf k}\left(1-\frac{\xi_{\bf k}}{E_{\bf k}}\right)+\frac{\lambda}{4\pi^2}\int_0^\infty dk_z\int_0^\lambda dk_\bot
\left[1-\frac{\xi_{\bf k}-\lambda^2/2}{\sqrt{(\xi_{\bf k}-\lambda^2/2)^2+\Delta^2}}\right].
\end{eqnarray}
The integrations over $k_z$ can be analytically carried out with the help of the elliptic functions, which helps us to obtain numerical solutions
with high accuracy.

%%%%%%%%%%%%%%%%%%%%%%%%%%%%%%%%%%%%%%%%%%%%%%%%%%%%%%%%%%%%%%%%%%%%%%%
\begin{figure}[!htb]
%\begin{center}
\includegraphics[width=7cm]{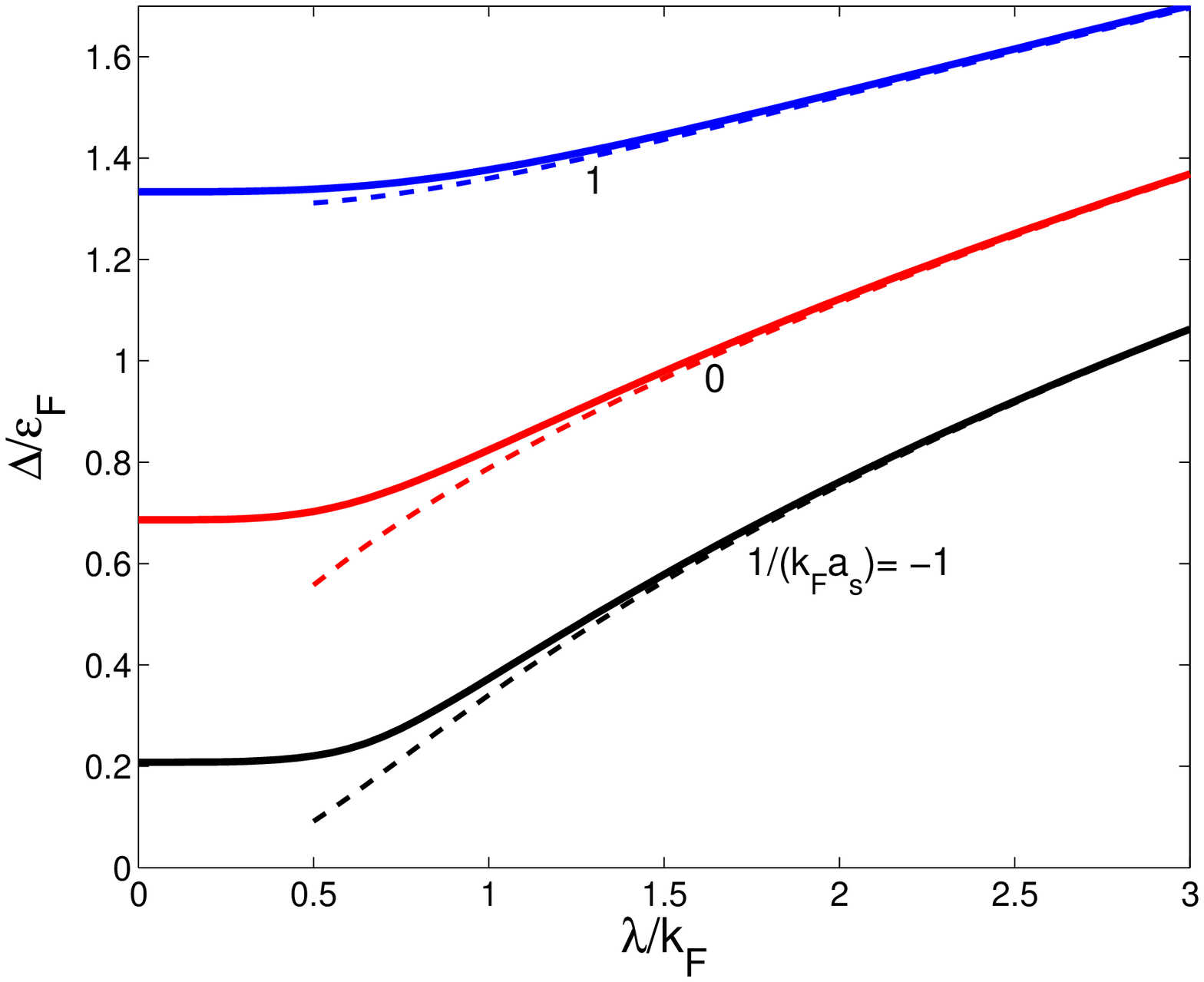}
\includegraphics[width=7cm]{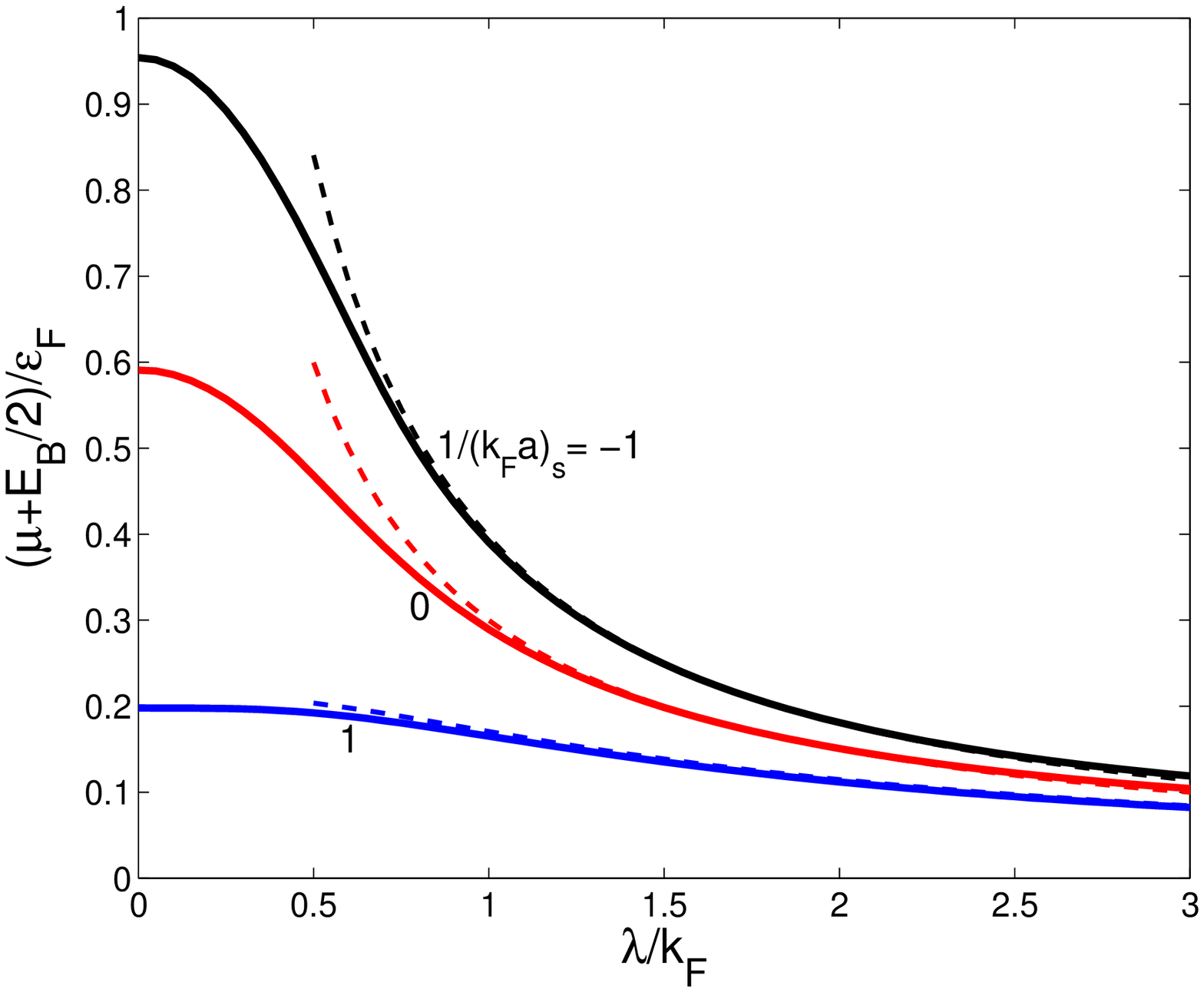}
\caption{(Color-online) The pairing gap $\Delta$ (divided by $\epsilon_{\rm F}$) and the quantity $\mu+E_{\rm B}/2$ (divided by $\epsilon_{\rm F}$)
as functions of $\lambda/k_{\rm F}$ for various values of $1/(k_{\rm F}a_s)$. The dashed lines correspond to the analytical results (\ref{EQ87})
and (\ref{EQ93}) for large spin-orbit coupling.\label{GAP3D}}
%\end{center}
\end{figure}
%%%%%%%%%%%%%%%%%%%%%%%%%%%%%%%%%%%%%%%%%%%%%%%%%%%%%%%%%%%%%%%%%%%%%%%%

The numerical results for $\Delta/\epsilon_{\rm F}$ and $\mu/\epsilon_{\rm F}$ are shown in Fig. \ref{GAP3D} . For large spin-orbit coupling, the
chemical potential becomes negative and $\Delta\ll|\mu|$. Therefore, the gap and number equations can be expanded in powers of $\Delta^2/|\mu|^2$.
To the leading order, the gap equation can be approximated as
\begin{eqnarray}
-\frac{1}{4\pi a_s}=\sum_{\bf k}\left[\frac{{\bf k}^2-2\mu}{({\bf k}^2-2\mu)^2-4\lambda^2{\bf k}_\bot^2}-\frac{1}{{\bf k}^2}\right]
\end{eqnarray}
which gives
\begin{eqnarray}
\mu=-\frac{E_{\rm B}}{2}=-\frac{\lambda^2}{2}[1+\eta(\kappa)].
\end{eqnarray}
Then the number equation becomes
\begin{eqnarray}
\label{numbereq}
n=2\Delta^2\sum_{\bf k}\frac{({\bf k}^2+E_{\rm B})^2+4\lambda^2{\bf k}_\bot^2}{[({\bf k}^2+E_{\rm B})^2-4\lambda^2{\bf k}_\bot^2]^2}
\end{eqnarray}
which yields
\begin{eqnarray}
\frac{\Delta}{\epsilon_{\rm F}}=\sqrt{\frac{16\eta(\kappa)}{3\pi \sqrt{1+\eta(\kappa)}}\frac{\lambda}{k_{\rm F}}}. \label{EQ87}
\end{eqnarray}
For large $\lambda/k_{\rm F}$, we have $\kappa\simeq0$. Using the result $\eta(0)=0.439$, we obtain the following asymptotical behaviors
\begin{eqnarray}
\frac{\Delta}{\epsilon_{\rm F}}\simeq0.788\sqrt{\frac{\lambda}{k_{\rm F}}},\ \ \
\frac{\mu}{\epsilon_{\rm F}}\simeq-1.44\left(\frac{\lambda}{k_{\rm F}}\right)^2.
\end{eqnarray}
The above result indicates that for $\lambda/k_{\rm F}\rightarrow\infty$, the properties of the system become independent of the interaction
parameter $1/(k_{\rm F}a_s)$.

Beyond the leading order of $\Delta^2/|\mu|^2$, the chemical potential at large $\lambda/k_{\rm F}$ can be expressed as $\mu=-E_{\rm B}/2
+\mu_{\rm B}/2$, where $\mu_{\rm B}\ll E_{\rm B}$ can be referred to as the chemical potential of the rashbons. To determine this chemical potential
as well as the interaction among the rashbons, we construct the Gross-Pitaevskii free energy of the rashbon condensate. To this end, we first derive
the Ginzburg-Landau free energy functional
\begin{eqnarray}
{\cal F}_{\rm{GL}}[\Delta(x)]=\int dx\Bigg[\Delta^\dagger(x)\left(a\frac{\partial}{\partial\tau}-b_\bot\mbox{\boldmath{$\nabla$}}_\bot^2
-b_\|\mbox{\boldmath{$\nabla$}}_\|^2\right)\Delta(x)+c|\Delta(x)|^2+\frac{1}{2}d|\Delta(x)|^4\Bigg]
\end{eqnarray}
according to the fact $\Delta\ll|\mu|$. The coefficients $a,b_\bot,b_\|$ can be obtained from the two-body vertex function $\Gamma^{-1}(Q)$ and $c,d$
can be obtained from the ground state energy $\Omega(\Delta)$. To the leading order of $\mu_{\rm B}/E_{\rm B}$, these coefficients can be evaluated as
\begin{eqnarray}
&&a=\sum_{\bf k}\frac{({\bf k}^2+E_{\rm B})^2+4\lambda^2{\bf k}_\bot^2}{[({\bf k}^2+E_{\rm B})^2-4\lambda^2{\bf k}_\bot^2]^2}
=\frac{1}{\lambda}\frac{\sqrt{1+\eta(\kappa)}}{8\pi\eta(\kappa)},\nonumber\\
&&b_\bot=\frac{1}{2M_{\rm B}^\bot}a,\ \ \ b_\|=\frac{1}{2M_{\rm B}^\|}a,\ \ \ c=-a\mu_{\rm B},\nonumber\\
&&d=2\sum_{\bf k}\frac{({\bf k}^2+E_{\rm B})[({\bf k}^2+E_{\rm B})^2+12\lambda^2{\bf k}_\bot^2]}{[({\bf k}^2+E_{\rm B})^2-4\lambda^2{\bf k}_\bot^2]^3}
=\frac{1}{\lambda^3}\frac{2+\eta(\kappa)}{16\pi\eta^2(\kappa)\sqrt{1+\eta(\kappa)}}.
\end{eqnarray}

Defining a new condensate wave function $\psiup(x)=\sqrt{a}\Delta(x)$, we obtain the Gross-Pitaevskii free energy functional
\begin{eqnarray}
{\cal F}_{\rm{GP}}[\psiup(x)]=\int dx\Bigg[\psiup^\dagger(x)\left(\frac{\partial}{\partial\tau}
-\frac{\mbox{\boldmath{$\nabla$}}_\bot^2}{2M_{\rm B}^\bot}-\frac{\mbox{\boldmath{$\nabla$}}_\|^2}{2M_{\rm B}^\|}\right)\psiup(x)
-\mu_{\rm B}|\psiup(x)|^2+\frac{1}{2}g|\psiup(x)|^4\Bigg],
\end{eqnarray}
The coupling $g$ describes the two-body repulsive interaction among the rashbons. It is given by
\begin{eqnarray}
g=\frac{d}{a^2}=\frac{4\pi}{\lambda}\frac{2+\eta(\kappa)}{[1+\eta(\kappa)]^{3/2}}.
\end{eqnarray}
For $\lambda\rightarrow\infty$, the coupling $g$ goes as $g\simeq1.11/\lambda$. Therefore, the system is a Bose-Einstein condensate of weakly repulsive
rashbons for large values of $\lambda/k_{\rm F}$. Minimizing the Gross-Pitaevskii free energy, we obtain $|\psiup|^2=\mu_{\rm B}/g$ and the rashbon
density $n_{\rm B}=n/2=|\psiup|^2$. This is consistent with Eq. (\ref{numbereq}), which gives $n=2a\Delta^2$. Therefore, the rashbon chemical potential
$\mu_{\rm B}$ can be expressed as $\mu_{\rm B}=gn/2$ or
\begin{eqnarray}
\frac{\mu_{\rm B}}{\epsilon_{\rm F}}=\frac{4[2+\eta(\kappa)]}{3\pi[1+\eta(\kappa)]^{3/2}}\left(\frac{\lambda}{k_{\rm F}}\right)^{-1}. \label{EQ93}
\end{eqnarray}
 At large $\lambda/k_{\rm F}$, this analytical result is in good agreement with the numerical results in Fig. \ref{GAP3D}. For
 $\lambda/k_{\rm F}\rightarrow\infty$, $\mu_{\rm B}/\epsilon_{\rm F}$ goes as $\mu_{\rm B}/\epsilon_{\rm F}\simeq 0.60(\lambda/k_{\rm F})^{-1}$.

%%%%%%%%%%%%%%%%%%%%%%%%%%%%%%%%%%%%%%%%%%%%%%%%%%%%%%%%%%%%%%%%%%%%%%%%%%%%%%%%%%%%%%%%%%%%%%%%%%%%%%%%%%%%%%
\subsection{Superfluid density}
\begin{figure}[!htb]
\begin{center}
\includegraphics[width=8cm]{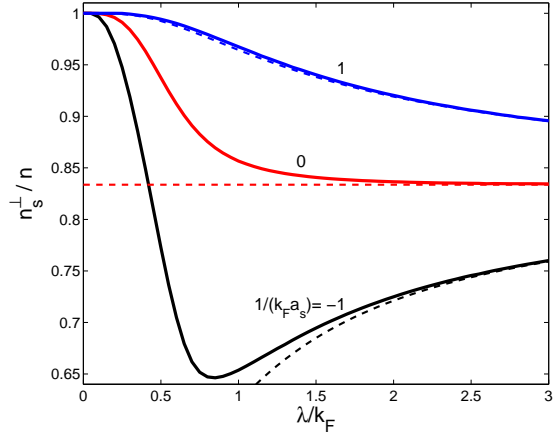}
\caption{(Color-online) The transverse superfluid density $n_s^\bot$ (divided by $n$) as a function of $\lambda/k_{\rm F}$ for various values of
$1/(k_{\rm F}a_s)$. The dashed lines correspond to the analytical result $2m/M_{\rm B}^\bot$.
 \label{NS3D}}
\end{center}
\end{figure}

%%%%%%%%%%%%%%%%%%%%%%%%%%%%%%%%%%%%%%%%%%%%%%%%%%%%%%%%%%%%%%%%%%%%%%%%%%%%%%%%%%%%%%%%%%%%%%%%%%%%%%%%%%%%%%
Now we discuss the superfluid densities, $n_s^\bot$ and $n_s^\|$, which are the key quantities to determine the Goldstone mode velocities.
For $h=0$, the longitudinal superfluid density $n_s^\|$ equals the total fermion density $n$. However, the transverse superfluid density does not.
It can be expressed as
\begin{eqnarray}
n_s^\bot=n-n_\lambda,
\end{eqnarray}
where the spin-orbit coupling induced normal fluid density $n_\lambda$ reads
\begin{eqnarray}
n_\lambda=\frac{\lambda}{8\pi^2}\int_0^\infty dk_z\int_0^\infty dk_\bot \sum_{\alpha=\pm}
\frac{\alpha}{E_{\bf k}^\alpha}\left(\xi_{\bf k}^\alpha+\frac{\Delta^2}{\xi_{\bf k}}\right).
\end{eqnarray}
Therefore, the transverse superfluid density $n_s^\bot$ is always smaller than $n$ for $\lambda\neq0$, as shown in Fig. \ref{NS3D}. Our result is
consistent with the result of the superfluid density for Rashba spin-orbit coupled Fermi superfluids first reported by Zhou and Zhang~\cite{SOC-Zhou}.
This behavior is in contrast to ordinary Fermi superfluids, where the superfluid density always equals the total density at zero temperature.

%%%%%%%%%%%%%%%%%%%%%%%%%%%%%%%%%%%%%%%%%%%%%%%%%%%%%%%%%%%%%%%%%%%%%%%
%%%%%%%%%%%%%%%%%%%%%%%%%%%%%%%%%%%%%%%%%%%%%%%%%%%%%%%%%%%%%%%%%%%%%%%%

To understand the fact $n_s^\bot<n$, we explore the behavior of $n_s^\bot$ at large spin-orbit coupling. To the leading order of $\Delta^2/|\mu|^2$,
$n_\lambda$ can be approximated as
\begin{eqnarray}
n_\lambda&\simeq&\frac{\lambda\Delta^2}{8\pi^2}\int_0^\infty dk_z\int_0^\infty dk_\bot\sum_{\alpha=\pm}
\frac{\alpha}{\xi_{\bf k}^\alpha}\left(\frac{1}{\xi_{\bf k}}-\frac{1}{2\xi_{\bf k}^\alpha}\right)\nonumber\\
&\simeq&2\Delta^2\sum_{\bf k}\frac{8\lambda^4{\bf k}_\bot^2}{({\bf k}^2+E_{\rm B})[({\bf k}^2+E_{\rm B})^2-4\lambda^2{\bf k}_\bot^2]^2}.
\end{eqnarray}
Together with Eq. (87) for $n$, we obtain
\begin{eqnarray}
\frac{n_\lambda}{n}\simeq1-\frac{2m}{M_{\rm B}^\bot},\ \ \ \ \ \ \frac{n_s^\bot}{n}\simeq\frac{2m}{M_{\rm B}^\bot}.
\end{eqnarray}
Therefore, at large $\lambda/k_{\rm F}$, the transverse superfluid density is suppressed by a factor $2m/M_{\rm B}^\bot<1$. For
$\lambda\rightarrow\infty$, we have $M_{\rm B}^\bot/(2m)\rightarrow 1.20$. This means that the transverse superfluid density approaches the limit
\begin{eqnarray}
\frac{n_s^\bot}{n}\rightarrow0.834\ \ \ {\rm for} \ \ \ \frac{\lambda}{k_{\rm F}}\rightarrow\infty.
\end{eqnarray}
In Fig. \ref{NS3D}, we show the result of $n_s^\bot/n$ for various values of $1/(k_{\rm F}a_s)$. At unitary, it approaches this limit very fast.

Then the physical picture becomes clear when we take a look at the phase stiffnesses $J_\bot$ and $J_\|$. At large spin-orbit coupling, we obtain
\begin{eqnarray}
J_\bot=\frac{n_s^\bot}{4m}\simeq\frac{n_{\rm B}}{M_{\rm B}^\bot},\ \ \ J_\|=\frac{n_s^\|}{4m}=\frac{n_{\rm B}}{M_{\rm B}^\|}.
\end{eqnarray}
These results show explicitly that at large $\lambda/k_{\rm F}$ we recover the phase stiffnesses for an anisotropic rashbon superfluid with density
$n_{\rm B}=n/2$ and anisotropic effective masses $M_{\rm B}^\bot>2m$ and $M_{\rm B}^\|=2m$.

On the other hand, the condensation density $n_0$, which is another important quantity for fermionic superfluidity, can be expressed as
\begin{eqnarray}
n_0=\frac{\Delta^2}{8}\sum_{\bf k}\sum_{\alpha=\pm}\frac{1}{(E_{\bf k}^\alpha)^2}.
\end{eqnarray}
At large spin-orbit coupling, we have $n_0\simeq (n/2)[1+O(\Delta^2/|\mu|^2)]$. This implies that the condensate fraction $2n_0/n$ approaches unity at
large $\lambda/k_{\rm F}$, consistent with the picture of Bose-Einstein condensation of weakly interacting rashbons. In contrast to the superfluid
density, we find numerically that $n_0$ is always an increasing function of $\lambda/k_{\rm F}$.

%%%%%%%%%%%%%%%%%%%%%%%%%%%%%%%%%%%%%%%%%%%%%%%%%%%%%%%%%%%%%%%%%%%%%%%%%%%%%%%%%%%%%%%%%%%%%%%%%%%%%%%%%%%%%%
\subsection{Collective modes}
%%%%%%%%%%%%%%%%%%%%%%%%%%%%%%%%%%%%%%%%%%%%%%%%%%%%%%%%%%%%%%%%%%%%%%%%%%%%%%%%%%%%%%%%%%%%%%%%%%%%%%%%%%%%%%
For $h=0$, the expression for the inverse collective mode propagator ${\bf M}(Q)$ becomes very simple. We have
\begin{eqnarray}
&&{\cal W}_1^{\alpha\beta}({\bf k},{\bf q})=\frac{1}{4}\left(1+\frac{\xi_{{\bf k}+{\bf p}}^\alpha\xi_{{\bf k}-{\bf p}}^\beta}
{E_{{\bf k}+{\bf p}}^\alpha E_{{\bf k}-{\bf p}}^\beta}\right){\cal T}_{\bf kq}^{\alpha\beta},\nonumber\\
&&{\cal W}_2^{\alpha\beta}({\bf k},{\bf q})=\frac{1}{4}\left(\frac{\xi_{{\bf k}+{\bf p}}^\alpha}{E_{{\bf k}+{\bf p}}^\alpha}
+\frac{\xi_{{\bf k}-{\bf p}}^\beta}{E_{{\bf k}-{\bf p}}^\beta}\right){\cal T}_{\bf kq}^{\alpha\beta},\nonumber\\
&&{\cal W}_3^{\alpha\beta}({\bf k},{\bf q})=\frac{\Delta^2}{4E_{{\bf k}+{\bf p}}^\alpha E_{{\bf k}-{\bf p}}^\beta}{\cal T}_{\bf kq}^{\alpha\beta},
\end{eqnarray}
where ${\cal T}_{\bf kq}^{\alpha\beta}$ has a nice property ${\cal T}_{\bf k0}^{\alpha\beta}=\delta_{\alpha\beta}$ for $h=0$. The expansion parameters
$A,B,D,R$ are simplified as
\begin{eqnarray}
&&A=\frac{\Delta^2}{2}\sum_{\alpha=\pm}\sum_{\bf k}\frac{1}{(E_{\bf k}^\alpha)^3},\ \ \ \
B=\frac{\Delta}{4}\sum_{\alpha=\pm}\sum_{\bf k}\frac{\xi_{\bf k}^\alpha}{(E_{\bf k}^\alpha)^3},\nonumber\\
&&D=\frac{1}{8}\sum_{\alpha=\pm}\sum_{\bf k}\frac{(\xi_{\bf k}^\alpha)^2}{(E_{\bf k}^\alpha)^5},\ \ \ \ \
R=\frac{\Delta^2}{8}\sum_{\alpha=\pm}\sum_{\bf k}\frac{1}{(E_{\bf k}^\alpha)^3}.
\end{eqnarray}

\begin{figure}[!htb]
\begin{center}
\includegraphics[width=8cm]{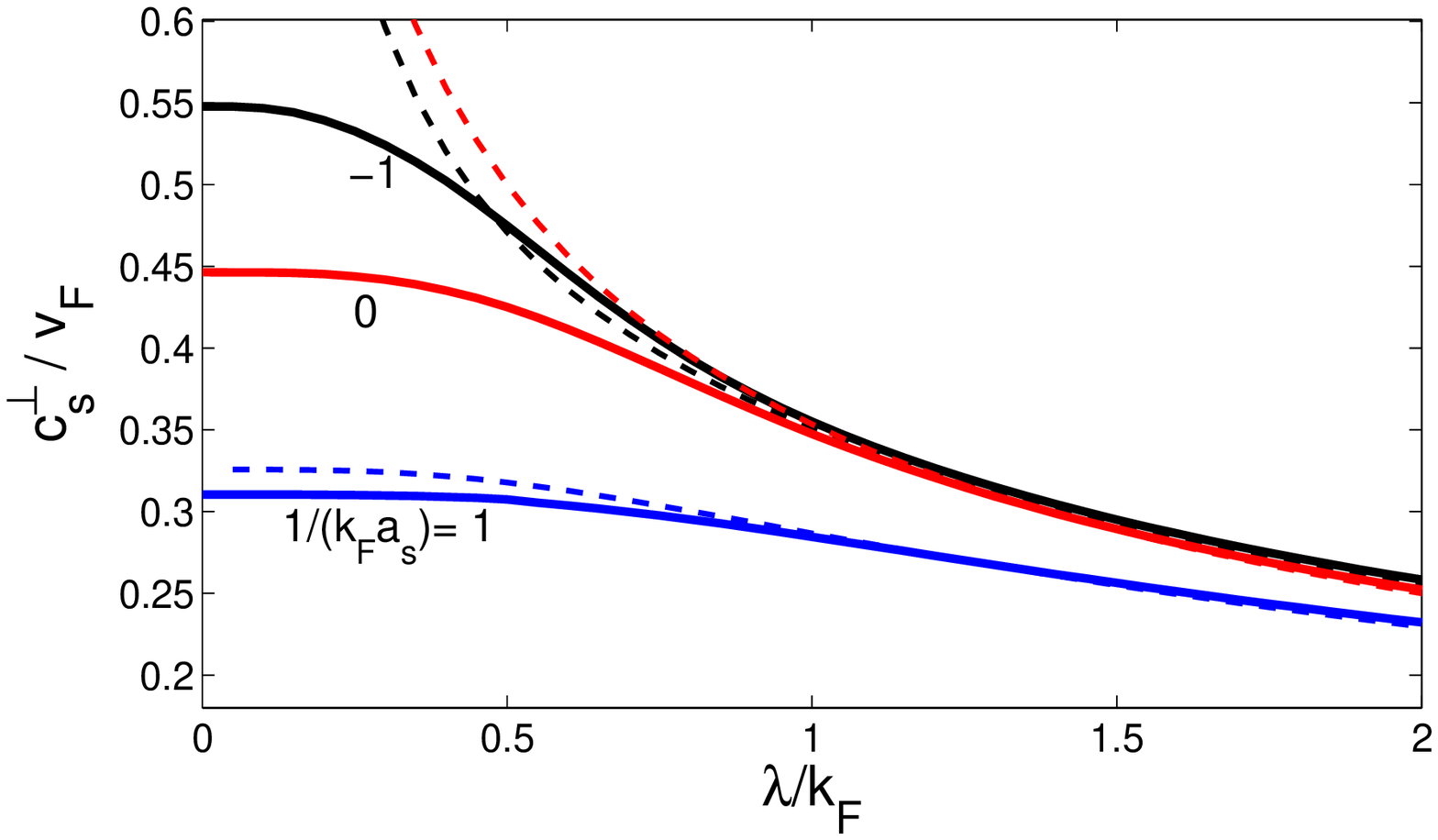}
\includegraphics[width=8cm]{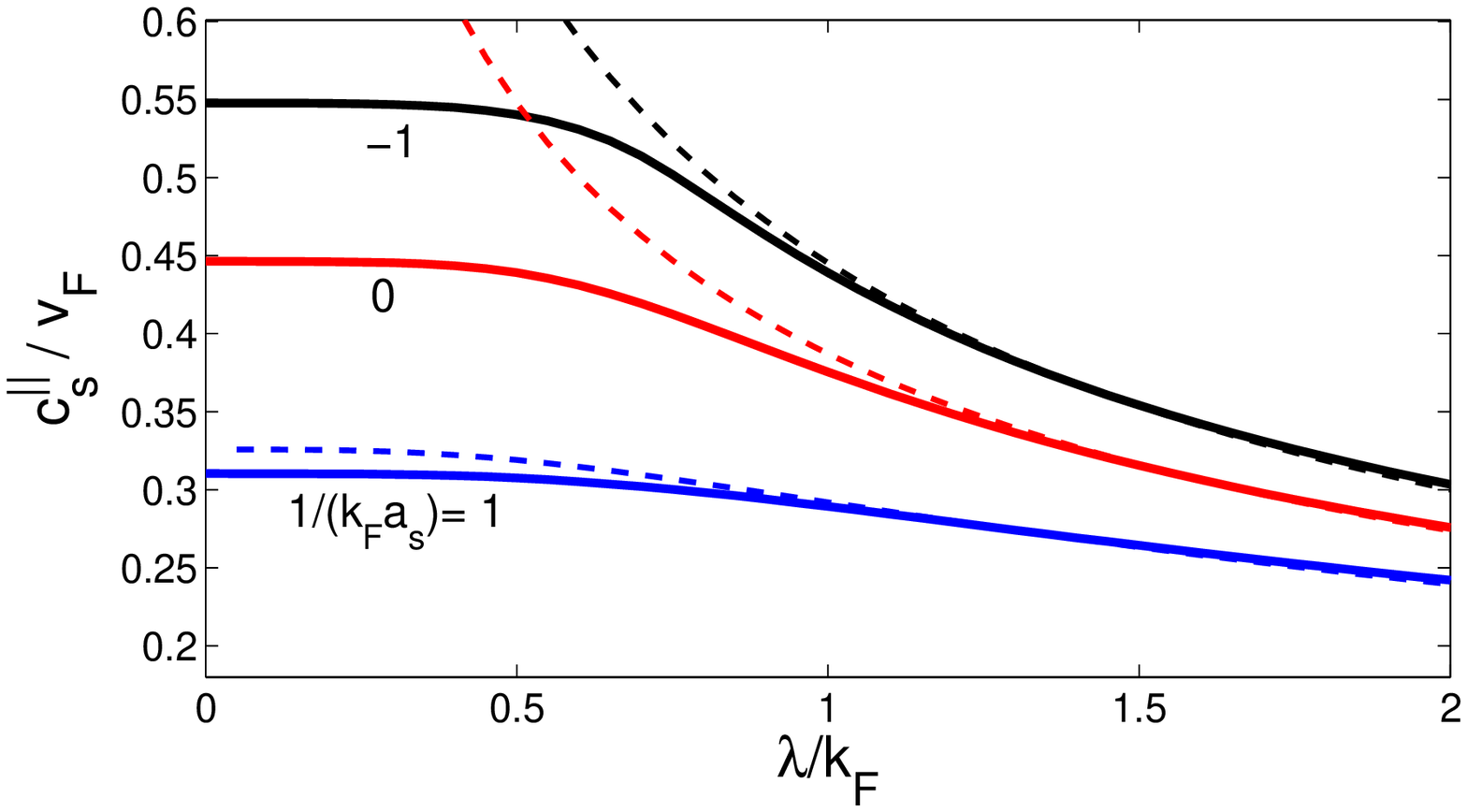}
\caption{(Color-online) The transverse and longitudinal velocities of the Goldstone mode, $c_s^\bot$ and $c_s^\|$ (divided by the Fermi velocity
$\upsilon_{\rm F}=k_{\rm F}/m$) as functions of $\lambda/k_{\rm F}$. The dashed lines correspond to the analytical results (\ref{EQ104}) for
large spin-orbit coupling.
 \label{CS3D}}
\end{center}
\end{figure}

Analytical results can be achieved at large spin-orbit coupling with the help of these simplified formulas. To the leading order of $\Delta^2/|\mu|^2$,
the parameters $A,B,D,R$ can be well approximated as
\begin{eqnarray}
A\simeq 4\Delta^2d,\ \ \  B\simeq 2\Delta a,\ \ \ D\simeq d,  \ \ \ R\simeq \Delta^2 d.
\end{eqnarray}
Using the expressions for $a$ and $d$, we find that at large $\lambda$, $B^2/A$ goes as $B^2/A\sim\lambda$ while $R$ goes as $R\sim \lambda^{-2}$.
Therefore, we have $B^2/A\gg R$ at large $\lambda$ and the amplitude-phase mixing term dominates the low-energy behavior of the collective modes.
Using the result $g=d/a^2$, we express the Goldstone mode velocities as
\begin{eqnarray}
c_s^\bot\simeq\sqrt{\frac{gn_{\rm B}}{M_{\rm B}^\bot}}=\sqrt{\frac{\mu_{\rm B}}{M_{\rm B}^\bot}},\ \ \ \ \ \ \
c_s^\|\simeq\sqrt{\frac{gn_{\rm B}}{M_{\rm B}^\|}}=\sqrt{\frac{\mu_{\rm B}}{M_{\rm B}^\|}}. \label{EQ104}
\end{eqnarray}
These results are just the Goldstone mode velocities of a weakly interacting Bose condensate, despite that the bosons possess anisotropic effective
masses. For $\lambda\rightarrow\infty$, the ratio $c_s^\bot/c_s^\|$ approaches a universal limit
\begin{eqnarray}
\frac{c_s^\bot}{c_s^\|}\rightarrow0.913.
\end{eqnarray}
In Fig. \ref{CS3D}, we show the numerical results for $c_s^\bot$ and $c_s^\|$ for various values of $1/(k_{\rm F}a_s)$. For
$\lambda/k_{\rm F}\rightarrow\infty$, they both go as $1/\sqrt{\lambda}$, independent of the interaction parameter $1/(k_{\rm F}a_s)$.

%%%%%%%%%%%%%%%%%%%%%%%%%%%%%%%%%%%%%%%%%%%%%%%%%%%%%%%%%%%%%%%%%%%%%%%
%%%%%%%%%%%%%%%%%%%%%%%%%%%%%%%%%%%%%%%%%%%%%%%%%%%%%%%%%%%%%%%%%%%%%%%%

Actually, to the leading order of $\Delta^2/|\mu|^2$,  the inverse boson propagator ${\bf M}(Q)$ [at $h=0$, ${\bf M}_{12}(Q)={\bf M}_{21}(Q)$]
can be approximated as
\begin{eqnarray}
&&{\bf M}_{11}(Q)={\bf M}_{22}(-Q)\simeq \Gamma^{-1}(Q)+2d\Delta^2,\nonumber\\
&&{\bf M}_{12}(Q)={\bf M}_{21}(Q)\simeq d\Delta^2.
\end{eqnarray}
At large spin-orbit coupling, the two-body vertex function can be well approximated as $\Gamma^{-1}(Q)\simeq-a[i\nu_n+\mu_{\rm B}-\omega_{\rm B}({\bf q})]$,
where $\omega_{\rm B}({\bf q})={\bf q}_\bot^2/(2M_{\rm B}^\bot)+{\bf q}_\|^2/(2M_{\rm B}^\|)$ is the rashbon dispersion. Using the result from the
Gross-Pitaevskii free energy, $\mu_{\rm B}=gn_{\rm B}$, the dispersion of the Goldstone mode can be expressed as
\begin{eqnarray}
\omega({\bf q})=\sqrt{\omega_{\rm B}({\bf q})\left[\omega_{\rm B}({\bf q})+2gn_{\rm B}\right]}.
\end{eqnarray}
This is nothing but the Bogoliubov excitation spectrum in a weakly interacting Bose condensate with anisotropic effective masses.

On the other hand, the mass gap of the Higgs mode can be estimated as
\begin{eqnarray}
M_{\rm H}\simeq\frac{2a}{d}=\frac{4\eta(\kappa)[1+\eta(\kappa)]}{2+\eta(\kappa)}\lambda^2.
\end{eqnarray}
For $\lambda\rightarrow\infty$, $M_{\rm H}$ goes as $M_{\rm H}\sim1.04\lambda^2$. We see that in the presence of the spin-orbit coupling,
the Higgs mode is also pushed up to the large characteristic energy scale ($\sim\lambda^2$) of the system.

%%%%%%%%%%%%%%%%%%%%%%%%%%%%%%%%%%%%%%%%%%%%%%%%%%%%%%%%%%%%%%%%%%%%%%%%%%%%%%%%%%%%%%%%%%%%%%%%%%%%%%%%%%%%%%
\section{Results for zero Zeeman field: 2D case}\label{s4}
%%%%%%%%%%%%%%%%%%%%%%%%%%%%%%%%%%%%%%%%%%%%%%%%%%%%%%%%%%%%%%%%%%%%%%%%%%%%%%%%%%%%%%%%%%%%%%%%%%%%%%%%%%%%%%

In two spatial dimensions, the two-body bound state exists for arbitrarily weak attraction. Here we show that the spin-orbit coupling effect 
enhances the binding energy. For zero Zeeman field, the two-body binding energy $E_{\rm B}$ is determined by the equation
\begin{eqnarray}
\sum_{\bf k}\left[\frac{{\bf k}^2+E_{\rm B}}{({\bf k}^2+E_{\rm B})^2-4\lambda^2{\bf k}^2}-\frac{1}{{\bf k}^2+\epsilon_{\rm B}}\right]=0.
\end{eqnarray}
The solution can also be written as $E_{\rm B}/\lambda^2=1+\eta(\kappa)$, where $\kappa=\ln(\lambda a_{2\rm D})$ and $\eta(\kappa)$ is determined by
\begin{eqnarray}
\frac{1}{\sqrt{\eta}}\arctan\frac{1}{\sqrt{\eta}}-\ln\frac{\sqrt{1+\eta}}{2}-\gamma=\kappa.
\end{eqnarray}
The ratio $E_{\rm B}/\epsilon_{\rm B}$ can be expressed as $E_{\rm B}/\epsilon_{\rm B}=e^{2\gamma+2\kappa}(1+\eta)/4$. The asymptotic behaviors of
$\eta(\kappa)$ are: (1) $\eta(\kappa)\rightarrow 4e^{-2\kappa-2\gamma}-1$ for $\kappa\rightarrow-\infty$; (2) $\eta(\kappa)\rightarrow \pi^2/(4\kappa^2)$
for $\kappa\rightarrow +\infty$. Therefore, we have $E_{\rm B}\rightarrow \epsilon_{\rm B}$ for weak spin-orbit coupling ($\kappa\rightarrow-\infty$)
and $E_{\rm B}\gg \epsilon_{\rm B}$ for strong spin-orbit coupling ($\kappa\rightarrow+\infty$).

\begin{figure}[!htb]
\begin{center}
\includegraphics[width=10cm]{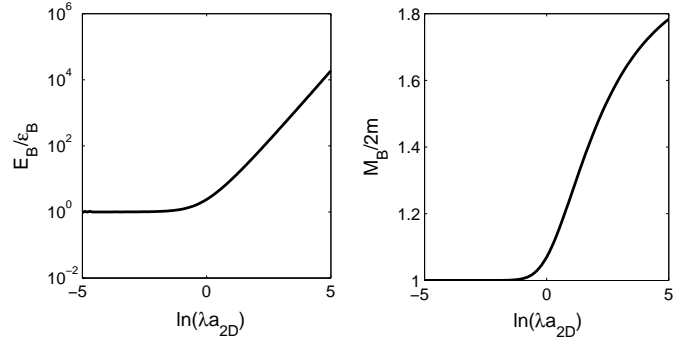}
\caption{The binding energy $E_{\rm B}$ (divided by $\epsilon_{\rm B}$) and the effective mass $M_{\rm B}$ (divided by $2m$)
as functions of $\ln(\lambda a_{2\rm D})$.
 \label{rashbon2D}}
\end{center}
\end{figure}

The effective mass $M_{\rm B}$ of the bound state is given by
\begin{eqnarray}
\left(1-\frac{2m}{M_{\rm B}}\right)\sum_{\bf k}\frac{({\bf k}^2+E_{\rm B})^2+4\lambda^2{\bf k}^2}
{[({\bf k}^2+E_{\rm B})^2-4\lambda^2{\bf k}^2]^2}=
\sum_{\bf k}\frac{8\lambda^4{\bf k}^2}{({\bf k}^2+E_{\rm B})[({\bf k}^2+E_{\rm B})^2-4\lambda^2{\bf k}^2]^2},
\end{eqnarray}
which gives
\begin{eqnarray}
\frac{2m}{M_{\rm B}}=1-\frac{1-(\eta-1)I(\eta)}{2(\eta+1)\left[1+I(\eta)\right]},
\end{eqnarray}
where
\begin{eqnarray}
I(\eta)=\frac{1}{2\sqrt{\eta}}\left(\frac{\pi}{2}-\arctan\frac{\eta-1}{2\sqrt{\eta}}\right).
\end{eqnarray}
For $\kappa\rightarrow+\infty$, $I(\eta)$ has the asymptotic behavior $I(\eta)\rightarrow \kappa$. For $\kappa\rightarrow-\infty$, we have
$I(\eta)\rightarrow1/\eta$. The $\kappa$-dependence of the binding energy and the effective mass is shown in Fig. \ref{rashbon2D}. Similar
to the 3D case, the effective mass $M_{\rm B}$ approaches $4m$ for $\kappa\rightarrow +\infty$.

%%%%%%%%%%%%%%%%%%%%%%%%%%%%%%%%%%%%%%%%%%%%%%%%%%%%%%%%%%%%%%%%%%%%%%%
%%%%%%%%%%%%%%%%%%%%%%%%%%%%%%%%%%%%%%%%%%%%%%%%%%%%%%%%%%%%%%%%%%%%%%%%

\begin{figure}[!htb]
%\begin{center}
\includegraphics[width=7cm]{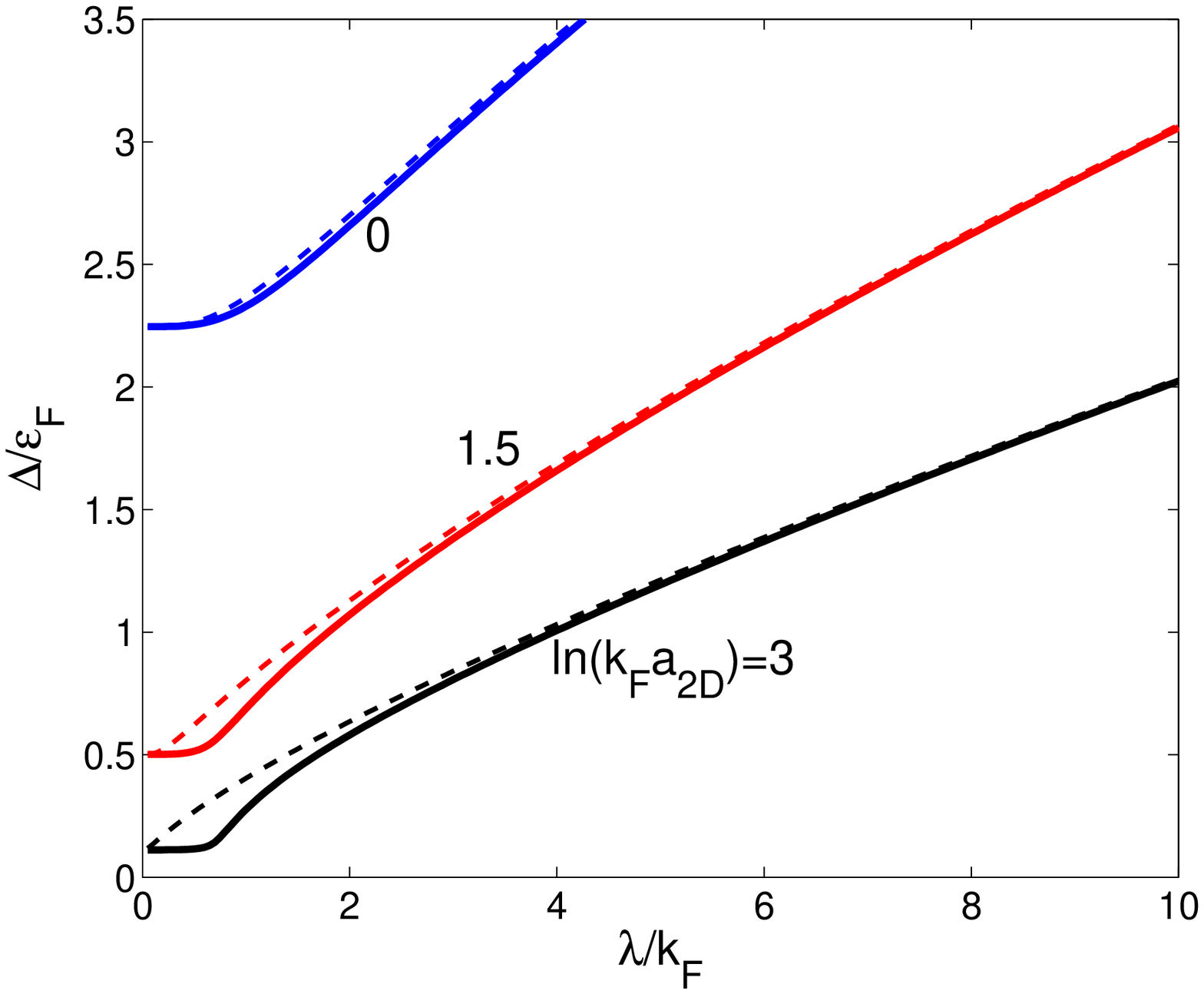}
\includegraphics[width=7cm]{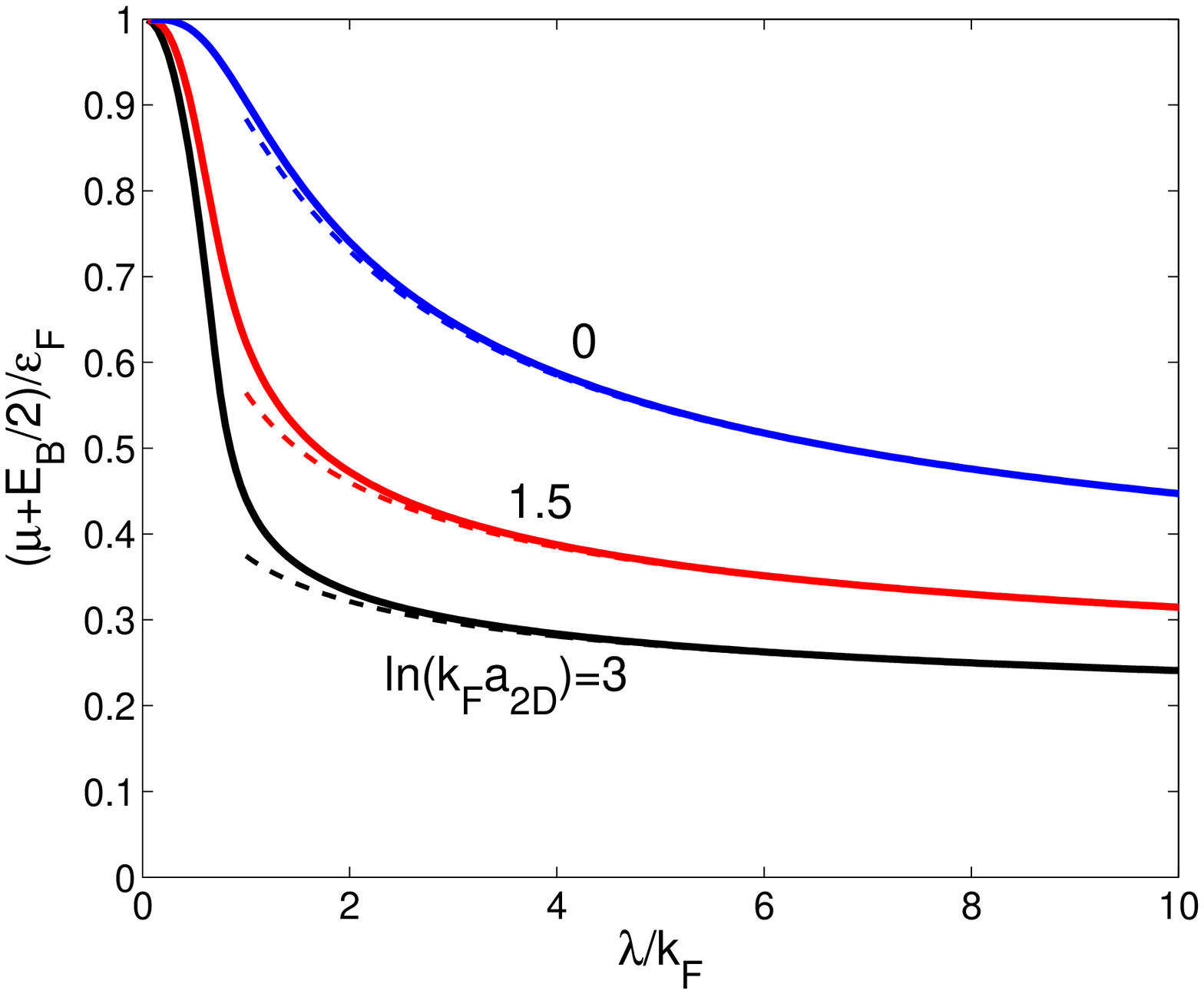}
\caption{(Color-online) The pairing gap $\Delta$ (divided by $\epsilon_{\rm F}$) and the quantity $\mu+E_{\rm B}/2$ (divided by $\epsilon_{\rm F}$)
in the 2D system as functions of $\lambda/k_{\rm F}$ for various values of the interaction parameter $\ln(k_{\rm F} a_{2\rm D})$. The dashed lines
correspond to the analytical results at strong spin-orbit coupling.
 \label{GAP2D}}
%\end{center}
\end{figure}

The order parameter $\Delta$ and the chemical potential $\mu$ are obtained from the gap and number equations. The 2D system is characterized
by two dimensionless parameters, $\ln(k_{\rm F}a_{2\rm D})$ and $\lambda/k_{\rm F}$, which represent the attractive strength and spin-orbit
coupling strength, respectively. For numerical calculations, it is convenient to employ the following analytical forms for the gap and number
equations~\cite{SOC-He},
\begin{eqnarray}
&&\ln\frac{\sqrt{\mu^2+\Delta^2}-\mu}{\epsilon_{\rm B}}=\lambda\int_0^\lambda\frac{dk}{\sqrt{(\xi_{\bf k}-\lambda^2/2)^2+\Delta^2}},\nonumber\\
&&2\epsilon_{\rm F}=\sqrt{\mu^2+\Delta^2}+\mu+\lambda\int_0^\lambda dk\left[1-\frac{\xi_{\bf k}-\lambda^2/2}
{\sqrt{(\xi_{\bf k}-\lambda^2/2)^2+\Delta^2}}\right].
\end{eqnarray}
For $\lambda=0$, they give the simple analytical results $\Delta=\sqrt{2\epsilon_{\rm B}\epsilon_{\rm F}}$ and $\mu=\epsilon_{\rm F}-\epsilon_{\rm B}/2$
~\cite{BCSBEC4}. At large spin-orbit coupling, we have $\mu<0$ and $\Delta\ll|\mu|$, which leads to the following analytical results
\begin{eqnarray}
\mu\simeq-\frac{E_{\rm B}}{2},\ \ \ \ \Delta\simeq\sqrt{\frac{n}{2a}}=\sqrt{\frac{\epsilon_{\rm F}}{2\pi a}},
\end{eqnarray}
where the 2D version of $a$ reads
\begin{eqnarray}
a=\sum_{\bf k}\frac{({\bf k}^2+E_{\rm B})^2+4\lambda^2{\bf k}^2}{[({\bf k}^2+E_{\rm B})^2-4\lambda^2{\bf k}^2]^2}
=\frac{1}{\lambda^2}\frac{1+I(\eta)}{4\pi\eta}.
\end{eqnarray}
Note that in 2D the density $n$ is related to the Fermi momentum $k_{\rm F}$ by $n=k_{\rm F}^2/(2\pi)=\epsilon_{\rm F}/\pi$. The numerical results for
$\Delta$ and $\mu$ are shown in Fig. \ref{GAP2D}. The results are in good agreement with the analytical results for large $\lambda/k_{\rm F}$. For the 2D
case we find that the quantity $\mu+E_{\rm B}/2$ goes down very slowly, unlike the 3D case where it goes as $1/\lambda$ at large $\lambda/k_{\rm F}$.
This is because for large $\lambda$, the rashbon-rashbon coupling $g$ goes as $g\sim 1/\ln(\lambda a_{\rm 2D})$ in 2D rather than $g\sim 1/\lambda$ as in 3D (see below).

%%%%%%%%%%%%%%%%%%%%%%%%%%%%%%%%%%%%%%%%%%%%%%%%%%%%%%%%%%%%%%%%%%%%%%%
%%%%%%%%%%%%%%%%%%%%%%%%%%%%%%%%%%%%%%%%%%%%%%%%%%%%%%%%%%%%%%%%%%%%%%%%

\begin{figure}[!htb]
\begin{center}
\includegraphics[width=8cm]{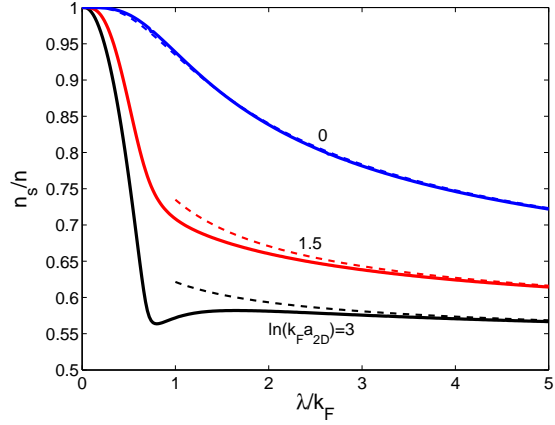}
\caption{The superfluid density $n_s$ (divided by $n$) in the 2D system as a function
of $\lambda/k_{\rm F}$ for various values of $\ln(k_{\rm F} a_{2\rm D})$.
The dashed lines are analytical results for large spin-orbit coupling.
 \label{NS2D}}
\end{center}
\end{figure}

For the 2D case, the superfluid density is isotropic. The superfluid density can also be written as $n_s=n-n_\lambda$, where the spin-orbit coupling
induced normal fluid density $n_\lambda$ reads
\begin{eqnarray}
n_\lambda=\frac{\lambda}{8\pi}\int_0^\infty dk \sum_{\alpha=\pm}\frac{\alpha}{E_{\bf k}^\alpha}
\left(\xi_{\bf k}^\alpha+\frac{\Delta^2}{\xi_{\bf k}}\right).
\end{eqnarray}
At large spin-orbit coupling $n_\lambda$ can be approximated as
\begin{eqnarray}
n_\lambda\simeq2\Delta^2\sum_{\bf k}\frac{8\lambda^4{\bf k}^2}{({\bf k}^2+E_{\rm B})[({\bf k}^2+E_{\rm B})^2-4\lambda^2{\bf k}^2]^2},
\end{eqnarray}
which leads to the result
\begin{eqnarray}
\frac{n_s}{n}\simeq\frac{2m}{M_{\rm B}},\ \ \ \ J\simeq\frac{n_{\rm B}}{M_{\rm B}}.
\end{eqnarray}
These are just the superfluid density and phase stiffness for the 2D rashbon condensate. In Fig. \ref{NS2D}, we show the results for the superfluid density.
At large spin-orbit coupling and/or attraction, the numerical results are in good agreement with the above analytical result.

%%%%%%%%%%%%%%%%%%%%%%%%%%%%%%%%%%%%%%%%%%%%%%%%%%%%%%%%%%%%%%%%%%%%%%%
%%%%%%%%%%%%%%%%%%%%%%%%%%%%%%%%%%%%%%%%%%%%%%%%%%%%%%%%%%%%%%%%%%%%%%%%

\begin{figure}[!htb]
\begin{center}
\includegraphics[width=8cm]{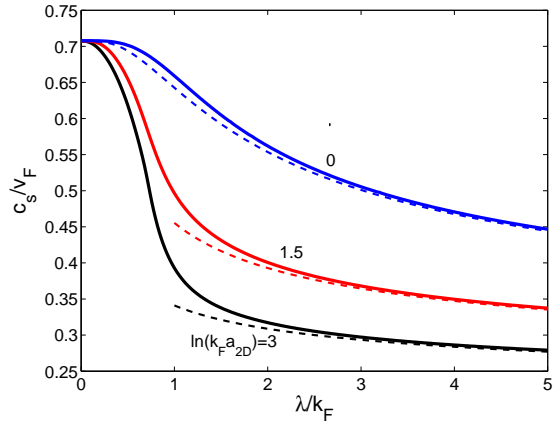}
\caption{(Color-online) The Goldstone mode velocity $c_s$ (divided by $\upsilon_{\rm F}=k_{\rm F}/m$)
in the 2D system as a function of $\lambda/k_{\rm F}$ for various values of $\ln(k_{\rm F} a_{2\rm D})$.
The dashed lines are analytical results for large spin-orbit coupling.
 \label{CS2D}}
\end{center}
\end{figure}

Meanwhile, for large spin-orbit coupling, the expansion parameters $A,B,D,R$ can be approximated as
\begin{eqnarray}
A\simeq 4\Delta^2d,\ \ \  B\simeq 2\Delta a,\ \ \ D\simeq d,  \ \ \ R\simeq \Delta^2 d,
\end{eqnarray}
where the 2D version of $d$ reads
\begin{eqnarray}
d=2\sum_{\bf k}\frac{({\bf k}^2+E_{\rm B})[({\bf k}^2+E_{\rm B})^2+12\lambda^2{\bf k}^2]}
{[({\bf k}^2+E_{\rm B})^2-4\lambda^2{\bf k}^2]^3}=
\frac{1}{\lambda^4}\frac{2+(1+\eta)^{-1}+3I(\eta)}{8\pi\eta^2}.
\end{eqnarray}
At large $\lambda$, $B^2/A$ goes as $B^2/A\sim \ln(\lambda a_{\rm 2D})$ while $R$ goes as $R\sim [\ln(\lambda a_{\rm 2D})/\lambda]^2$.
Therefore, the amplitude-phase mixing term $B^2/A$ also dominates in 2D. Finally, the Goldstone mode velocity $c_s$ can be expressed as
\begin{eqnarray}
c_s\simeq\sqrt{\frac{gn_{\rm B}}{M_{\rm B}}},
\end{eqnarray}
where the 2D version of the coupling $g=d/a^2$ reads
\begin{eqnarray}
g=2\pi\frac{2+(1+\eta)^{-1}+3I(\eta)}{[1+I(\eta)]^2}.
\end{eqnarray}
In Fig. \ref{CS2D}, we show the numerical results of $c_s$ and compare it with the analytical result. We note that the rashbon-rashbon coupling $g$ is
dimensionless and at large spin-orbit coupling it goes as
\begin{equation}
g\simeq\frac{6\pi}{\ln(\lambda a_{2\rm D})}.
\end{equation}
This is in contrast to the 3D case where $g\sim 1/\lambda$. The rashbon chemical potential $\mu_{\rm B}=2\mu+E_{\rm B}$ reads
$\mu_{\rm B}=gn_{\rm B}=gn/2$. In Fig. \ref{GAP2D}, we show the analytical results for $\mu_{\rm B}/2$ by dashed lines and compare them with the quantity
$\mu+E_{\rm B}/2$. They are in good agreement at large $\lambda/k_{\rm F}$. The large-$\lambda$ behavior of the coupling $g$ explains why the quantity
$\mu+E_{\rm B}/2$ decreases slower than the 3D case.

%%%%%%%%%%%%%%%%%%%%%%%%%%%%%%%%%%%%%%%%%%%%%%%%%%%%%%%%%%%%%%%%%%%%%%%
%%%%%%%%%%%%%%%%%%%%%%%%%%%%%%%%%%%%%%%%%%%%%%%%%%%%%%%%%%%%%%%%%%%%%%%%

%%%%%%%%%%%%%%%%%%%%%%%%%%%%%%%%%%%%%%%%%%%%%%%%%%%%%%%%%%%%%%%%%%%%%%%%%%%%%%%%%%%%%%%%%%%%%%%%%%%%%%%%%%%%%%
\section{Results for finite Zeeman field}\label{s5}
%%%%%%%%%%%%%%%%%%%%%%%%%%%%%%%%%%%%%%%%%%%%%%%%%%%%%%%%%%%%%%%%%%%%%%%%%%%%%%%%%%%%%%%%%%%%%%%%%%%%%%%%%%%%%%
Now we turn to the case of nonzero Zeeman field $h$.  In the absence of spin-orbit coupling, it is known that the BCS superfluidity is completely
destroyed when the Zeeman field $h$ is large enough. In the weak coupling limit, there exists a first-order phase transition from the BCS state to
the normal state at $h_{\rm CC}=\Delta_0/\sqrt{2}$ where $\Delta_0$ is the gap at $h=0$, which is referred to as the Chandrasekhar-Clogston limit.
Further theoretical studies showed that the inhomogeneous Fulde-Ferrell-Larkin-Ovchinnikov state can survive in a narrow window between $h_{\rm CC}$
and $h_{\rm FFLO}=0.754\Delta_0$. The Zeeman field effects on fermionic superfluidity in the whole BCS-BEC crossover regime have been experimentally
studied in recent years~\cite{Imexp}. Two-component Fermi gases with population imbalance ($N_\uparrow\neq N_\downarrow$) are realized to simulate
the Zeeman field effect. Around the unitary point ($1/(k_{\rm F}a_s)=0$), the phase separation between the superfluid and normal phases has been
observed in accordance with the first-order phase transition. Despite the rich phase structure in the BCS-BEC crossover, one finds that the fermionic
superfluidity is completely destroyed at large enough Zeeman field in the whole BCS-BEC crossover regime.
%%%%%%%%%%%%%%%%%%%%%%%%%%%%%%%%%%%%%%%%%%%%%%%%%%%%%%%%%%%%%%%%%%%%%%%
\begin{figure}[!htb]
\begin{center}
\includegraphics[width=10cm]{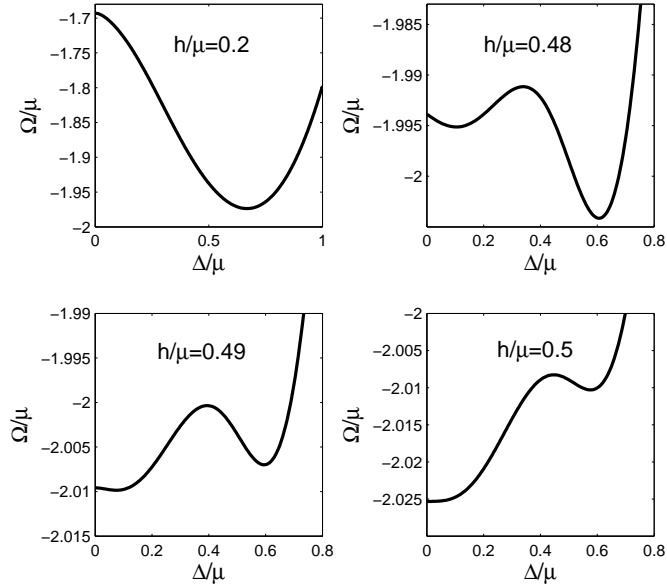}
\caption{Evolution of the grand potential $\Omega(\Delta)$ with varying Zeeman field $h$ for a small spin-orbit coupling $\lambda/\sqrt{2\mu}=0.1$.
 In the calculations we set $\epsilon_{\rm B}/\mu=0.2$. \label{Potential01}}
\end{center}
\end{figure}
%%%%%%%%%%%%%%%%%%%%%%%%%%%%%%%%%%%%%%%%%%%%%%%%%%%%%%%%%%%%%%%%%%%%%%%%

%%%%%%%%%%%%%%%%%%%%%%%%%%%%%%%%%%%%%%%%%%%%%%%%%%%%%%%%%%%%%%%%%%%%%%%
\begin{figure}[!htb]
\begin{center}
\includegraphics[width=10cm]{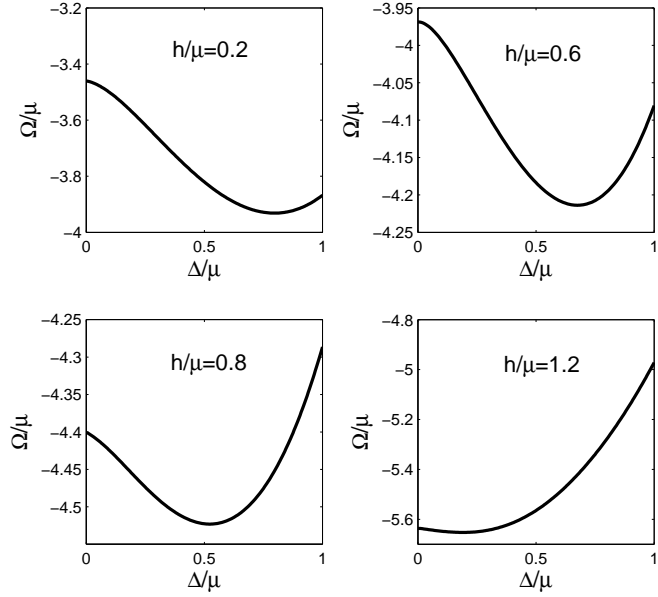}
\caption{Evolution of the grand potential $\Omega(\Delta)$ with varying Zeeman field $h$ for a strong spin-orbit coupling $\lambda/\sqrt{2\mu}=0.5$.
 In the calculations we set $\epsilon_{\rm B}/\mu=0.2$. \label{Potential02}}
\end{center}
\end{figure}
%%%%%%%%%%%%%%%%%%%%%%%%%%%%%%%%%%%%%%%%%%%%%%%%%%%%%%%%%%%%%%%%%%%%%%%%

%%%%%%%%%%%%%%%%%%%%%%%%%%%%%%%%%%%%%%%%%%%%%%%%%%%%%%%%%%%%%%%%%%%%%%%%%%%%%%%%%%%%%%%%%%%%%%%%%%%%%%%%%%%%%%
\subsection{Fate of the first-order phase transition}
%%%%%%%%%%%%%%%%%%%%%%%%%%%%%%%%%%%%%%%%%%%%%%%%%%%%%%%%%%%%%%%%%%%%%%%%%%%%%%%%%%%%%%%%%%%%%%%%%%%%%%%%%%%%%%

In the presence of spin-orbit coupling, the phase structure at $h\neq0$ becomes qualitatively different. There are two crucial observations as shown in the previous studies~\cite{SOC-Gong,SOC-Iskin,SOC-Yi,TSC07}. First, in the presence of spin-orbit coupling, the order parameter $\Delta$ never vanishes even for large $h$. The superfluid phase at large enough $h$ is a topological superfluid. Second, the first-order phase transition is eliminated at large spin-orbit coupling. Instead, the order parameter varies smoothly with the Zeeman field $h$. Therefore, at strong spin-orbit coupling, there exists only topological quantum phase transitions from the normal superfluid phase to some topological superfluid phases.

The elimination of the first-order phase transition at strong spin-orbit coupling has been systematically studied in some previous papers~ \cite{SOC-Gong,SOC-Iskin,SOC-Yi,TSC07}. For the sake of completeness, here we briefly show how the spin-orbit coupling eliminates the first-order phase transition, using the 2D case as an example. For the 3D case, the conclusion is the same \cite{SOC-Gong,SOC-Iskin,SOC-Yi}.
To study the first-order phase transition, we have to study the property of the grand potential $\Omega(\Delta;\mu,h)$. For the 2D case, it reads
\begin{eqnarray}
\Omega(\Delta;\mu,h)=\sum_{\bf k}\left(\frac{\Delta^2}{{\bf k}^2+\epsilon_{\rm B}}-\frac{E_{\bf k}^++E_{\bf k}^-}{2}+\xi_{\bf k}\right).
\end{eqnarray}
In Fig. \ref{Potential01} and Fig. \ref{Potential02}, we show the evolution of the grand potential curve with varying Zeeman field for weak and strong spin-orbit couplings, respectively. For nonzero spin-orbit coupling, we find that there is always a minimum at $\Delta\neq0$ no matter how large the Zeeman field is, in contrast to the case of vanishing spin-orbit coupling, where the minimum shifts to $\Delta=0$ at large Zeeman field.
For weak spin-orbit coupling (Fig. \ref{Potential01}), the first order phase transition persists, as we expected. The difference from the case of vanishing spin-orbit coupling is that the order parameter $\Delta$ jumps from a large value to a small value. When the total particle number $n$ instead of the chemical potential $\mu$ is fixed, the mixed phase (phase separation) appears in a range of the Zeeman field in accordance of the first-order phase transition. For strong spin-orbit coupling (Fig. \ref{Potential02}), the potential curve always shows only one minimum at nonzero $\Delta$, which means that there exists no longer first-order phase transition. Therefore, the first-order phase transition becomes eliminated at strong enough spin-orbit coupling.

In this paper, we are interested in the properties of the collective modes across the topological quantum phase transition. Therefore, we focus on strong spin-orbit coupling where the first-order phase transition is eliminated. In the following, we will first study the 2D case,
since in 2D the results of the bulk superfluid properties and the collective modes are most pronouncedly associated with the quantum phase transition.

%%%%%%%%%%%%%%%%%%%%%%%%%%%%%%%%%%%%%%%%%%%%%%%%%%%%%%%%%%%%%%%%%%%%%%%%%%%%%%%%%%%%%%%%%%%%%%%%%%%%%%%%%%%%%%
\subsection{2D system}
%%%%%%%%%%%%%%%%%%%%%%%%%%%%%%%%%%%%%%%%%%%%%%%%%%%%%%%%%%%%%%%%%%%%%%%%%%%%%%%%%%%%%%%%%%%%%%%%%%%%%%%%%%%%%%

%%%%%%%%%%%%%%%%%%%%%%%%%%%%%%%%%%%%%%%%%%%%%%%%%%%%%%%%%%%%%%%%%%%%%%%%%%%%%%%%%%%%%%%%%%%%%%%%%%%%%%%%%%%%%%
\subsubsection{Quantum phase transition}
%%%%%%%%%%%%%%%%%%%%%%%%%%%%%%%%%%%%%%%%%%%%%%%%%%%%%%%%%%%%%%%%%%%%%%%%%%%%%%%%%%%%%%%%%%%%%%%%%%%%%%%%%%%%%%
First, we show that for the 2D system, there exist two different superfluid phases at $h\neq0$ distinguished by the quantity
~\cite{TSC01,TSC02,TSC03,TSC04,TSC05,TSC06,TSC07,TSC08,TSC09}
\begin{eqnarray}
{\cal C}_0=\mu^2+\Delta^2-h^2.
\end{eqnarray}
The superfluid phases with ${\cal C}_0>0$ and ${\cal C}_0<0$ are linked by a quantum phase transition.

For $h\neq0$, the upper quasiparticle branch $E_{\bf k}^+$ is fully gapped, while the lower excitation spectrum $E_{\bf k}^-$ can have zeros at some
critical Zeeman field. To show this, we employ the following identity
\begin{eqnarray}
(E_{\bf k}^+)^2(E_{\bf k}^-)^2=\left(E_{\bf k}^2-h^2-\lambda^2{\bf k}^2\right)^2+4\lambda^2{\bf k}^2\Delta^2.
\end{eqnarray}
Since we are considering the superfluid phases with $\Delta\neq0$ for $\lambda\neq0$, the only possible zero for $E_{\bf k}^-$ is located at
${\bf k}=0$. This zero appears only when the quantity ${\cal C}_0$ is precisely zero or the Zeeman field $h$ equals the critical value
$h_c=\sqrt{\mu^2+\Delta^2}$. For ${\cal C}_0\neq0$ or $h\neq h_c$, the fermionic excitations are fully gapped. At the critical point $h=h_c$,
the lower branch $E_{\bf k}^-$ has a linear dispersion near ${\bf k}=0$, i.e.,
\begin{eqnarray}
E_{\bf k}^-=\upsilon_c|{\bf k}|+O(|{\bf k}|^2),\ \ \ \ \ {\bf k}\rightarrow0,
\end{eqnarray}
where the velocity $\upsilon_c$ can be determined as
\begin{eqnarray}
\upsilon_c=\frac{\lambda \Delta}{\sqrt{\mu^2+\Delta^2}}.
\end{eqnarray}

The existence of such a gapless fermionic spectrum causes singularities of the thermodynamic functions at the critical point ${\cal C}_0=0$.
To be specific, we consider the thermodynamic potential $\Omega(\mu,h)\equiv\Omega(\mu,h,\Delta(\mu,h))$, where
\begin{eqnarray}
\Omega(\mu,h,\Delta)=\sum_{\bf k}\left(\frac{\Delta^2}{{\bf k}^2+\epsilon_{\rm B}}-\frac{E_{\bf k}^++E_{\bf k}^-}{2}+\xi_{\bf k}\right).
\end{eqnarray}
To obtain the thermodynamic potential $\Omega(\mu,h)$, the superfluid order parameter $\Delta(\mu,h)$, which is regarded as an implicit function of
$\mu$ and $h$, should be determined by the gap equation
\begin{eqnarray}
\sum_{\bf k}\left[\sum_{\alpha=\pm}\left(1+\alpha\frac{h^2}{\zeta_{\bf k}}\right)
\frac{1}{4E_{\bf k}^\alpha}-\frac{1}{{\bf k}^2+\epsilon_{\rm B}}\right]=0.
\end{eqnarray}
We now demonstrate that the infrared singularities caused by the gapless fermionic spectrum show up at the fourth derivatives of the thermodynamic
function $\Omega(\mu,h)$ with respect to the thermodynamic variables $\mu$ and $h$. To this end, we consider the following two susceptibilities
\begin{eqnarray}
\chi_{\mu\mu}=-\frac{\partial^2\Omega(\mu,h)}{\partial \mu^2}, \ \ \ \ \chi_{hh}=-\frac{\partial^2\Omega(\mu,h)}{\partial h^2}.
\end{eqnarray}
$\chi_{\mu\mu}$ is related to the isothermal compressibility and $\chi_{hh}$ is the spin susceptibility. To obtain their explicit expressions, we need
the derivatives $\partial\Delta(\mu,h)/\partial\mu$ and $\partial\Delta(\mu,h)/\partial h$. With the help of the gap equation
$\partial \Omega(\mu,h,\Delta)/\partial\Delta=0$, we obtain
\begin{eqnarray}
\frac{\partial\Delta(\mu,h)}{\partial \mu}=\frac{1}{A}\frac{\partial n(\mu,h,\Delta)}{\partial\Delta},\ \ \ \ \ \
\frac{\partial\Delta(\mu,h)}{\partial h}=\frac{1}{A}\frac{\partial \delta n(\mu,h,\Delta)}{\partial\Delta}.
\end{eqnarray}
Here $A=\partial^2\Omega(\mu,h,\Delta)/\partial \Delta^2$ is one of the expansion parameters obtained in Sec. \ref{s2}, $n$ is the total density, and
$\delta n=n_\uparrow-n_\downarrow$ is the spin polarization. We note that the delta-function term in $A$ vanishes automatically. The explicit
expressions of $n$ and $\delta n$ can be evaluated as
\begin{eqnarray}
n(\mu,h,\Delta)&=&\frac{1}{2}\sum_{\bf k}\sum_{\alpha=\pm}
\left[1-\left(1+\alpha\frac{\eta_{\bf k}^2}{\zeta_{\bf k}}\right)\frac{\xi_{\bf k}}{E_{\bf k}^\alpha}\right],\nonumber\\
\delta n(\mu,h,\Delta)&=&\frac{1}{2}\sum_{\bf k}\sum_{\alpha=\pm}\frac{h}{E_{\bf k}^\alpha}
\left(1+\alpha\frac{E_{\bf k}^2}{\zeta_{\bf k}}\right).
\end{eqnarray}
Finally, the two susceptibilities can be evaluated as
\begin{eqnarray}
\chi_{\mu\mu}&=&\frac{\partial n(\mu,h,\Delta)}{\partial \mu}+\frac{1}{A}\left(\frac{\partial n(\mu,h,\Delta)}{\partial \Delta}\right)^2,\nonumber\\
\chi_{hh}&=&\frac{\partial \delta n(\mu,h,\Delta)}{\partial h}+\frac{1}{A}\left(\frac{\partial \delta n(\mu,h,\Delta)}{\partial \Delta}\right)^2.
\end{eqnarray}

Using the following results
\begin{eqnarray}
\frac{\partial E_{\bf k}^-}{\partial \Delta}=\frac{\Delta}{E_{\bf k}^-}\left(1-\frac{h^2}{\zeta_{\bf k}}\right),\ \ \
\frac{\partial E_{\bf k}^-}{\partial \mu}=-\frac{\xi_{\bf k}}{E_{\bf k}^-}\left(1-\frac{\eta_{\bf k}^2}{\zeta_{\bf k}}\right),\ \ \
\frac{\partial E_{\bf k}^-}{\partial h}=\frac{h}{E_{\bf k}^-}\left(1-\frac{E_{\bf k}^2}{\zeta_{\bf k}}\right),
\end{eqnarray}
we find that the expressions of $\chi_{\mu\mu}$ and $\chi_{hh}$ contain integrals of the following type
\begin{eqnarray}
{\cal I}_{ij}\sim \int_0^\infty kdk \frac{{\cal Q}_i{\cal Q}_j}{(E_{\bf k}^-)^3},
\end{eqnarray}
where
\begin{eqnarray}
{\cal Q}_1=1-\frac{h^2}{\zeta_{\bf k}},\ \ \ \ \
{\cal Q}_2=1-\frac{\eta_{\bf k}^2}{\zeta_{\bf k}},\ \ \ \ \
{\cal Q}_3=1-\frac{E_{\bf k}^2}{\zeta_{\bf k}}.
\end{eqnarray}
At $h=h_c$, the integrals ${\cal I}_{ij}$ ($i,j=1,2,3$) are infrared safe since the quantities ${\cal Q}_i$ go as $k^2$ for $k\rightarrow 0$. Therefore,
the susceptibilities $\chi_{\mu\mu}$ and $\chi_{hh}$ are continuous across the critical point $h=h_c$. Then we consider the $l$-th derivative of the
susceptibilities with respect to $\mu$ or $h$. We find that the $l$-th derivative contains terms of which the infrared behavior goes as
\begin{eqnarray}
\int_0^\epsilon kdk \frac{k^{4-2l}}{k^3}=\int_0^\epsilon dk k^{2-2l}.
\end{eqnarray}
For $l=2$, infrared divergences show up, which means that the fourth derivative of the thermodynamic potential $\Omega(\mu,h)$ becomes divergent at the
critical point. Therefore, the third derivative of the thermodynamic potential is discontinuous across the critical point, which means that the
susceptibilities themselves are continuous but not smooth. Based on these observations, we conclude that the 2D system undergoes a third order
quantum phase transition at $h=h_c$ or ${\cal C}_0=0$ even though the superfluid order parameter does not undergo characteristic change.

On the other hand, we consider the so-called topological invariant ${\cal N}$ associated with the fermion Green's function
${\cal G}(i\omega,{\bf k})$. For the present 2D system it is defined as~\cite{TSC06,TopoN01,TopoN02}
\begin{eqnarray}
{\cal N}=\int\frac{d^2{\bf k}d\omega}{8\pi^2}\Bigg[{\rm Tr}\left({\cal G}\frac{\partial {\cal G}^{-1}}{\partial k_x}{\cal G}
\frac{\partial {\cal G}^{-1}}{\partial k_y}{\cal G}\frac{\partial {\cal G}^{-1}}{\partial \omega}\right)
-{\rm Tr}\left({\cal G}\frac{\partial {\cal G}^{-1}}{\partial k_y}{\cal G}\frac{\partial {\cal G}^{-1}}{\partial k_x}{\cal G}
\frac{\partial {\cal G}^{-1}}{\partial \omega}\right)\Bigg].
\end{eqnarray}
Using the explicit form of the fermion Green's function, one can show that ${\cal N}=0$ for ${\cal C}_0>0$ and ${\cal N}=1$ for ${\cal C}_0<0$.
Therefore, the superfluid phase with ${\cal C}_0<0$ or $h>h_c$ is topologically nontrivial. The phase transition at ${\cal C}_0=0$ or $h=h_c$ is a
topological quantum phase transition. The superfluid phase with ${\cal C}_0<0$ can be called a topological superfluid (TSF), while the topologically
trivial phase with ${\cal C}_0>0$ is a normal superfluid phase (SF).

%%%%%%%%%%%%%%%%%%%%%%%%%%%%%%%%%%%%%%%%%%%%%%%%%%%%%%%%%%%%%%%%%%%%%%%%%%%%%%%%%%%%%%%%%%%%%%%%%%%%%%%%%%%%%%
\subsubsection{Homogeneous system}
%%%%%%%%%%%%%%%%%%%%%%%%%%%%%%%%%%%%%%%%%%%%%%%%%%%%%%%%%%%%%%%%%%%%%%%%%%%%%%%%%%%%%%%%%%%%%%%%%%%%%%%%%%%%%%

\begin{figure}[!htb]
%\begin{center}
\includegraphics[width=7cm]{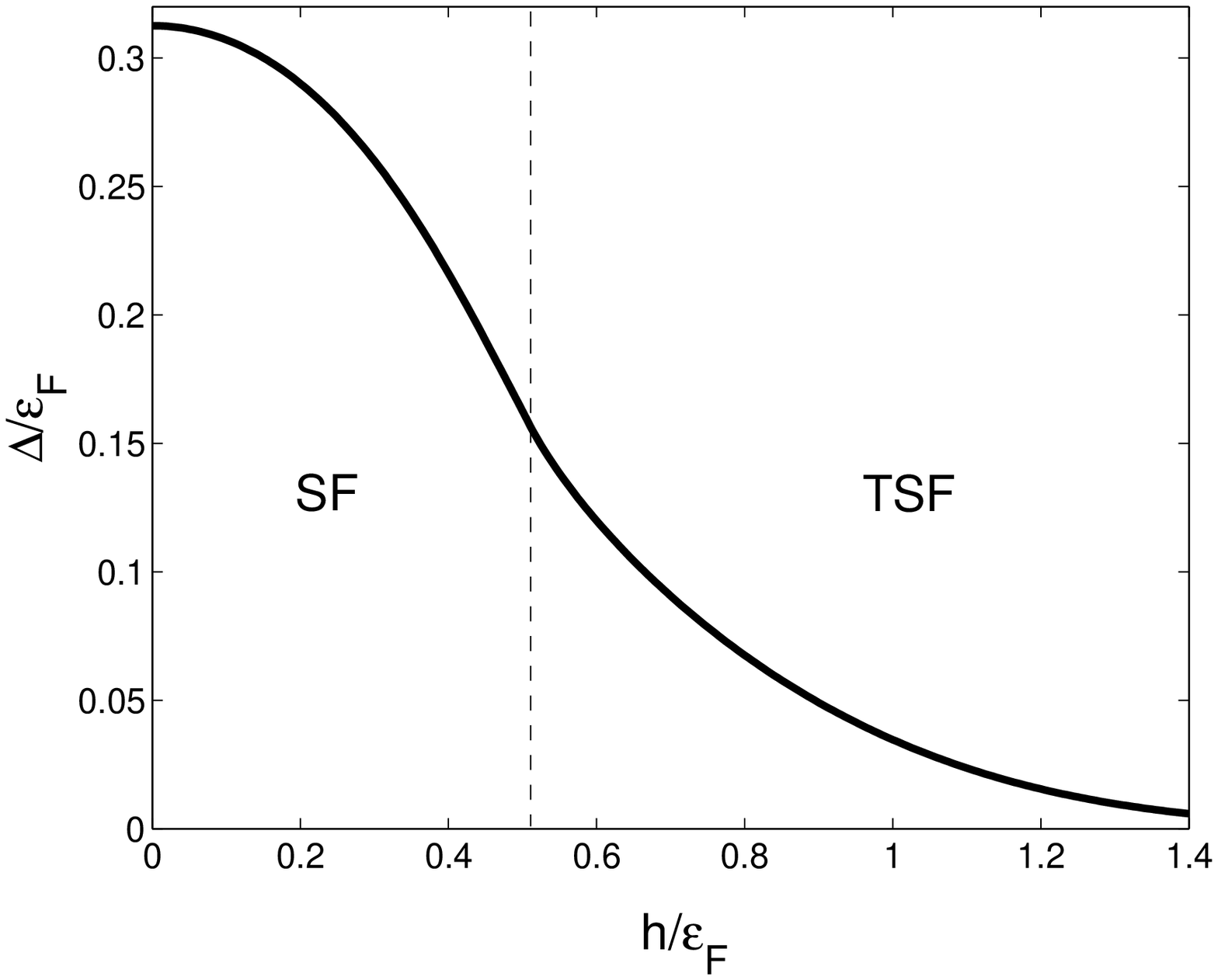}
\includegraphics[width=7cm]{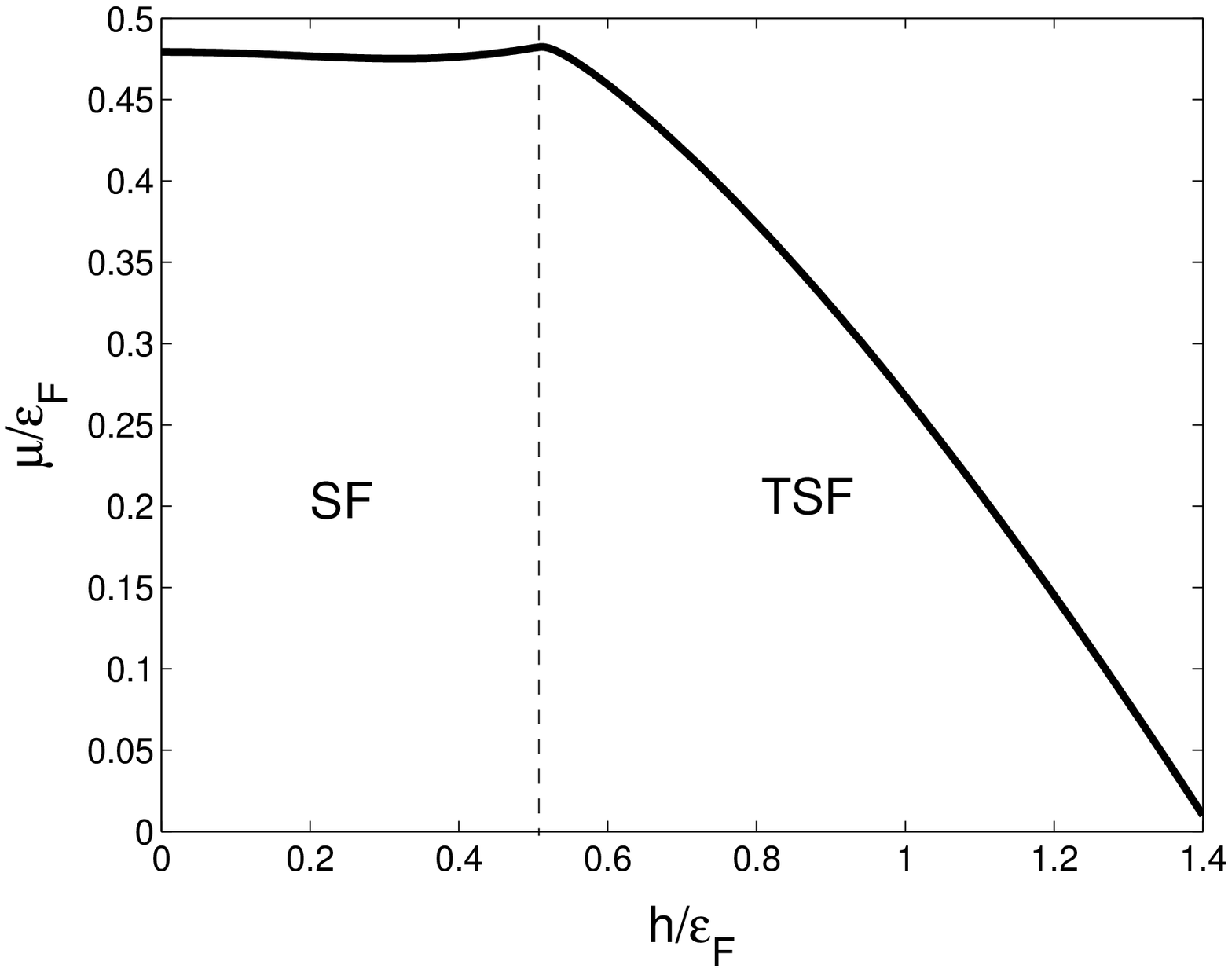}
\caption{The pairing gap $\Delta$ (divided by $\epsilon_{\rm F}$) and the chemical potential $\mu$ (divided by $\epsilon_{\rm F}$) as functions
of the Zeeman field $h/\epsilon_{\rm F}$. The dashed lines denote the critical Zeeman field $h_c=\sqrt{\mu^2+\Delta^2}$. In this plot, we take
$\ln(k_{\rm F}a_{2\rm D})=2$ and $\lambda/k_{\rm F}=0.5$. The value $h_c$ for this case is $h_c\simeq 0.51\epsilon_{\rm F}$.
 \label{GAPH2D}}
%\end{center}
\end{figure}

Now we turn to study the homogeneous system where the total density $n=k_{\rm F}^2/(2\pi)$ is fixed. We are interested in the topological quantum phase transition and the properties of the collective modes across the topological quantum phase transition. Therefore, we focus on the case of strong spin-orbit coupling strength $\lambda\sim O(k_{\rm F})$ which can be realized in the cold atom experiments~\cite{SOCF01,SOCF02}. As we have shown previously, the first-order phase transition has been eliminated and the order parameter goes smoothly with increasing Zeeman field. In the numerical calculations, we take the attractive coupling parameter as $\ln(k_{\rm F}a_{2\rm D})=2$. Change of this value does not lead to qualitatively different results. In Fig. \ref{GAPH2D}, we show typical numerical results for $\Delta$ and $\mu$ at $\lambda/k_{\rm F}=0.5$. Increasing the spin-orbit coupling strength does not change our results qualitatively. The pairing gap $\Delta$ drops down
smoothly with increased Zeeman field. At large $h$, we find numerically that $\Delta$ goes as $\Delta\sim 1/h^2$. The chemical potential $\mu$ becomes negative and goes as $\mu\simeq -h+b$ with $b\ll h$.

%%%%%%%%%%%%%%%%%%%%%%%%%%%%%%%%%%%%%%%%%%%%%%%%%%%%%%%%%%%%%%%%%%%%%%%
%%%%%%%%%%%%%%%%%%%%%%%%%%%%%%%%%%%%%%%%%%%%%%%%%%%%%%%%%%%%%%%%%%%%%%%%

%%%%%%%%%%%%%%%%%%%%%%%%%%%%%%%%%%%%%%%%%%%%%%%%%%%%%%%%%%%%%%%%%%%%%%%
\begin{figure}[!htb]
\begin{center}
\includegraphics[width=7.5cm]{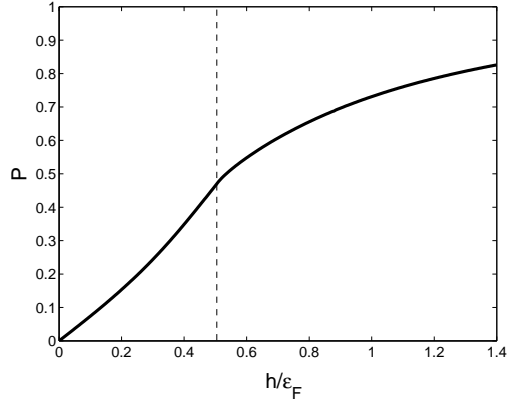}
\caption{The spin polarization $P=(n_\uparrow-n_\downarrow)/n$ as a function of $h/\epsilon_{\rm F}$.
 \label{NP2D}}
\end{center}
\end{figure}
%%%%%%%%%%%%%%%%%%%%%%%%%%%%%%%%%%%%%%%%%%%%%%%%%%%%%%%%%%%%%%%%%%%%%%%%

%%%%%%%%%%%%%%%%%%%%%%%%%%%%%%%%%%%%%%%%%%%%%%%%%%%%%%%%%%%%%%%%%%%%%%%
\begin{figure}[!htb]
\begin{center}
\includegraphics[width=8.5cm]{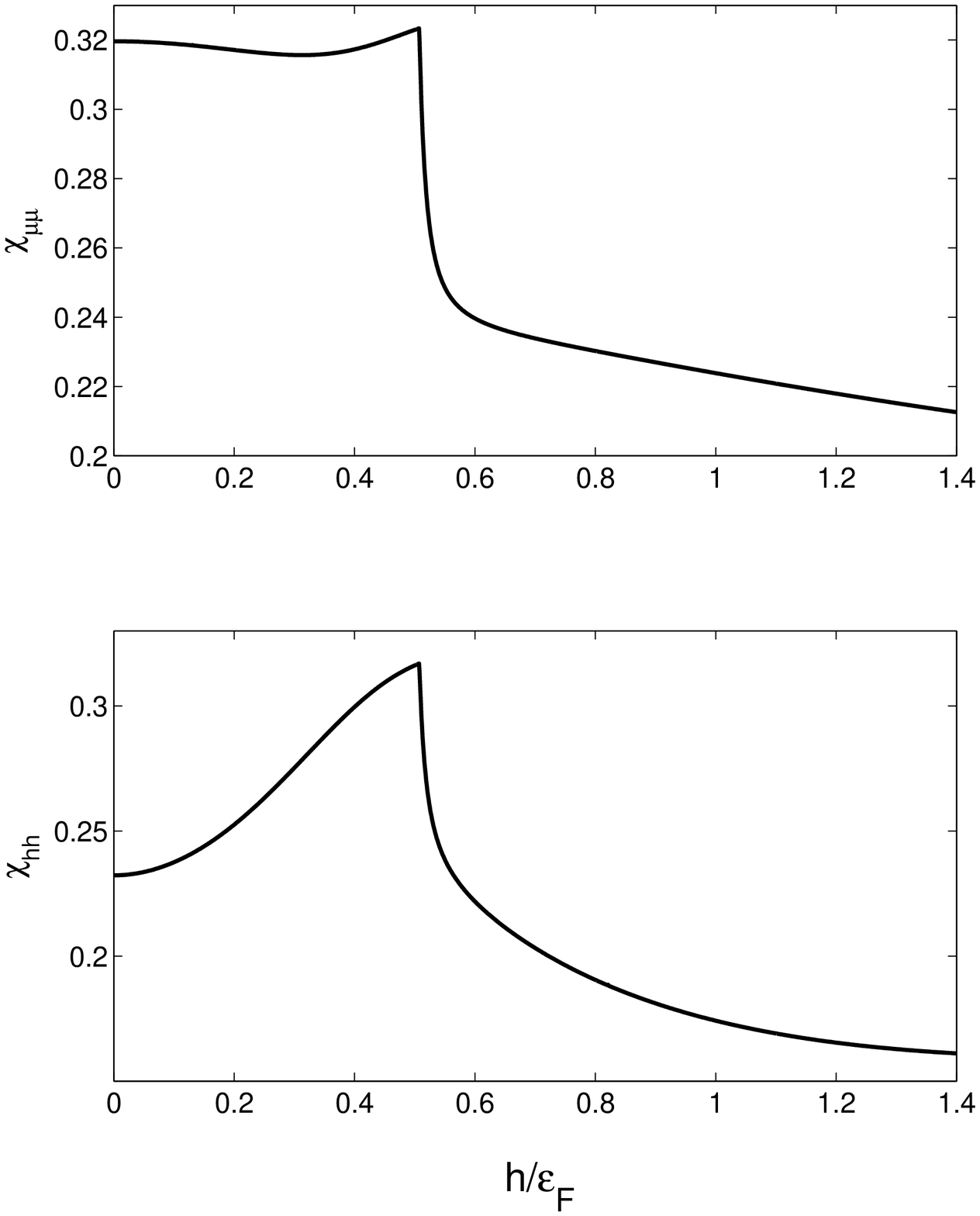}
\caption{The susceptibilities $\chi_{\mu\mu}$ and $\chi_{hh}$ as functions of $h/\epsilon_{\rm F}$.
 \label{CHI2D}}
\end{center}
\end{figure}
%%%%%%%%%%%%%%%%%%%%%%%%%%%%%%%%%%%%%%%%%%%%%%%%%%%%%%%%%%%%%%%%%%%%%%%%

For the present set of the attractive strength $\ln(k_{\rm F}a_{2\rm D})=2$ and the spin-orbit coupling strength $\lambda/k_{\rm F}=0.5$, we find
that the quantum phase transition occurs at $h=h_c\simeq0.51\epsilon_{\rm F}$. The chemical potential goes smoothly across the phase transition.
It reaches a maximum at $h=h_c$ and then drops down in the topological superfluid phase. The spin polarization $\delta n=n_\uparrow-n_\downarrow$
becomes nonzero once the Zeeman field is turned on. In Fig. \ref{NP2D}, we show the the spin polarization $P=\delta n/n$ as a function of $h$. We
find that $P$ is a smooth function of $h$, in accordance with the fact that the phase transition is of third order.  We note that the convexities
of the $\Delta$-$h$ and $P$-$h$ curves change at the critical field $h=h_c$. In Fig. \ref{CHI2D}, we show the susceptibilities $\chi_{\mu\mu}$ and
$\chi_{hh}$ as functions of $h/\epsilon_{\rm F}$. The susceptibilities are nonanalytical at the critical point $h=h_c$, as we expected. Note that
the spin susceptibility $\chi_{hh}$ is nonzero even at $h=0$. This is an important spin-orbit-coupling effect on fermionic
superfluidity/superconductivity~\cite{Gorkov}.

%%%%%%%%%%%%%%%%%%%%%%%%%%%%%%%%%%%%%%%%%%%%%%%%%%%%%%%%%%%%%%%%%%%%%%%
\begin{figure}[!htb]
\begin{center}
\includegraphics[width=7.5cm]{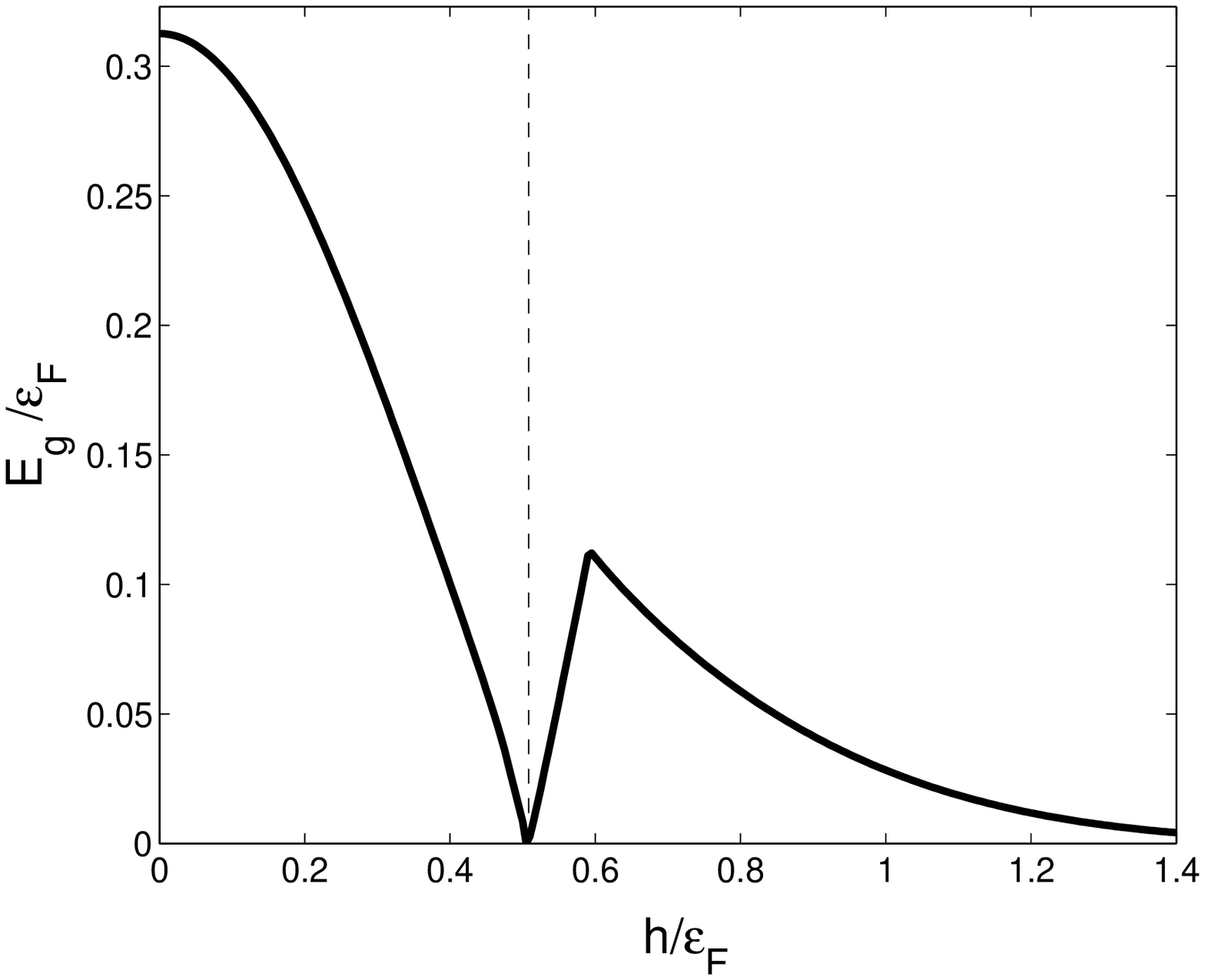}
\caption{The bulk excitation gap $E_{\rm g}$ (divided by $\epsilon_{\rm F}$) as a function of $h/\epsilon_{\rm F}$.
 \label{EGH2D}}
\end{center}
\end{figure}
%%%%%%%%%%%%%%%%%%%%%%%%%%%%%%%%%%%%%%%%%%%%%%%%%%%%%%%%%%%%%%%%%%%%%%%%

%%%%%%%%%%%%%%%%%%%%%%%%%%%%%%%%%%%%%%%%%%%%%%%%%%%%%%%%%%%%%%%%%%%%%%%
\begin{figure}[!htb]
\begin{center}
\includegraphics[width=7.5cm]{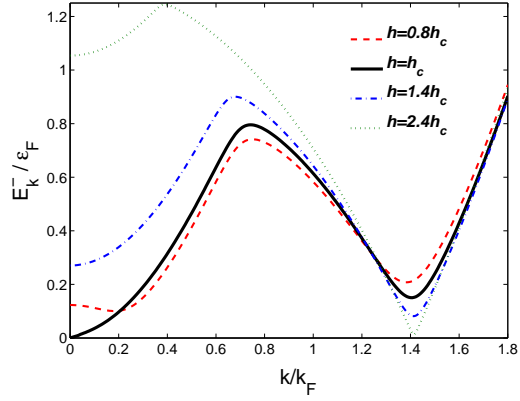}
\caption{(Color-online) The dispersion of the lower fermionic excitation $E_{\bf k}^-$ for various values of $h/h_c$.
 \label{DIS2D}}
\end{center}
\end{figure}
%%%%%%%%%%%%%%%%%%%%%%%%%%%%%%%%%%%%%%%%%%%%%%%%%%%%%%%%%%%%%%%%%%%%%%%%

The bulk single-particle excitation gap $E_g$ is defined as
\begin{eqnarray}
E_{\rm g}=\min_{\bf k}\{E_{\bf k}^+, E_{\bf k}^-\}=\min_{\bf k}\{E_{\bf k}^-\}.
\end{eqnarray}
For $h\neq0$, it does not equal the superfluid order parameter $\Delta$. In Fig. \ref{EGH2D}, we show the bulk excitation gap $E_{\rm g}$ as a function
of $h$. It vanishes only at the critical point $h=h_c$. In the SF phase ($h<h_c$), it is a decreasing function of $h$. In the TSF phase ($h>h_c$), it is
nonmonotonic and shows a maximum at $h\gtrsim h_c$. In Fig. \ref{DIS2D}, we show the dispersion of the lower fermionic excitation $E_{\bf k}^-$ for
various values of $h/h_c$. We find that the excitation gap $E_{\rm g}$ is located at different momenta for $h<h_c$ and $h>h_c$. For $h<h_c$, the gap is
located at low momentum, while for $h\gg h_c$, the momentum moves to $k\simeq\sqrt{2}k_{\rm F}$. This can be understood as follows. For $h\gg h_c$, the
pairing gap $\Delta$ is vanishingly small and the chemical potential $\mu$ becomes negative. To the leading order of $\Delta/h$ and $\Delta/|\mu|$, the
quasiparticle dispersions can be approximated as
\begin{eqnarray}
E_{\bf k}^\alpha\simeq\sqrt{(\xi_{\bf k}+\alpha\eta_{\bf k})^2+\Delta^2\left(1+\alpha\frac{h^2}{\xi_{\bf k}\eta_{\bf k}}\right)}.
\end{eqnarray}
Therefore, at $h\gg h_c$, the upper branch $E_{\bf k}^+$ has a large gap $\simeq h-\mu$ and essentially plays no role. The minimum of the lower branch
$E_{\bf k}^-$ is located at the ``Fermi surface" $k=\tilde{k}_{\rm F}$ determined by $\xi_{\bf k}-\eta_{\bf k}=0$. We have
\begin{eqnarray}
\tilde{k}_{\rm F}=\sqrt{2\left(\lambda^2+\mu+\sqrt{\lambda^4+2\lambda^2\mu+h^2}\right)}.
\end{eqnarray}
On the other hand, the total density $n$ can be well approximated as
\begin{eqnarray}
n\simeq\sum_{\bf k}\Theta(\eta_{\bf k}-\xi_{\bf k})=\frac{\tilde{k}_{\rm F}^2}{4\pi},
\end{eqnarray}
which leads to the result $\tilde{k}_{\rm F}\simeq \sqrt{2}k_{\rm F}$. Therefore, at $h\gg h_c$, the excitation gap $E_{\rm g}$ of the lower branch is
opened at the Fermi surface $k=\tilde{k}_{\rm F}$. Near the Fermi surface, the dispersion $E_{\bf k}^-$ can be approximated as
\begin{eqnarray}
\label{largehspec}
E_{\bf k}^-\simeq\sqrt{\tilde{\upsilon}_{\rm F}^2(k-\tilde{k}_{\rm F})^2+E_{\rm g}^2},
\end{eqnarray}
where the Fermi velocity $\tilde{\upsilon}_{\rm F}=\partial\xi_{\bf k}^-/\partial k|_{k=\tilde{k}_{\rm F}}$ reads
\begin{eqnarray}
\tilde{\upsilon}_{\rm F}=\sqrt{2}\upsilon_{\rm F}\left(1-\frac{\lambda^2}{\sqrt{2\lambda^2k_{\rm F}^2+h^2}}\right)
\end{eqnarray}
and the excitation gap $E_{\rm g}$ is given by
\begin{eqnarray}
E_{\rm g}=\Delta\sqrt{\frac{2\lambda^2k_{\rm F}^2}{2\lambda^2k_{\rm F}^2+h^2}}.
\end{eqnarray}

It is interesting to understand the above discussion in the helicity representation, i.e., Eqs. (\ref{helicity1})-(\ref{helicity5}). For $h\gg h_c$,
only the lower band with dispersion $\xi_{\bf k}^-$ has a Fermi surface $k=\tilde{k}_{\rm F}$ and carries the total fermion density $n$. The upper
band plays essentially no role and only the intraband pairing near the Fermi surface of the lower band is available. At large Zeeman field, the
interband pair potential $\Delta_{\rm s}({\bf k})$ can be safely dropped and the quasiparticle dispersion $E_{\bf k}^-$ can be well approximated as
\begin{eqnarray}
E_{\bf k}^-\simeq\sqrt{\left(\xi_{\bf k}-\eta_{\bf k}\right)^2+|\Delta_{\rm t}({\bf k})|^2}.
\end{eqnarray}
Near the Fermi surface $k=\tilde{k}_{\rm F}$, it gives exactly Eq. (\ref{largehspec}). This means that, for $h\rightarrow\infty$, the system can be
mapped to the $p_x+ip_y$ superfluid state in spinless Fermi gases~\cite{PIP}. Here the complex $p$-wave pairing occurs near the Fermi surface
$k=\tilde{k}_{\rm F}$ of the lower band.  However,  if the Zeeman field is not large enough, the mapping to a spinless $p_x+ip_y$ superfluid state is
no longer legitimate.

%%%%%%%%%%%%%%%%%%%%%%%%%%%%%%%%%%%%%%%%%%%%%%%%%%%%%%%%%%%%%%%%%%%%%%%%%%%%%%%%%%%%%%%%%%%%%%%%%%%%%%%%%%%%%%
\subsubsection{Collective modes}
\begin{figure}[!htb]
\begin{center}
\includegraphics[width=8cm]{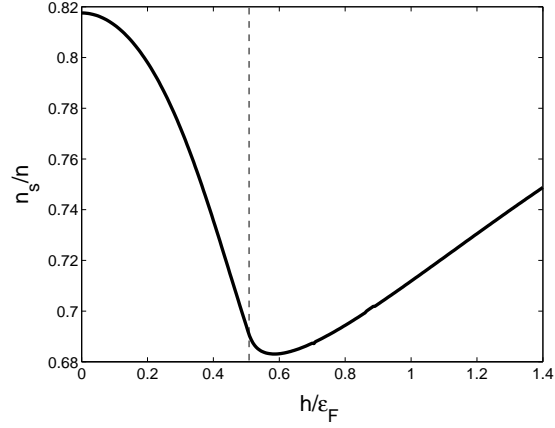}
\caption{The superfluid density $n_s$ (divided by the total density $n$)
as a function of $h/\epsilon_{\rm F}$.
 \label{NSH2D}}
\end{center}
\end{figure}

%%%%%%%%%%%%%%%%%%%%%%%%%%%%%%%%%%%%%%%%%%%%%%%%%%%%%%%%%%%%%%%%%%%%%%%%%%%%%%%%%%%%%%%%%%%%%%%%%%%%%%%%%%%%%%
Since the fermionic excitations are fully gapped except at the critical point $h=h_c$, the superfluid density can be expressed as
\begin{eqnarray}
n_s=n-\sum_{\alpha=\pm}\sum_{\bf k}\frac{\lambda^2}{2E_{\bf k}^\alpha}
\Bigg[\left(1-\frac{\lambda^2{\bf k}^2\xi_{\bf k}^2}{2\zeta_{\bf k}^2}\right)
+\alpha\left(1+\frac{h^2E_{\bf k}^2}{\zeta_{\bf k}^2}
+\frac{\lambda^2{\bf k}^2h^2\Delta^2}{\zeta_{\bf k}^2E_{\bf k}^2}\right)\frac{E_{\bf k}^2}{2\zeta_{\bf k}}\Bigg].
\end{eqnarray}
Analyzing the infrared behavior of the integrals at $h=h_c$, we find that $n_s$ goes smoothly across the quantum phase transition. In Fig. \ref{NSH2D},
we show the numerical result for the superfluid density. In the SF phase, $n_s$ is suppressed by the Zeeman field, as we expected. However, in the TSF
phase, it turns to be enhanced by the Zeeman field! To understand this surprising behavior, we first take a look at the case $h\gg h_c$. In this case,
$\Delta$ is very small compared with $h$ and $|\mu|$ and the superfluid density can be well approximated as
\begin{eqnarray}
n_s&\simeq&n-\frac{\lambda^2}{2}\sum_{\bf k}\frac{1}{\eta_{\bf k}}\left(1+\frac{h^2}{\eta_{\bf k}^2}\right)
\Theta(\tilde{k}_{\rm F}-|{\bf k}|)\nonumber\\
&=&n\left(1-\frac{\lambda^2}{\sqrt{2\lambda^2k_{\rm F}^2+h^2}}\right).
\end{eqnarray}
Therefore, for $h\rightarrow\infty$, we have $n_s\rightarrow n$. As we have shown above, for $h\gg h_c$, the topological superfluid state can be mapped
to a spinless $p_x+ip_y$ superfluid state where weak $p$-wave pairing occurs at the Fermi surface $k=\tilde{k}_{\rm F}$. Therefore, the fermion pairing
in the TSF phase feels less stress from the Zeeman field than in the normal superfluid phase. At large enough Zeeman field, since the lower band carries
almost the total fermion density $n$, we naturally have $n_s\rightarrow n$.

%%%%%%%%%%%%%%%%%%%%%%%%%%%%%%%%%%%%%%%%%%%%%%%%%%%%%%%%%%%%%%%%%%%%%%%
%%%%%%%%%%%%%%%%%%%%%%%%%%%%%%%%%%%%%%%%%%%%%%%%%%%%%%%%%%%%%%%%%%%%%%%%

%%%%%%%%%%%%%%%%%%%%%%%%%%%%%%%%%%%%%%%%%%%%%%%%%%%%%%%%%%%%%%%%%%%%%%%
\begin{figure}[!htb]
\begin{center}
\includegraphics[width=9cm]{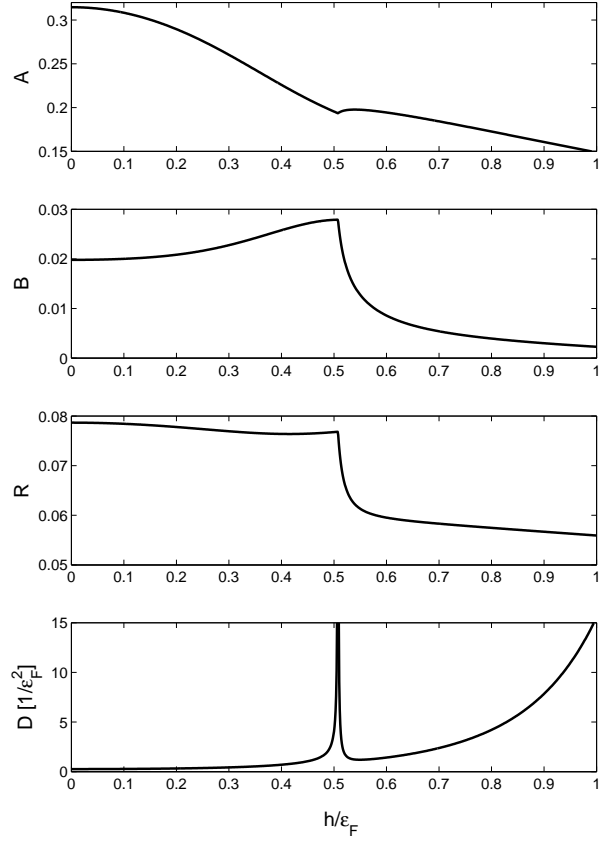}
\caption{The expansion parameters  $A,B,R,D$  as functions
of $h/\epsilon_{\rm F}$.
 \label{ABRD2D}}
\end{center}
\end{figure}
%%%%%%%%%%%%%%%%%%%%%%%%%%%%%%%%%%%%%%%%%%%%%%%%%%%%%%%%%%%%%%%%%%%%%%%%

%%%%%%%%%%%%%%%%%%%%%%%%%%%%%%%%%%%%%%%%%%%%%%%%%%%%%%%%%%%%%%%%%%%%%%%
\begin{figure}[!htb]
\begin{center}
\includegraphics[width=8cm]{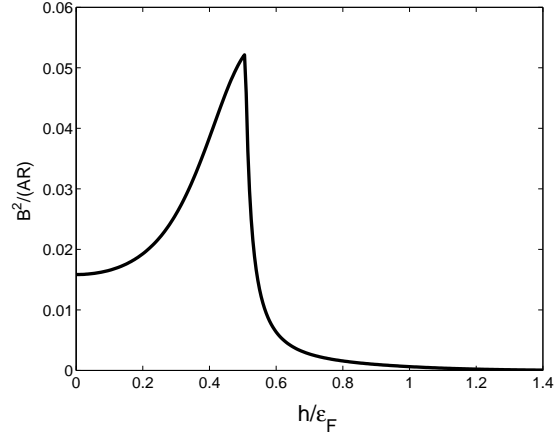}
\caption{The ratio $B^2/(AR)$ which represents the mixing strength between the
phase and amplitude modes as a function of $h/\epsilon_{\rm F}$.
 \label{MIX2D}}
\end{center}
\end{figure}
%%%%%%%%%%%%%%%%%%%%%%%%%%%%%%%%%%%%%%%%%%%%%%%%%%%%%%%%%%%%%%%%%%%%%%%%

%%%%%%%%%%%%%%%%%%%%%%%%%%%%%%%%%%%%%%%%%%%%%%%%%%%%%%%%%%%%%%%%%%%%%%%
\begin{figure}[!htb]
\begin{center}
\includegraphics[width=8cm]{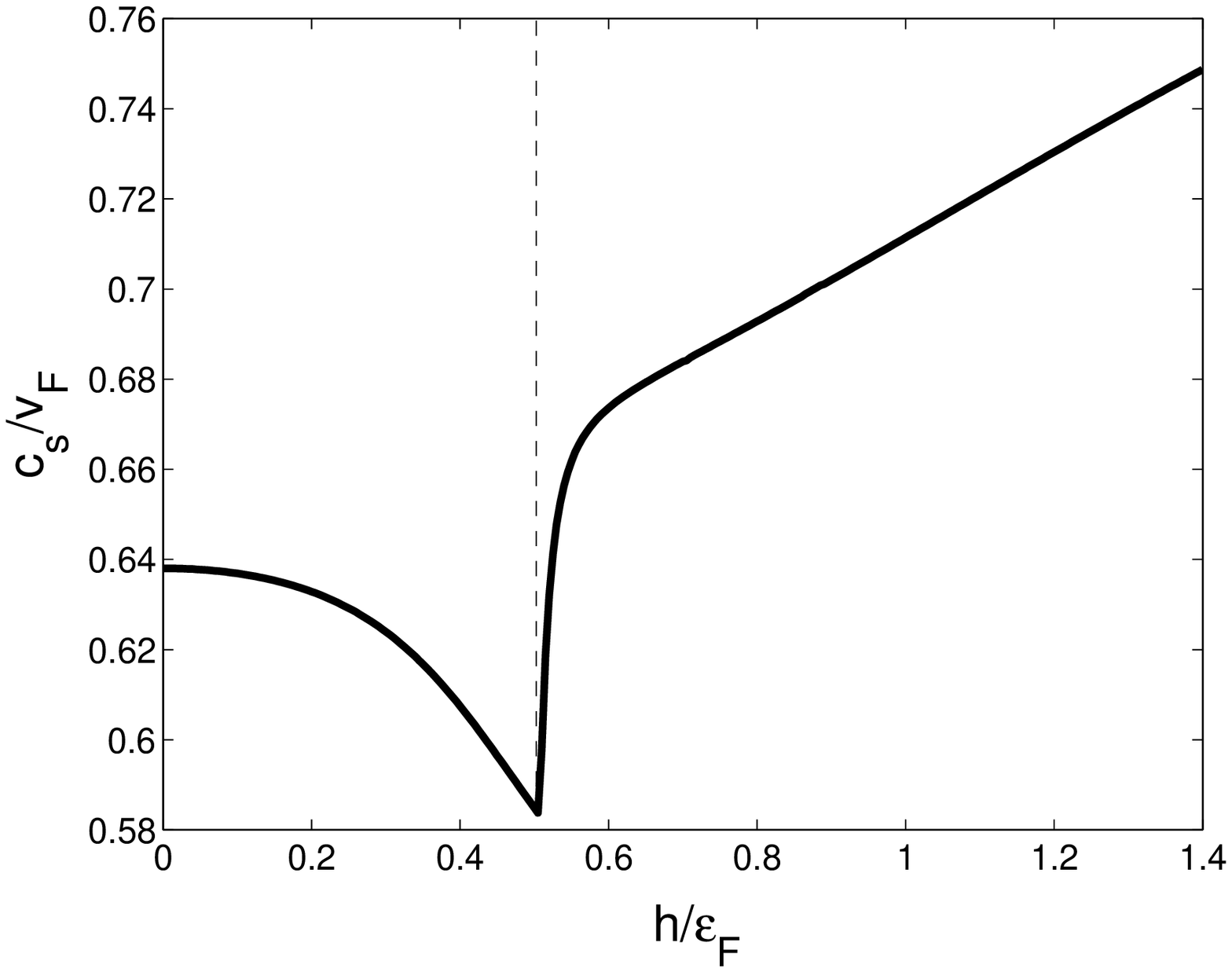}
\caption{The sound velocity $c_s$ (divided by the Fermi velocity $\upsilon_{\rm F}=k_{\rm F}/m$  for the
noninteracting Fermi gas in the absence of  spin-orbit coupling) as a function of $h/\epsilon_{\rm F}$.
 \label{CSH2D}}
\end{center}
\end{figure}
%%%%%%%%%%%%%%%%%%%%%%%%%%%%%%%%%%%%%%%%%%%%%%%%%%%%%%%%%%%%%%%%%%%%%%%%

Next, the behavior of the expansion parameters $A,B,D,R$ across the critical point $h=h_c$ can be summarized as follows: (1) $A$, $B$ and $R$ are
continuous but nonanalytical; (2) $D$ is divergent. These conclusions can be obtained by analyzing the infrared behavior of the momentum integrals.
The numerical results for these parameters are shown in Fig. \ref{ABRD2D}. The ratio $B^2/(AR)$ which represents the mixing between the phase and
amplitude modes are shown in Fig. \ref{MIX2D}. We find that the mixing is strongest near the critical point, while it becomes negligible for
$h\gg h_c$. Since the parameters $A,B$ and $R$ are nonanalytical at $h=h_c$, the Goldstone mode velocity $c_s$ also behaves nonanalytically across
the quantum phase transition. In Fig. \ref{CSH2D}, we show the numerical result of $c_s$ as a function of the Zeeman field. It is suppressed by the
Zeeman field in the SF phase ($h<h_c$), as we expected. However, for $h>h_c$, it goes up rapidly and is always enhanced by the Zeeman field. Therefore,
the low-energy collective mode, i.e., the Goldstone sound mode, is a sensitive probe of the quantum phase transition in 2D spin-orbit coupled Fermi
superfluids. For $h\gg h_c$, the quantities $A,B,R$ are dominated by the integrals over the lower band. They can be well approximated as
\begin{eqnarray}
&&A\simeq\frac{\Delta^2}{2}\sum_{\bf k}\frac{1}{(E_{\bf k}^-)^3}\left(1-\frac{h^2}{\zeta_{\bf k}}\right)^2,\nonumber\\
&&B\simeq\frac{\Delta}{4}\sum_{\bf k}\frac{\xi_{\bf k}}{(E_{\bf k}^-)^3}
\left(1-\frac{\lambda^2{\bf k}^2}{\zeta_{\bf k}}-\frac{h^2E_{\bf k}^2}{\zeta_{\bf k}^2}\right),\nonumber\\
&&R\simeq\frac{\Delta^2}{8}\sum_{\bf k}\frac{1}{(E_{\bf k}^-)^3}\frac{\lambda^2{\bf k}^2\xi_{\bf k}^2}{\zeta_{\bf k}^2}.
\end{eqnarray}
Since all integrands are peaked at the Fermi surface $k=\tilde{k}_{\rm F}$, we can set $k=\tilde{k}_{\rm F}$ for the integrands except
$E_{\bf k}^-\simeq\sqrt{\tilde{\upsilon}_{\rm F}^2(k-\tilde{k}_{\rm F})^2+E_{\rm g}^2}$. Meanwhile, we have $\zeta_{\bf k}\simeq
\xi_{\bf k}\eta_{\bf k}\simeq \eta_{\bf k}^2$ at $k=\tilde{k}_{\rm F}$ since $\Delta$ goes as $\Delta\sim1/h^2$. Then we obtain
\begin{eqnarray}
&&\frac{B^2}{A}\simeq\frac{1}{64\pi}\frac{\Delta^2}{\lambda^2k_{\rm F}^2}\frac{h^4}{(2\lambda^2k_{\rm F}^2+h^2)^2}
\left(1-\frac{\lambda^2}{\sqrt{2\lambda^2k_{\rm F}^2+h^2}}\right)^{-1},\nonumber\\
&&R\simeq\frac{1}{8\pi}\left(1-\frac{\lambda^2}{\sqrt{2\lambda^2k_{\rm F}^2+h^2}}\right)^{-1}.
\end{eqnarray}
Therefore, for $h\rightarrow\infty$, we have $B^2/A\rightarrow 0$ and $R\rightarrow 1/(8\pi)$. Since the phase-amplitude mixing becomes negligible,
the Goldstone mode velocity is given by $c_s=\sqrt{n_s/(4R)}$. Combining the fact $n_s\rightarrow n$ for $h\rightarrow\infty$, we obtain
\begin{eqnarray}
c_s\rightarrow \upsilon_{\rm F}\ \ \ {\rm for} \ \ \ h\rightarrow\infty.
\end{eqnarray}
This result can be reexpressed as $c_s\simeq \tilde{\upsilon}_{\rm F}/\sqrt{2}$. It is just the sound velocity of a weakly coupled 2D Fermi superfluid,
which is consistent with the fact that the system can be mapped to a weakly coupled $p_x+ip_y$ superfluid.

On the other hand, the divergence of $D$ at the quantum phase transition indicates that $M_{\rm H}$ vanishes, that is, the massive Higgs mode gets
softened near the critical point. In general, the Higgs mode is a resonance. Therefore, we study the nature of its spectral density
$\rho(\omega, {\bf q})=-(1/\pi){\rm Im}\Gamma_{\rm H}(\omega+i\epsilon, {\bf q})$, where the Higgs mode propagator $\Gamma_{\rm H}(Q)$ is the massive
eigenmode obtained by diagonalizing the matrix ${\bf M}(Q)$. Here we briefly discuss the case of ${\bf q}=0$. From the explicit form of the collective
mode propagator, we find that the spectral density arises when the frequency $\omega$ exceeds the threshold $\omega_{\rm th}=2E_{\rm g}$, which is just
the two-particle continuum $E_{\rm c}({\bf q})=\min_{\bf k}\{E_{{\bf k}+{\bf q}/2}^-+E_{{\bf k}-{\bf q}/2}^-\}$ at ${\bf q}=0$. Near the quantum phase
transition point $h=h_c$, the threshold $\omega_{\rm th}$ tends to zero, while the pairing gap $\Delta$ (as well as the critical temperature $T_c$)
remains relatively large. This indicates that the Higgs mode spectrum arises in the low-frequency regime near the critical point. It may be interesting
for the future experimental search of the Higgs mode in fermionic superfluids.

%%%%%%%%%%%%%%%%%%%%%%%%%%%%%%%%%%%%%%%%%%%%%%%%%%%%%%%%%%%%%%%%%%%%%%%
\begin{figure}[!htb]
%\begin{center}
\includegraphics[width=7cm]{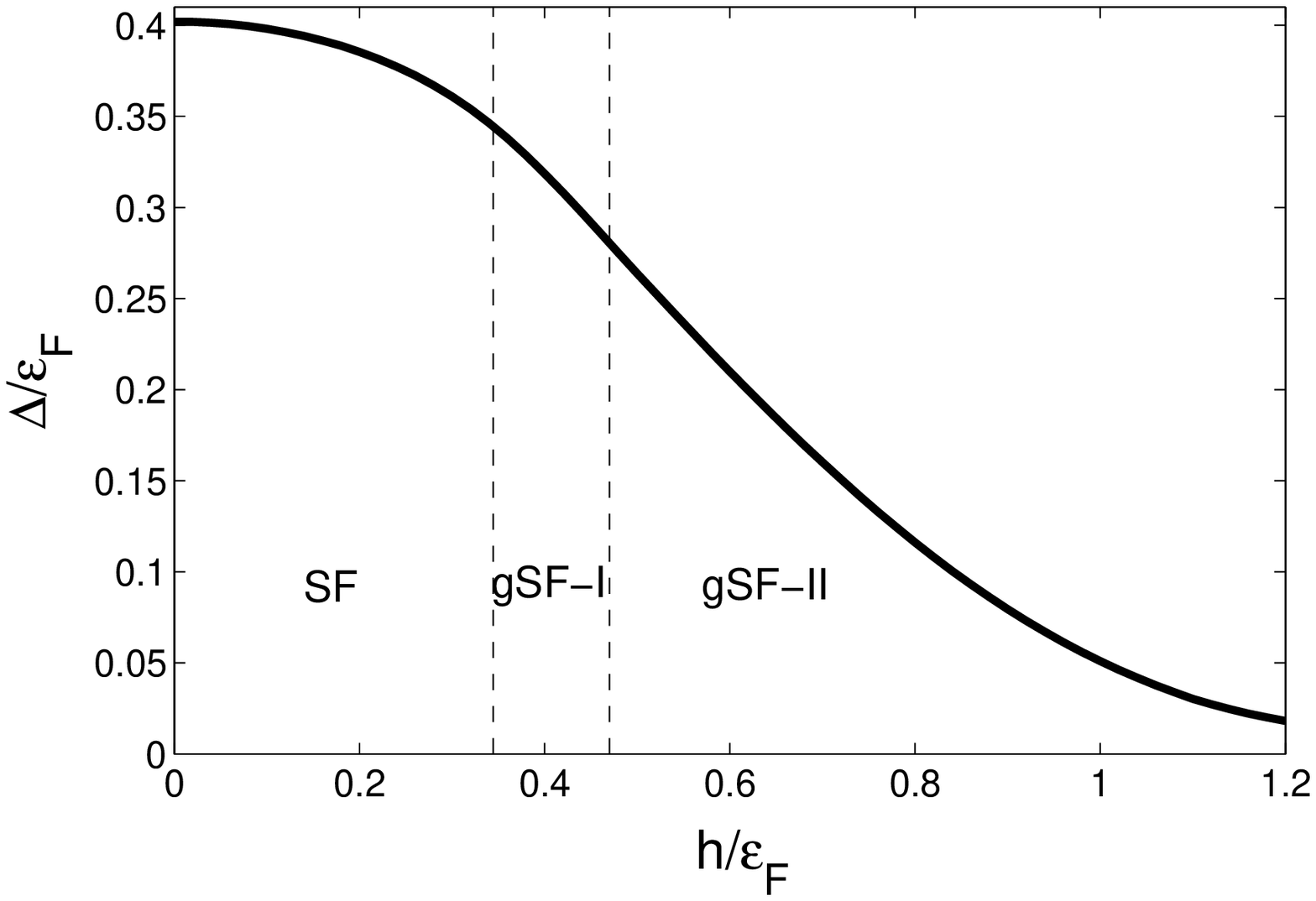}
\includegraphics[width=7cm]{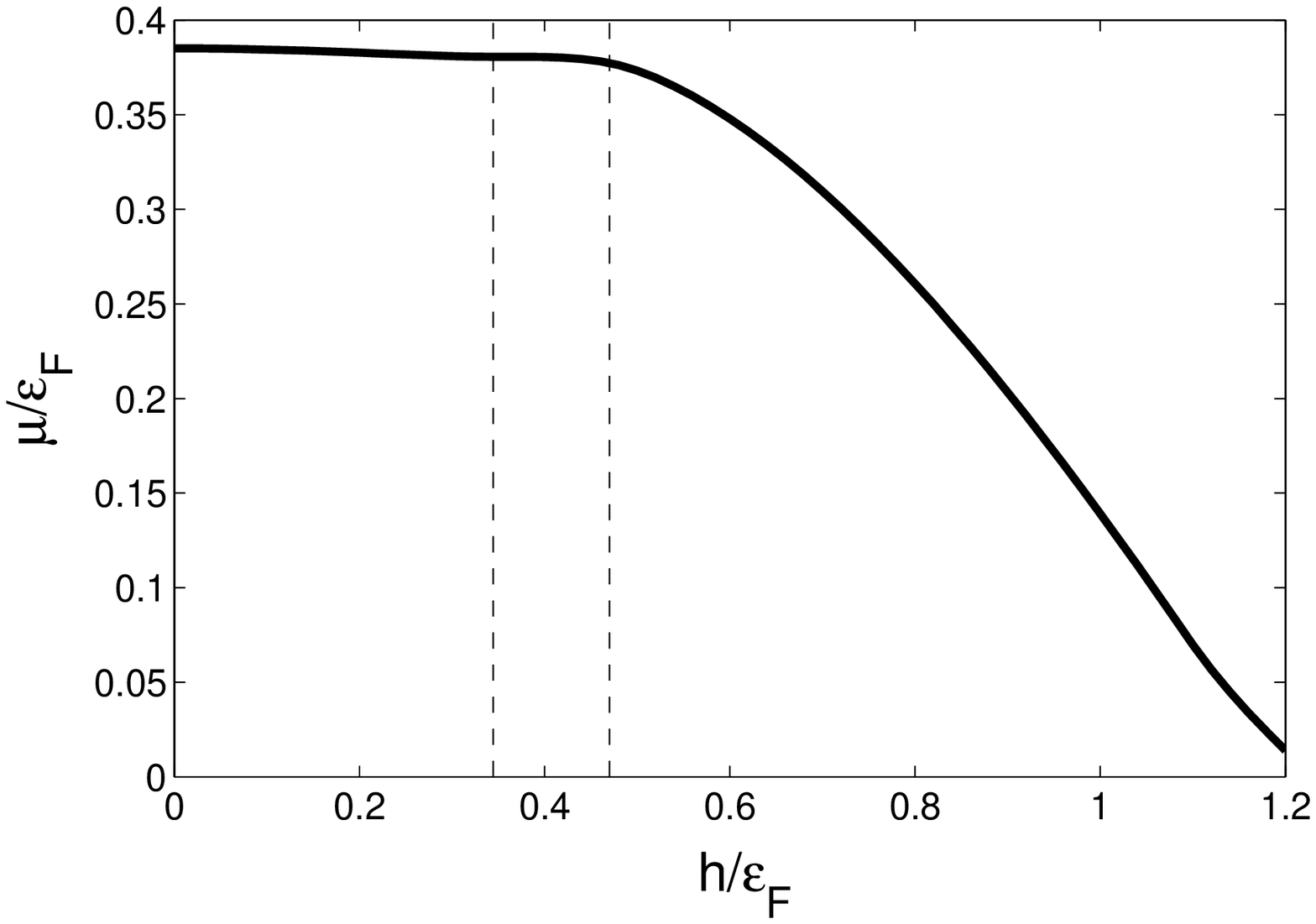}
\caption{The pairing gap $\Delta$ and the chemical potential $\mu$ for the 3D system as functions of the Zeeman field.
In this plot we take $1/(k_{\rm F}a_s)=-0.5$ and $\lambda/k_{\rm F}=0.5$. The two critical fields in this case read
$h_{c1}\simeq 0.34\epsilon_{\rm F}$ and $h_{c2}\simeq 0.47\epsilon_{\rm F}$.
 \label{GAPH3D}}
%\end{center}
\end{figure}
%%%%%%%%%%%%%%%%%%%%%%%%%%%%%%%%%%%%%%%%%%%%%%%%%%%%%%%%%%%%%%%%%%%%%%%%

%%%%%%%%%%%%%%%%%%%%%%%%%%%%%%%%%%%%%%%%%%%%%%%%%%%%%%%%%%%%%%%%%%%%%%%
\begin{figure}[!htb]
\begin{center}
\includegraphics[width=8cm]{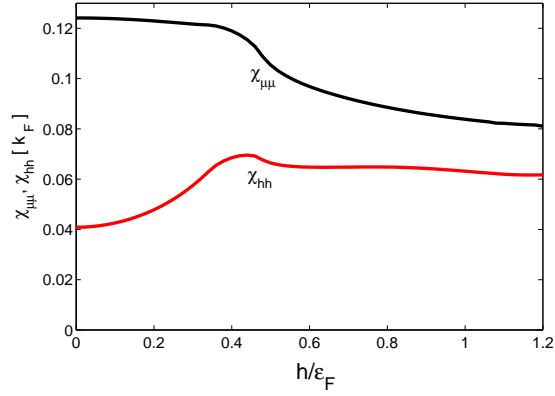}
\caption{(Color-online) The susceptibilities $\chi_{\mu\mu}$ and $\chi_{hh}$ for the 3D system as functions of $h/\epsilon_{\rm F}$.
 \label{CHI3D}}
\end{center}
\end{figure}
%%%%%%%%%%%%%%%%%%%%%%%%%%%%%%%%%%%%%%%%%%%%%%%%%%%%%%%%%%%%%%%%%%%%%%%%

%%%%%%%%%%%%%%%%%%%%%%%%%%%%%%%%%%%%%%%%%%%%%%%%%%%%%%%%%%%%%%%%%%%%%%%
\begin{figure}[!htb]
\begin{center}
\includegraphics[width=9cm]{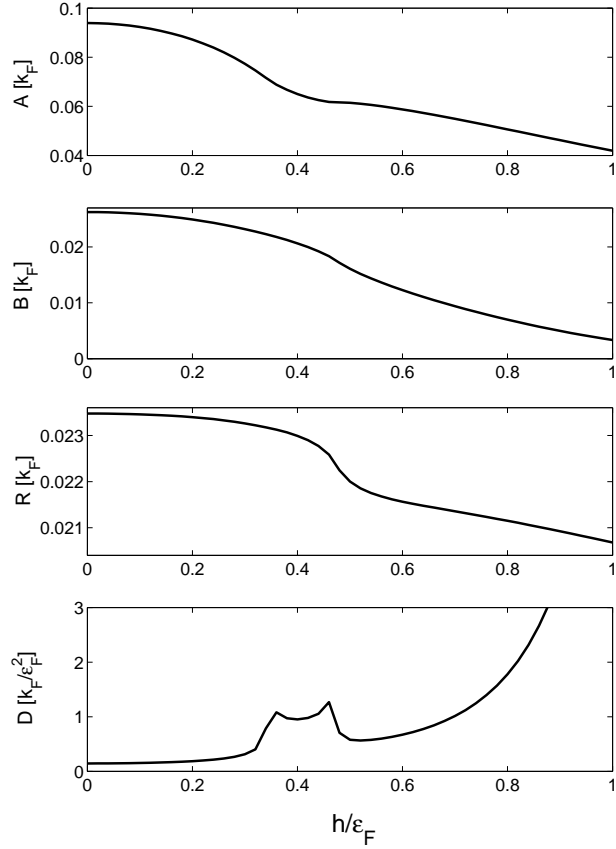}
\caption{The expansion parameters $A,B,R$ and $D$ for the 3D system as functions of $h/\epsilon_{\rm F}$.
 \label{ABRD3D}}
\end{center}
\end{figure}
%%%%%%%%%%%%%%%%%%%%%%%%%%%%%%%%%%%%%%%%%%%%%%%%%%%%%%%%%%%%%%%%%%%%%%%%

%%%%%%%%%%%%%%%%%%%%%%%%%%%%%%%%%%%%%%%%%%%%%%%%%%%%%%%%%%%%%%%%%%%%%%%
\begin{figure}[!htb]
\begin{center}
\includegraphics[width=8cm]{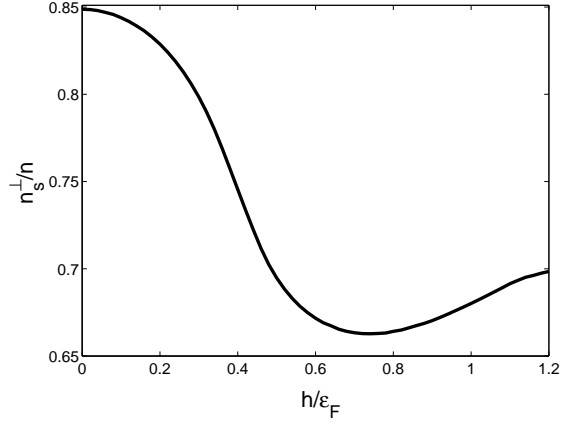}
\caption{The transverse superfluid density $n_s^\bot$ (divided by the total density $n$)
for the 3D system as a function of $h/\epsilon_{\rm F}$.
 \label{NSH3D}}
\end{center}
\end{figure}
%%%%%%%%%%%%%%%%%%%%%%%%%%%%%%%%%%%%%%%%%%%%%%%%%%%%%%%%%%%%%%%%%%%%%%%%

%%%%%%%%%%%%%%%%%%%%%%%%%%%%%%%%%%%%%%%%%%%%%%%%%%%%%%%%%%%%%%%%%%%%%%%
\begin{figure}[!htb]
\begin{center}
\includegraphics[width=8.5cm]{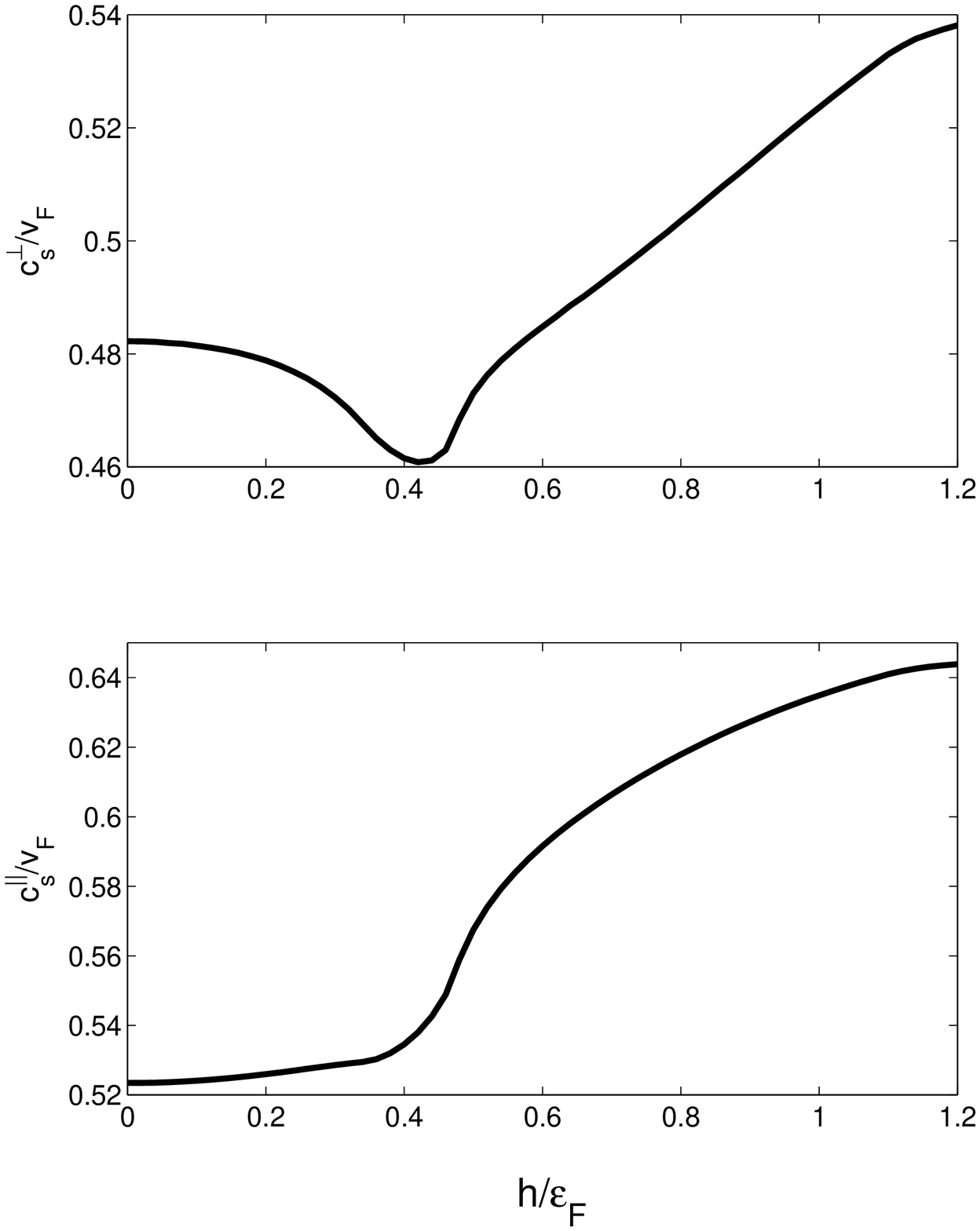}
\caption{The transverse and longitudinal velocities of the Goldstone mode, $c_s^\bot$ and $c_s^\|$, as functions of $h/\epsilon_{\rm F}$.
 \label{CSH3D}}
\end{center}
\end{figure}
%%%%%%%%%%%%%%%%%%%%%%%%%%%%%%%%%%%%%%%%%%%%%%%%%%%%%%%%%%%%%%%%%%%%%%%%

%%%%%%%%%%%%%%%%%%%%%%%%%%%%%%%%%%%%%%%%%%%%%%%%%%%%%%%%%%%%%%%%%%%%%%%%%%%%%%%%%%%%%%%%%%%%%%%%%%%%%%%%%%%%%%
\subsection{3D system}
%%%%%%%%%%%%%%%%%%%%%%%%%%%%%%%%%%%%%%%%%%%%%%%%%%%%%%%%%%%%%%%%%%%%%%%%%%%%%%%%%%%%%%%%%%%%%%%%%%%%%%%%%%%%%%
The three-dimensional case is different from the 2D case. The superfluid phase at large $h$ is a gapless superfluid state where the lower
quasiparticle branch has gapless nodes~\cite{SOC-Gong}. This difference is due to the existence of the $k_z$ degree of freedom in the 3D system.
From the identity
\begin{eqnarray}
(E_{\bf k}^+)^2(E_{\bf k}^-)^2=(E_{\bf k}^2-h^2-\lambda^2{\bf k}_\bot^2)^2+4\lambda^2{\bf k}_\bot^2\Delta^2,
\end{eqnarray}
we find that the lower branch $E_{\bf k}^-$ has some zeros located at the $k_z$-axis (${\bf k}_\bot=0$). The values of $k_z$ of these zeros are
determined by
\begin{eqnarray}
\left(k_z^2/2-\mu\right)^2+\Delta^2=h^2.
\end{eqnarray}
These zeros are isolated and are called Fermi points. There exist three possible phases according to the number of Fermi points: (1) for $h<\Delta$,
we have a fully gapped phase (SF) without Fermi points; (2) for $\mu>0$ and $\Delta<h<\sqrt{\mu^2+\Delta^2}$, we have a gapless phase (called gSF-I)
with four Fermi points given by $({\bf k}_\bot, k_z)=({\bf 0},\pm[2(\mu+\sqrt{h^2-\Delta^2})]^{1/2})$ and
$({\bf k}_\bot, k_z)=({\bf 0},\pm[2(\mu-\sqrt{h^2-\Delta^2})]^{1/2})$; (3) for $h>\sqrt{\mu^2+\Delta^2}$, we have a gapless phase (called gSF-II) with
two Fermi points given by $({\bf k}_\bot, k_z)=({\bf 0},\pm[2(\mu+\sqrt{h^2-\Delta^2})]^{1/2})$. One can show that the quasiparticle dispersion
$E_{\bf k}^-$ is linear around each Fermi point. Therefore, each Fermi point behaves like a Dirac point and is topologically protected with a topological
charge $N_a=\pm1$~\cite{SOC-Gong}. The three superfluid phases (SF, gSF-I, gSF-II) are therefore lined by two topological quantum phase transitions.
For convenience, in the following we denote the two critical Zeeman fields as $h_{c1}=\Delta$ and $h_{c2}=\sqrt{\mu^2+\Delta^2}$.

Since the gapless quasiparticle spectrum has only Fermi points rather than Fermi surfaces, the delta-function terms in the expressions of $J_\bot, J_\|$
and $A$ vanish automatically. This is quite different from the case of vanishing spin-orbit coupling. For that case, the quasiparticle spectrum in the
gapless phase has gapless Fermi surfaces and delta-function terms have large negative contributions due to the large density of state at the gapless
Fermi surfaces~\cite{NSI01,NSI02,NSI03,NSI04,NSI05,NSI06,NSI07,Gubankova,Lamacraft}. As a result, the gapless superfluid state in the absence of
spin-orbit coupling is unstable in a large regime of the BCS-BEC crossover~\cite{NSI02}, because $A$ and the superfluid density are negative there.
For large enough spin-orbit coupling, the gapless phases become stable, since these negative contributions are totally removed. However, the cost is that
we now have only Fermi points rather than Fermi surfaces.

In Fig. \ref{GAPH3D}, we show the results of the pairing gap $\Delta$ and the chemical potential $\mu$ for attraction strength
$1/(k_{\rm F}a_s)=-0.5$ and spin-orbit coupling $\lambda/k_{\rm F}=0.5$. Similar to the 2D case, the pairing gap never vanishes at large Zeeman
field, and we do have two critical Zeeman fields $h_{c1}\simeq 0.34\epsilon_{\rm F}$ and $h_{c2}\simeq 0.47\epsilon_{\rm F}$ which separate the three
superfluid phases. For stronger attraction strength and/or spin-orbit coupling, the chemical potential $\mu$ becomes smaller or even negative at $h=0$.
In this case, the gSF-I phase with four Fermi points does not appear. In Fig. \ref{CHI3D}, we show the results for the susceptibilities $\chi_{\mu\mu}$
and $\chi_{hh}$ which are defined in the same way as the 2D case. We find that they behave smoothly across the quantum phase transitions, which
indicates that the quantum phase transition in the 3D case is of higher than third order. The reason can be understood as follows: due to the existence
of the $k_z$ degree of freedom, the infrared singularities at ${\bf k}_\bot\rightarrow 0$ are weakened by the integrals over $k_z$. Therefore, the
expansion parameters $A, B, R$ behave smoothly across the quantum phase transitions, while $D$ does not diverge but exhibits nonanalytical behavior.
The numerical results of these parameters are shown in Fig. \ref{ABRD3D}. At large $h$, we also have $B\rightarrow 0$ and $B^2/A\ll R$, which means that
the phase and amplitude modes decouple.

Since the gapless quasiparticle spectrum $E_{\bf k}^-$ has only Fermi points, the delta-function terms vanish. We have $n_s^\|=n$ for arbitrary $h$ and
\begin{eqnarray}
n_s^\bot=n-\sum_{\alpha=\pm}\sum_{\bf k}\frac{\lambda^2}{2E_{\bf k}^\alpha}\Bigg[\bigg(1-\frac{\lambda^2{\bf k}_\bot^2\xi_{\bf
k}^2}{2\zeta_{\bf k}^2}\bigg)+\alpha\bigg(\frac{E_{\bf
k}^2}{2\zeta_{\bf k}}+h^2\frac{E_{\bf
k}^4+\lambda^2{\bf k}_\bot^2\Delta^2}{2\zeta_{\bf
k}^3}\bigg)\Bigg].
\end{eqnarray}
The numerical result for $n_s^\bot$ is shown in Fig. \ref{NSH3D}. Its behavior is similar to the 2D superfluid density: it decreases at small $h$ and
then turns to increase at large $h$. Together with the behavior of the expansion parameters $A,B$ and $R$, the results of the Goldstone mode velocities
$c_s^\bot$ and $c_s^\|$ are shown in Fig. \ref{CSH3D}. The nonmonotonic behavior of the transverse velocity $c_s^\bot$ is similar to the
2D case. The longitudinal velocity $c_s^\|$ is always an increasing function of $h$, since we have $n_s^\|=n$ for arbitrary $h$. Unlike the 2D case,
these velocities do not exhibit nonanalyticities at the quantum phase transitions.

%%%%%%%%%%%%%%%%%%%%%%%%%%%%%%%%%%%%%%%%%%%%%%%%%%%%%%%%%%%%%%%%%%%%%%%%%%%%%%%%%%%%%%%%%%%%%%%%%%%%%%%%%%%%%%
\section{Summary}\label{s6}
%%%%%%%%%%%%%%%%%%%%%%%%%%%%%%%%%%%%%%%%%%%%%%%%%%%%%%%%%%%%%%%%%%%%%%%%%%%%%%%%%%%%%%%%%%%%%%%%%%%%%%%%%%%%%%
In summary, we have investigated some bulk superfluid properties and the properties of the collective modes in attractive Fermi gases with Rashba
spin-orbit coupling. Our main results can be summarized as follows. \\ (A) For zero Zeeman field, we studied some bulk superfluid properties and the collective modes associated with the crossover from the ordinary
BCS/BEC superfluidity to the Bose-Einstein condensation of rashbons. The novel bound state, rashbon, exists even for negative $s$-wave scattering
length and possesses an effective mass which is generally larger than twice of the fermion mass. The behavior of the superfluid density and the
sound velocity manifests the rashbon BEC state at large spin-orbit coupling. Especially, we showed that for $\lambda\gg k_{\rm F}$, the behavior of
these quantities is universal, that is, independent of the attraction strength denoted by the parameter $1/(k_{\rm F}a_s)$. We also derived the
free energy which describes the weakly interacting rashbon condensate and determined the rashbon-rashbon coupling $g$. For large spin-orbit coupling,
this coupling $g$ goes as $g\sim 1/\lambda$ for the 3D case, while it goes as $g\sim 1/\ln(\lambda a_{\rm 2D})$ for the 2D case.
\\ (B) For nonzero Zeeman field, we studied the quantum phase transitions and properties of the collective modes across the phase transitions. For the
2D case, we found that the susceptibilities $\chi_{\mu\mu}$ and $\chi_{hh}$ as well as some other thermodynamic quantities behave nonanalytically
across the quantum phase transition, which indicates that the phase transition is of third order. The singularities of the thermodynamic functions
originate from the infrared divergence caused by the gapless fermionic spectrum at the quantum phase transition. As a result, the properties of the
collective modes also exhibit nonanalytical behavior at the quantum phase transition. The superfluid density and the sound velocity behave
nonmonotonically. In the normal superfluid phase, they are suppressed by the Zeeman field, as we expect from the fact that the Zeeman field serves
as a stress on the Cooper pairing. However, in the topological superfluid phase, they turn to be enhanced by the Zeeman field. Especially, we find
analytically that $n_s\rightarrow n$ and $c_s\rightarrow \upsilon_{\rm F}$ in the limit $h\rightarrow\infty$. This unusual phenomenon can be understood
from the fact that the system can be mapped to a spinless $p_x+ip_y$ superfluid state so that the fermion pairing does not feel stress from the
Zeeman field. For the 3D case, we found that the singularities at the quantum phase transitions are weakened. However, the behavior of the transverse
superfluid density and the transverse sound velocity is similar to their 2D counterparts.

Spin-orbit coupled atomic Fermi gases have been realized at ShanXi University~\cite{SOCF01} and at the Massachusetts Institute of Technology (MIT)
~\cite{SOCF02}, by using $^{40}$K atoms and $^{6}$Li atoms, respectively. These experiments realized spin-orbit coupled Fermi gases with equal Rashba
and Dresselhaus spin-orbit couplings. While the pure Rashba spin-orbit coupling is hopeful to be realized in the future experiments, it is interesting
to extend our studies to the present experimental systems. On the other hand, the collective mode spectrum can be measured by using the Bragg
scattering~\cite{Bragg01,Bragg02} and there are some proposals for the measurements of the superfluid density~\cite{NSexp01,NSexp02}. We hope our
theoretical predictions can be tested in the future experiments of spin-orbit coupled Fermi gases.

\section*{Acknowledgments} L. He acknowledges the support from the Helmholtz International Center for FAIR within the framework of the
LOEWE program (Landes- offensive zur Entwicklung Wissenschaftlich-{\"O}konomischer Exzellenz) launched by the State of Hesse. X.-G. Huang
is supported by Indiana University Bloomington.

\appendix

%%%%%%%%%%%%%%%%%%%%%%%%%%%%%%%%%%%%%%%%%%%%%%%%%%%%%%%%%%%%%%%%%%%%%%%
\section {Evaluating the collective mode propagator}
\label{app1}
%%%%%%%%%%%%%%%%%%%%%%%%%%%%%%%%%%%%%%%%%%%%%%%%%%%%%%%%%%%%%%%%%%%%%%%
In this Appendix we evaluate the explicit form of ${\bf M}(Q)$ for arbitrary spin-orbit coupling $\lambda$ and Zeeman field $h$. To evaluate the
diagonal element ${\bf M}_{11}(Q)$, we decompose the diagonal elements of the fermion propagator as follows
\begin{eqnarray}
{\cal G}_{11}(i\omega_n,{\bf k})=\sum_{s,\alpha=\pm}\frac{{\cal C}_{11}^{s,\alpha}({\bf k})}{i\omega_n-sE_{\bf k}^\alpha},\ \ \ \ \
\ \ \  {\cal G}_{22}(i\omega_n,{\bf k})=\sum_{s,\alpha=\pm}\frac{{\cal C}_{22}^{s,\alpha}({\bf k})}{i\omega_n-sE_{\bf k}^\alpha},
\end{eqnarray}
where the quantities ${\cal C}_{11}^{s,\alpha}({\bf k})$ and ${\cal C}_{22}^{s,\alpha}({\bf k})$ are given by
\begin{eqnarray}
&&{\cal C}_{11}^{s,\alpha}({\bf k})=\sum_{\gamma=\pm\alpha}\frac{sE_{\bf k}^\alpha+\xi_{\bf k}^\gamma}{2sE_{\bf k}^\alpha}
\frac{[(E_{\bf k}^\alpha)^2-(\xi_{\bf k}^{-\gamma})^2]{\cal P}_{\bf k}^\gamma(-h)-\Delta^2{\cal P}_{\bf k}^\gamma(h)}
{(E_{\bf k}^\alpha)^2-(E_{\bf k}^{-\alpha})^2},\nonumber\\
&&{\cal C}_{22}^{s,\alpha}({\bf k})=\sum_{\gamma=\pm\alpha}\frac{sE_{\bf k}^\alpha-\xi_{\bf k}^\gamma}{2sE_{\bf k}^\alpha}
\frac{[(E_{\bf k}^\alpha)^2-(\xi_{\bf k}^{-\gamma})^2]{\cal P}_{\bf k}^\gamma(h)-\Delta^2{\cal P}_{\bf k}^\gamma(-h)}
{(E_{\bf k}^\alpha)^2-(E_{\bf k}^{-\alpha})^2}.
\end{eqnarray}
Therefore, ${\bf M}_{11}(Q)$ and ${\bf M}_{22}(Q)$ can be evaluated as
\begin{eqnarray}
&&{\bf M}_{11}(\omega,{\bf q})={\bf M}_{22}(-\omega,{\bf q})=\nonumber\\
&&\frac{1}{U}-\frac{1}{2}\sum_{s,t=\pm}\sum_{\alpha,\beta=\pm}\sum_{\bf k}
\frac{f(sE_{{\bf k}+{\bf p}}^\alpha)-f(tE_{{\bf k}-{\bf p}}^\beta)}
{\omega-sE_{{\bf k}+{\bf p}}^\alpha+tE_{{\bf k}-{\bf p}}^\beta}
{\rm Tr}\left[{\cal C}_{11}^{s,\alpha}({\bf k}+{\bf p}){\cal C}_{22}^{t,\beta}({\bf k}-{\bf p})\right].
\end{eqnarray}
Here and in the following ${\bf p}={\bf q}/2$ for convenience. To
calculate the trace in the spin space, we use the following
properties of the projectors
\begin{eqnarray}
{\rm Tr}\left[{\cal P}_{{\bf k}+{\bf p}}^\alpha(h){\cal P}_{{\bf k}-{\bf p}}^\beta(h)\right]
={\rm Tr}\left[{\cal P}_{{\bf k}+{\bf p}}^\alpha(-h){\cal P}_{{\bf k}-{\bf p}}^\beta(-h)\right]
=\frac{1}{2}+\alpha\beta\frac{\lambda^2({\bf k}_\bot^2-{\bf p}_\bot^2)+h^2}
{2\eta_{{\bf k}+{\bf p}}\eta_{{\bf k}-{\bf p}}},\nonumber\\
{\rm Tr}\left[{\cal P}_{{\bf k}+{\bf p}}^\alpha(h){\cal P}_{{\bf k}-{\bf p}}^\beta(-h)\right]
={\rm Tr}\left[{\cal P}_{{\bf k}+{\bf p}}^\alpha(-h){\cal P}_{{\bf k}-{\bf p}}^\beta(h)\right]
=\frac{1}{2}+\alpha\beta\frac{\lambda^2({\bf k}_\bot^2-{\bf p}_\bot^2)-h^2}
{2\eta_{{\bf k}+{\bf p}}\eta_{{\bf k}-{\bf p}}}.
\end{eqnarray}
Finally the result can be rearranged as
\begin{eqnarray}
{\bf M}_{11}(\omega,{\bf q})&=&\frac{1}{U}+\frac{1}{2}\sum_{\alpha,\beta=\pm}\sum_{\bf k}
\left[\frac{{\cal W}_{++}^{\alpha\beta}({\bf k},{\bf q})}{\omega-E_{{\bf k}+{\bf p}}^\alpha-E_{{\bf k}-{\bf p}}^\beta}
-\frac{{\cal W}_{--}^{\alpha\beta}({\bf k},{\bf q})}{\omega+E_{{\bf k}+{\bf p}}^\alpha+E_{{\bf k}-{\bf p}}^\beta}\right]\nonumber\\
&\times&\left[1-f(E_{{\bf k}+{\bf p}}^\alpha)-f(E_{{\bf k}-{\bf p}}^\beta)\right]\nonumber\\
&+&\frac{1}{2}\sum_{\alpha,\beta=\pm}\sum_{\bf k}\left[\frac{{\cal W}_{-+}^{\alpha\beta}({\bf k},{\bf q})}
{\omega+E_{{\bf k}+{\bf p}}^\alpha-E_{{\bf k}-{\bf p}}^\beta}
-\frac{{\cal W}_{+-}^{\alpha\beta}({\bf k},{\bf q})}{\omega-E_{{\bf k}+{\bf p}}^\alpha+E_{{\bf k}-{\bf p}}^\beta}\right]\nonumber\\
&\times&\left[f(E_{{\bf k}+{\bf p}}^\alpha)-f(E_{{\bf k}-{\bf p}}^\beta)\right]
\end{eqnarray}
where the function ${\cal W}_{st}^{\alpha\beta}({\bf k},{\bf q})$ is defined as
\begin{eqnarray}
{\cal W}_{st}^{\alpha\beta}({\bf k},{\bf q})&=&\frac{1}{4}\sum_{e,f=\pm}
\left(1+s\frac{\xi_{{\bf k}+{\bf p}}^{e\alpha}}{E_{{\bf k}+{\bf p}}^\alpha}\right)
\left(1+t\frac{\xi_{{\bf k}-{\bf p}}^{f\beta}}{E_{{\bf k}-{\bf p}}^\beta}\right){\cal D}^{ef}_{\alpha\beta}({\bf k},{\bf q})
\end{eqnarray}
with
\begin{eqnarray}
{\cal D}^{ef}_{\alpha\beta}({\bf k},{\bf q})=
\left(\frac{\varphi^e_{{\bf k}+{\bf p}}\varphi^f_{{\bf k}-{\bf p}}}{\zeta_{{\bf k}+{\bf p}}\zeta_{{\bf k}-{\bf p}}}{\cal R}_{\bf kq}^+
-ef\frac{\Delta^2+2\alpha\varphi^e_{{\bf k}+{\bf p}}+2\beta\varphi^f_{{\bf k}-{\bf p}}}{8\zeta_{{\bf k}+{\bf p}}\zeta_{{\bf k}-{\bf p}}}
\frac{h^2\Delta^2}{\eta_{{\bf k}+{\bf p}}\eta_{{\bf k}-{\bf p}}}\right).
\end{eqnarray}
The functions $\varphi_{\bf k}^\pm$ and ${\cal R}_{\bf kq}^{\pm}$
are defined as
\begin{eqnarray}
\varphi_{\bf k}^\pm=\frac{1}{2}(\zeta_{\bf k}\pm\xi_{\bf k}\eta_{\bf k}),\ \ \ \ \ \ \
{\cal R}_{\bf kq}^{\pm}=\frac{1}{2}\left[1\pm\alpha\beta\frac{\lambda^2({\bf k}_\bot^2-{\bf p}_\bot^2)-h^2}
{\eta_{{\bf k}+{\bf p}}\eta_{{\bf k}-{\bf p}}}\right].
\end{eqnarray}

We can decompose ${\bf M}_{11}(\omega,{\bf q})$ as ${\bf M}_{11}(\omega,{\bf q})={\bf M}_{11}^+(\omega,{\bf q})+{\bf M}_{11}^-(\omega,{\bf q})$, where
${\bf M}_{11}^+(\omega,{\bf q})$ and ${\bf M}_{11}^-(\omega,{\bf q})$ are even and odd functions of $\omega$, respectively. Their explicit forms read
\begin{eqnarray}
{\bf M}_{11}^+(\omega,{\bf q})&=&\frac{1}{U}+\frac{1}{2}\sum_{\alpha,\beta=\pm}\sum_{\bf k}{\cal W}_1^{\alpha\beta}({\bf k},{\bf q})
\left[\frac{1}{\omega-E_{{\bf k}+{\bf p}}^\alpha-E_{{\bf k}-{\bf p}}^\beta}
-\frac{1}{\omega+E_{{\bf k}+{\bf p}}^\alpha+E_{{\bf k}-{\bf p}}^\beta}\right]\nonumber\\
&\times&\left[1-f(E_{{\bf k}+{\bf p}}^\alpha)-f(E_{{\bf k}-{\bf p}}^\beta)\right]\nonumber\\
&+&\frac{1}{2}\sum_{\alpha,\beta=\pm}\sum_{\bf k}{\cal U}_1^{\alpha\beta}({\bf k},{\bf q})
\left[\frac{1}{\omega+E_{{\bf k}+{\bf p}}^\alpha-E_{{\bf k}-{\bf p}}^\beta}
-\frac{1}{\omega-E_{{\bf k}+{\bf p}}^\alpha+E_{{\bf k}-{\bf p}}^\beta}\right]\nonumber\\
&\times&\left[f(E_{{\bf k}+{\bf p}}^\alpha)-f(E_{{\bf k}-{\bf p}}^\beta)\right]
\end{eqnarray}
and
\begin{eqnarray}
{\bf M}_{11}^-(\omega,{\bf q})&=&\frac{1}{2}\sum_{\alpha,\beta=\pm}\sum_{\bf k}{\cal W}_2^{\alpha\beta}({\bf k},{\bf q})
\left[\frac{1}{\omega-E_{{\bf k}+{\bf p}}^\alpha-E_{{\bf k}-{\bf p}}^\beta}
+\frac{1}{\omega+E_{{\bf k}+{\bf p}}^\alpha+E_{{\bf k}-{\bf p}}^\beta}\right]\nonumber\\
&\times&\left[1-f(E_{{\bf k}+{\bf p}}^\alpha)-f(E_{{\bf k}-{\bf p}}^\beta)\right]\nonumber\\
&+&\frac{1}{2}\sum_{\alpha,\beta=\pm}\sum_{\bf k}{\cal U}_2^{\alpha\beta}({\bf k},{\bf q})
\left[\frac{1}{\omega+E_{{\bf k}+{\bf p}}^\alpha-E_{{\bf k}-{\bf p}}^\beta}
+\frac{1}{\omega-E_{{\bf k}+{\bf p}}^\alpha+E_{{\bf k}-{\bf p}}^\beta}\right]\nonumber\\
&\times&\left[f(E_{{\bf k}+{\bf p}}^\alpha)-f(E_{{\bf k}-{\bf p}}^\beta)\right].
\end{eqnarray}
The functions ${\cal W}_i^{\alpha\beta}({\bf k},{\bf q})$ and ${\cal U}_i^{\alpha\beta}({\bf k},{\bf q})$ ($i=1,2$) read
\begin{eqnarray}
{\cal W}_1^{\alpha\beta}({\bf k},{\bf q})&=&\frac{1}{4}\sum_{e,f=\pm}
\left(1+\frac{\xi_{{\bf k}+{\bf p}}^{e\alpha}\xi_{{\bf k}-{\bf p}}^{f\beta}}
{E_{{\bf k}+{\bf p}}^\alpha E_{{\bf k}-{\bf p}}^\beta}\right){\cal D}^{ef}_{\alpha\beta}({\bf k},{\bf q}),\nonumber\\
{\cal U}_1^{\alpha\beta}({\bf k},{\bf q})&=&\frac{1}{4}\sum_{e,f=\pm}
\left(1-\frac{\xi_{{\bf k}+{\bf p}}^{e\alpha}\xi_{{\bf k}-{\bf p}}^{f\beta}}
{E_{{\bf k}+{\bf p}}^\alpha E_{{\bf k}-{\bf p}}^\beta}\right){\cal D}^{ef}_{\alpha\beta}({\bf k},{\bf q}),\nonumber\\
{\cal W}_2^{\alpha\beta}({\bf k},{\bf q})&=&\frac{1}{4}\sum_{e,f=\pm}
\left(\frac{\xi_{{\bf k}+{\bf p}}^{e\alpha}}{E_{{\bf k}+{\bf p}}^\alpha}
+\frac{\xi_{{\bf k}-{\bf p}}^{f\beta}}{E_{{\bf k}-{\bf p}}^\beta}\right){\cal D}^{ef}_{\alpha\beta}({\bf k},{\bf q}),\nonumber\\
{\cal U}_2^{\alpha\beta}({\bf k},{\bf q})&=&-\frac{1}{4}\sum_{e,f=\pm}
\left(\frac{\xi_{{\bf k}+{\bf p}}^{e\alpha}}{E_{{\bf k}+{\bf p}}^\alpha}
-\frac{\xi_{{\bf k}-{\bf p}}^{f\beta}}{E_{{\bf k}-{\bf p}}^\beta}\right){\cal D}^{ef}_{\alpha\beta}({\bf k},{\bf q}).
\end{eqnarray}

For the off-diagonal fermion propagators, we have
\begin{eqnarray}
{\cal G}_{12}(i\omega_n,{\bf k})=\sum_{s,\alpha=\pm}\frac{{\cal C}_{12}^{s,\alpha}({\bf k})}{i\omega_n-sE_{\bf k}^\alpha},
\ \ \ \ \ \ \ \
{\cal G}_{21}(i\omega_n,{\bf k})=\sum_{s,\alpha=\pm}\frac{{\cal C}_{21}^{s,\alpha}({\bf k})}{i\omega_n-sE_{\bf k}^\alpha},
\end{eqnarray}
where the quantities ${\cal C}_{12}^{s,\alpha}({\bf k})$ and ${\cal C}_{21}^{s,\alpha}({\bf k})$ are given by
\begin{eqnarray}
{\cal C}_{12}^{s,\alpha}({\bf k})=-\sum_{\gamma=\pm}\frac{\Delta}{2sE_{\bf k}^\alpha}
\frac{[(E_{\bf k}^\alpha)^2-(\xi_{\bf k}^{-\gamma})^2-\Delta^2]{\cal P}_{\bf k}^\gamma(h)
-2h(sE_{\bf k}^\alpha+\xi_{\bf k}^{-\gamma})\sigma_z{\cal P}_{\bf k}^\gamma(h)}
{(E_{\bf k}^\alpha)^2-(E_{\bf k}^{-\alpha})^2},\nonumber\\
=-\sum_{\gamma=\pm}\frac{\Delta}{2sE_{\bf k}^\alpha}
\frac{[(E_{\bf k}^\alpha)^2-(\xi_{\bf k}^{-\gamma})^2-\Delta^2]{\cal P}_{\bf k}^\gamma(-h)
-2h(sE_{\bf k}^\alpha-\xi_{\bf k}^{-\gamma}){\cal P}_{\bf k}^\gamma(-h)\sigma_z}
{(E_{\bf k}^\alpha)^2-(E_{\bf k}^{-\alpha})^2},\nonumber\\
{\cal C}_{21}^{s,\alpha}({\bf k})=-\sum_{\gamma=\pm}\frac{\Delta}{2sE_{\bf k}^\alpha}
\frac{[(E_{\bf k}^\alpha)^2-(\xi_{\bf k}^{-\gamma})^2-\Delta^2]{\cal P}_{\bf k}^\gamma(h)
-2h(sE_{\bf k}^\alpha+\xi_{\bf k}^{-\gamma}){\cal P}_{\bf k}^\gamma(h)\sigma_z}
{(E_{\bf k}^\alpha)^2-(E_{\bf k}^{-\alpha})^2},\nonumber\\
=-\sum_{\gamma=\pm}\frac{\Delta}{2sE_{\bf k}^\alpha}
\frac{[(E_{\bf k}^\alpha)^2-(\xi_{\bf k}^{-\gamma})^2-\Delta^2]{\cal P}_{\bf k}^\gamma(-h)
-2h(sE_{\bf k}^\alpha-\xi_{\bf k}^{-\gamma})\sigma_z{\cal P}_{\bf k}^\gamma(-h)}
{(E_{\bf k}^\alpha)^2-(E_{\bf k}^{-\alpha})^2}.
\end{eqnarray}
Therefore ${\bf M}_{12}(Q)$ and ${\bf M}_{21}(Q)$ can be evaluated as
\begin{eqnarray}
{\bf M}_{12}(\omega,{\bf q})=-\frac{1}{2}\sum_{s,t=\pm}\sum_{\alpha,\beta=\pm}
\sum_{\bf k}\frac{f(sE_{{\bf k}+{\bf p}}^\alpha)-f(tE_{{\bf k}-{\bf p}}^\beta)}
{\omega-sE_{{\bf k}+{\bf p}}^\alpha+tE_{{\bf k}-{\bf p}}^\beta}
{\rm Tr}\left[{\cal C}_{12}^{s,\alpha}({\bf k}+{\bf p}){\cal C}_{12}^{t,\beta}({\bf k}-{\bf p})\right],\nonumber\\
{\bf M}_{21}(\omega,{\bf q})=-\frac{1}{2}\sum_{s,t=\pm}\sum_{\alpha,\beta=\pm}
\sum_{\bf k}\frac{f(sE_{{\bf k}+{\bf p}}^\alpha)-f(tE_{{\bf k}-{\bf p}}^\beta)}
{\omega-sE_{{\bf k}+{\bf p}}^\alpha+tE_{{\bf k}-{\bf p}}^\beta}
{\rm Tr}\left[{\cal C}_{21}^{s,\alpha}({\bf k}+{\bf p}){\cal C}_{21}^{t,\beta}({\bf k}-{\bf p})\right].
\end{eqnarray}
Completing the traces, we find that they can be expressed as
\begin{eqnarray}
&&{\bf M}_{12}(\omega,{\bf q})={\bf M}_{12}^+(\omega,{\bf q})+i{\bf M}_{12}^-(\omega,{\bf q}),\nonumber\\
&&{\bf M}_{21}(\omega,{\bf q})={\bf M}_{12}^+(\omega,{\bf q})-i{\bf M}_{12}^-(\omega,{\bf q}),
\end{eqnarray}
where ${\bf M}_{12}^+(\omega,{\bf q})$ and ${\bf M}_{12}^-(\omega,{\bf q})$ read
\begin{eqnarray}
{\bf M}_{12}^+(\omega,{\bf q})&=&-\frac{1}{2}\sum_{\alpha,\beta=\pm}\sum_{\bf k}{\cal W}_3^{\alpha\beta}({\bf k},{\bf q})
\left[\frac{1}{\omega-E_{{\bf k}+{\bf p}}^\alpha-E_{{\bf k}-{\bf p}}^\beta}
-\frac{1}{\omega+E_{{\bf k}+{\bf p}}^\alpha+E_{{\bf k}-{\bf p}}^\beta}\right]\nonumber\\
&\times&\left[1-f(E_{{\bf k}+{\bf p}}^\alpha)-f(E_{{\bf k}-{\bf p}}^\beta)\right]\nonumber\\
&+&\frac{1}{2}\sum_{\alpha,\beta=\pm}\sum_{\bf k}{\cal U}_3^{\alpha\beta}({\bf k},{\bf q})
\left[\frac{1}{\omega+E_{{\bf k}+{\bf p}}^\alpha-E_{{\bf k}-{\bf p}}^\beta}
-\frac{1}{\omega-E_{{\bf k}+{\bf p}}^\alpha+E_{{\bf k}-{\bf p}}^\beta}\right]\nonumber\\
&\times&\left[f(E_{{\bf k}+{\bf p}}^\alpha)-f(E_{{\bf k}-{\bf p}}^\beta)\right]
\end{eqnarray}
and
\begin{eqnarray}
{\bf M}_{12}^-(\omega,{\bf q})&=&-\frac{1}{2}\sum_{\alpha,\beta=\pm}\sum_{\bf k}{\cal W}_4^{\alpha\beta}({\bf k},{\bf q})
\left[\frac{1}{\omega-E_{{\bf k}+{\bf p}}^\alpha-E_{{\bf k}-{\bf p}}^\beta}
-\frac{1}{\omega+E_{{\bf k}+{\bf p}}^\alpha+E_{{\bf k}-{\bf p}}^\beta}\right]\nonumber\\
&\times&\left[1-f(E_{{\bf k}+{\bf p}}^\alpha)-f(E_{{\bf k}-{\bf p}}^\beta)\right]\nonumber\\
&+&\frac{1}{2}\sum_{\alpha,\beta=\pm}\sum_{\bf k}{\cal U}_4^{\alpha\beta}({\bf k},{\bf q})
\left[\frac{1}{\omega+E_{{\bf k}+{\bf p}}^\alpha-E_{{\bf k}-{\bf p}}^\beta}
-\frac{1}{\omega-E_{{\bf k}+{\bf p}}^\alpha+E_{{\bf k}-{\bf p}}^\beta}\right]\nonumber\\
&\times&\left[f(E_{{\bf k}+{\bf p}}^\alpha)-f(E_{{\bf k}-{\bf p}}^\beta)\right].
\end{eqnarray}
The functions ${\cal W}_i^{\alpha\beta}({\bf k},{\bf q})$ and ${\cal U}_i^{\alpha\beta}({\bf k},{\bf q})$ ($i=3,4$) read
\begin{eqnarray}
{\cal W}_3^{\alpha\beta}({\bf k},{\bf q})&=&\frac{\Delta^2}{8E_{{\bf k}+{\bf p}}^\alpha E_{{\bf k}-{\bf p}}^\beta}
\Bigg[1+\alpha\beta\frac{(\xi_{{\bf k}+{\bf p}}\xi_{{\bf k}-{\bf p}}-h^2)\lambda^2({\bf k}_\bot^2-{\bf p}_\bot^2)}
{\zeta_{{\bf k}+{\bf p}}\zeta_{{\bf k}-{\bf p}}}\nonumber\\
&&+\alpha\beta h^2\frac{\alpha\zeta_{{\bf k}+{\bf p}}+\beta\zeta_{{\bf k}-{\bf p}}
-E_{{\bf k}+{\bf p}}^\alpha E_{{\bf k}-{\bf p}}^\beta+h^2}
{\zeta_{{\bf k}+{\bf p}}\zeta_{{\bf k}-{\bf p}}}\Bigg],\nonumber\\
{\cal U}_3^{\alpha\beta}({\bf k},{\bf q})&=&\frac{\Delta^2}{8E_{{\bf k}+{\bf p}}^\alpha E_{{\bf k}-{\bf p}}^\beta}
\Bigg[1+\alpha\beta\frac{(\xi_{{\bf k}+{\bf p}}\xi_{{\bf k}-{\bf p}}-h^2)\lambda^2({\bf k}_\bot^2-{\bf p}_\bot^2)}
{\zeta_{{\bf k}+{\bf p}}\zeta_{{\bf k}-{\bf p}}}\nonumber\\
&&+\alpha\beta h^2\frac{\alpha\zeta_{{\bf k}+{\bf p}}+\beta\zeta_{{\bf k}-{\bf p}}
+E_{{\bf k}+{\bf p}}^\alpha E_{{\bf k}-{\bf p}}^\beta+h^2}
{\zeta_{{\bf k}+{\bf p}}\zeta_{{\bf k}-{\bf p}}}\Bigg],\nonumber\\
{\cal W}_4^{\alpha\beta}({\bf k},{\bf q})&=&{\cal U}_4^{\alpha\beta}({\bf k},{\bf q})=\alpha\beta\frac{\Delta^2}{4E_{{\bf
k}+{\bf p}}^\alpha E_{{\bf k}-{\bf
p}}^\beta}\frac{h\lambda^2 p_xp_y(k_x^2-k_y^2)}{\zeta_{{\bf k}+{\bf p}}\zeta_{{\bf k}-{\bf
p}}}.
\end{eqnarray}

%%%%%%%%%%%%%%%%%%%%%%%%%%%%%%%%%%%%%%%%%%%%%%%%%%%%%%%%%%%%%%%%%%%%%%%
\section {Evaluating the Expansion Parameters}
\label{app2}
%%%%%%%%%%%%%%%%%%%%%%%%%%%%%%%%%%%%%%%%%%%%%%%%%%%%%%%%%%%%%%%%%%%%%%%
In this Appendix, we evaluate the expansion parameters $A,B,D,R$ and the phase stiffness $J_\bot,J_\|$ in the zero temperature limit. First, $A$ can
be evaluated as
\begin{equation}
A=\frac{1}{2}\sum_{\alpha=\pm}\sum_{\bf k}\left[\frac{\Delta^2}{(E_{\bf k}^\alpha)^3}\left(1+\alpha\frac{h^2}{\zeta_{\bf k}}\right)^2
+\alpha\frac{h^4\Delta^2}{E_{\bf k}^\alpha\zeta_{\bf k}^3}-2\frac{\Delta^2}{(E_{\bf k}^\alpha)^2}
\left(1+\alpha\frac{h^2}{\zeta_{\bf k}}\right)^2\delta(E_{\bf k}^\alpha)\right],
\end{equation}
where the delta function comes from the terms proportional to $f(E_{\bf k+p}^\alpha)-f(E_{\bf k-p}^\alpha)$. Taking the derivatives with respect to
$\omega$, we obtain
\begin{eqnarray}
&&B=2\Delta\sum_{\alpha,\beta=\pm}\sum_{\bf k}\frac{{\cal W}_2^{\alpha,\beta}({\bf k},{\bf 0})}
{(E_{\bf k}^\alpha+E_{\bf k}^\beta)^2},\nonumber\\
&&D=2\sum_{\alpha,\beta=\pm}\sum_{\bf k}\frac{{\cal W}_1^{\alpha\beta}({\bf k},{\bf 0})-{\cal W}_3^{\alpha\beta}({\bf k},{\bf 0})}
{(E_{\bf k}^\alpha+E_{\bf k}^\beta)^3},\nonumber\\
&&R=2\Delta^2\sum_{\alpha,\beta=\pm}\sum_{\bf k}\frac{{\cal W}_1^{\alpha\beta}({\bf k},{\bf 0})+{\cal W}_3^{\alpha\beta}({\bf k},{\bf 0})}
{(E_{\bf k}^\alpha+E_{\bf k}^\beta)^3}.
\end{eqnarray}
Completing the summation over $\alpha,\beta=\pm$, we obtain
\begin{eqnarray}
B&=&\Delta\sum_{\bf k}\frac{\xi_{\bf k}}{(E_{\bf k}^++E_{\bf k}^-)^2}\left[\left(\frac{1}{E_{\bf k}^+}+\frac{1}{E_{\bf k}^-}\right)
\frac{h^2E_{\bf k}^2}{\zeta_{\bf k}^2}+\left(\frac{1}{E_{\bf k}^+}-\frac{1}{E_{\bf k}^-}\right)\frac{h^2}{\zeta_{\bf k}}\right]\nonumber\\
&+&\frac{\Delta}{4}\sum_{\alpha\pm}\sum_{\bf k}\frac{\xi_{\bf k}}{(E_{\bf k}^\alpha)^3}
\left(1+\alpha\frac{\lambda^2{\bf k}_\bot^2}{\zeta_{\bf k}}-\frac{h^2E_{\bf k}^2}{\zeta_{\bf k}^2}\right),\nonumber\\
D&=&\frac{1}{8}\sum_{\alpha=\pm}\sum_{\bf k}\left[\frac{\xi_{\bf k}^2+\eta_{\bf k}^2+2\alpha\zeta_{\bf k}}{(E_{\bf k}^\alpha)^5}
\frac{\lambda^2{\bf k}_\bot^2\xi_{\bf k}^2}{\zeta_{\bf k}^2}+\frac{\Delta^2}{(E_{\bf k}^\alpha)^5}
\frac{\lambda^2{\bf k}_\bot^2h^2}{\zeta_{\bf k}^2}\right]\nonumber\\
&+&\sum_{\bf k}\frac{1}{(E_{\bf k}^++E_{\bf k}^-)^3}\frac{h^2\xi_{\bf k}^2}{\zeta_{\bf k}^2}
\left(1+\frac{E_{\bf k}^2-\eta_{\bf k}^2}{E_{\bf k}^+E_{\bf k}^-}\right),\nonumber\\
R&=&\Delta^2\sum_{\bf k}\frac{1}{(E_{\bf k}^++E_{\bf k}^-)^3}\frac{h^2E_{\bf k}^2}{\zeta_{\bf k}^2}
\left(1+\frac{E_{\bf k}^2-\eta_{\bf k}^2}{E_{\bf k}^+E_{\bf k}^-}+\frac{2\lambda^2{\bf k}_\bot^2\Delta^2}
{E_{\bf k}^+E_{\bf k}^-E_{\bf k}^2}\right)\nonumber\\
&+&\frac{\Delta^2}{8}\sum_{\alpha=\pm}\sum_{\bf k}\frac{1}{(E_{\bf k}^\alpha)^3}\frac{\lambda^2{\bf k}_\bot^2\xi_{\bf k}^2}{\zeta_{\bf k}^2}.
\end{eqnarray}

To evaluate the phase stiffnesses $J_\bot$ and $J_\|$ or the superfluid densities $n_s^\bot$ and $n_s^\|$, we assume that the order parameter takes the form
$\langle\Phi\rangle=\Delta e^{2i{\bf q}\cdot {\bf r}}$ where $\Delta$ corresponds to the saddle point solution. The superfluid density is defined as
the response of the thermodynamic potential to an infinitesimal momentum ${\bf q}$, that is,
\begin{eqnarray}
\Omega({\bf q})=\Omega({\bf 0})+\frac{1}{2}n_s^\bot{\bf q}_\bot^2+\frac{1}{2}n_s^\|{\bf q}_\|^2+O({\bf q}^4).
\end{eqnarray}
On the other hand, the thermodynamic potential for arbitrary ${\bf q}$ can be evaluated by making a phase transformation
$\psi\rightarrow e^{i{\bf q}\cdot {\bf r}}\psi$. We have
\begin{eqnarray}
\Omega({\bf q})=\Omega({\bf 0})+\frac{1}{2}\sum_{K}\sum_{l=1}^\infty\frac{1}{l}{\rm Tr}\left[{\cal G}(i\omega_n,{\bf k})\Sigma({\bf q})\right]^l,
\end{eqnarray}
where
\begin{eqnarray}
\Sigma({\bf q})=\frac{1}{2}\left({\bf q}_\bot^2+{\bf q}_\|^2\right)\tau_3
+({\bf k}_\bot\cdot{\bf q}_\bot+{\bf k}_\|\cdot{\bf q}_\|)\tau_0
+\lambda\mbox{\boldmath{$\sigma$}}_\bot\cdot{\bf q}_\bot\tau_0.
\end{eqnarray}
Here $\tau_i$ ($i=1,2,3$) and $\tau_0$ are the Pauli matrices and the identity matrix in the Nambu-Gor'kov space.

Then we can obtain the explicit form of the phase stiffness using the fermion Green's function ${\cal G}(i\omega_n,{\bf k})$ at the saddle point via
the derivative expansion. Obviously, only the $l=1$ and $l=2$ terms contribute. We note that the $l=1$ contribution is just identical to the total
density $n$. After some algebras, we obtain
\begin{eqnarray}
n_s^\bot&=&n+\sum_{\bf k}\frac{1}{\beta}\sum_n \left[\frac{{\bf k}_\bot^2}{4}X(i\omega_n,{\bf k})
+\frac{\lambda^2}{2}Y(i\omega_n,{\bf k})+\frac{\lambda^2{\bf k}_\bot^2}{2}Z(i\omega_n,{\bf k})\right],\nonumber\\
n_s^\|&=&n+\sum_{\bf k}\frac{1}{\beta}\sum_n \frac{{\bf k}_\|^2}{2}X(i\omega_n,{\bf k}).
\end{eqnarray}
The functions $X, Y$ and $Z$ are derived from Green's function ${\cal G}$. $X$ is related to the trace ${\rm Tr}[{\cal G}{\cal G}]$. The result of
$X$ reads
\begin{eqnarray}
X(i\omega_n,{\bf k})&=&\frac{1}{[(i\omega_n)^2-(E_{\bf k}^+)^2]^2[(i\omega_n)^2-(E_{\bf k}^-)^2]^2}\nonumber\\
&\times&\Bigg\{[(i\omega_n+h+\xi_{\bf k})\Lambda_-+2\lambda^2{\bf k}_\bot^2(\xi_{\bf k}+h)]^2\nonumber\\
&&+[(i\omega_n-h+\xi_{\bf k})\Lambda_++2\lambda^2{\bf k}_\bot(\xi_{\bf k}-h)]^2\nonumber\\
&&+[(i\omega_n+h-\xi_{\bf k})\Lambda_--2\lambda^2{\bf k}_\bot^2(\xi_{\bf k}-h)]^2\nonumber\\
&&+[(i\omega_n-h-\xi_{\bf k})\Lambda_+-2\lambda^2{\bf k}_\bot^2(\xi_{\bf k}+h)]^2\nonumber\\
&&+2\lambda^2{\bf k}_\bot^2[2(\xi_{\bf k}-h)(i\omega_n+h+\xi_{\bf k})+\Lambda_+]\nonumber\\
&&\times[2(\xi_{\bf k}+h)(i\omega_n-h+\xi_{\bf k})+\Lambda_-]\nonumber\\
&&+2\lambda^2{\bf k}_\bot^2[2(\xi_{\bf k}+h)(i\omega_n+h-\xi_{\bf k})-\Lambda_+]\nonumber\\
&&\times[2(\xi_{\bf k}-h)(i\omega_n-h-\xi_{\bf k})-\Lambda_-]\nonumber\\
&&+2\Delta^2[\Lambda_+^2+\Lambda_-^2+8\lambda^2{\bf k}_\bot^2(\xi_{\bf k}^2+h^2)]\Bigg\}
\end{eqnarray}
Here $\Lambda_\pm=(i\omega_n\pm h)^2-\xi_{\bf k}^2-\Delta^2-\lambda^2{\bf k}_\bot^2$. $Y$ is related to the trace
${\rm Tr}[{\cal G}\sigma_i{\cal G}\sigma_i]$ ($i=x,y$). Due to the angle integration, the nonzero contribution reads
\begin{eqnarray}
Y(i\omega_n,{\bf k})&=&\frac{2}{[(i\omega_n)^2-(E_{\bf k}^+)^2]^2[(i\omega_n)^2-(E_{\bf k}^-)^2]^2}\nonumber\\
&\times&\Bigg\{[(i\omega_n+h+\xi_{\bf k})\Lambda_-+2\lambda^2{\bf k}_\bot^2(\xi_{\bf k}+h)]\nonumber\\
&&\times[(i\omega_n-h+\xi_{\bf k})\Lambda_++2\lambda^2{\bf k}_\bot(\xi_{\bf k}-h)]\nonumber\\
&&+[(i\omega_n+h-\xi_{\bf k})\Lambda_--2\lambda^2{\bf k}_\bot^2(\xi_{\bf k}-h)]\nonumber\\
&&\times[(i\omega_n-h-\xi_{\bf k})\Lambda_+-2\lambda^2{\bf k}_\bot^2(\xi_{\bf k}+h)]+2\Delta^2\Lambda_+\Lambda_-\Bigg\}.
\end{eqnarray}
$Z$ is related to the trace ${\rm Tr}[{\cal G}{\cal G}\sigma_ik_i]$ ($i=x,y$). We have
\begin{eqnarray}
Z(i\omega_n,{\bf k})&=&\frac{1}{[(i\omega_n)^2-(E_{\bf k}^+)^2]^2[(i\omega_n)^2-(E_{\bf k}^-)^2]^2}\nonumber\\
&\times&\Bigg\{[(i\omega_n+h+\xi_{\bf k})\Lambda_-+(i\omega_n-h+\xi_{\bf k})\Lambda_++4\lambda^2{\bf k}_\bot^2\xi_{\bf k}]\nonumber\\
&&\times[4\xi_{\bf k}i\omega_n+4(\xi_{\bf k}^2-h^2)+\Lambda_++\Lambda_-]\nonumber\\
&&+[(i\omega_n+h-\xi_{\bf k})\Lambda_-+(i\omega_n-h-\xi_{\bf k})\Lambda_+-4\lambda^2{\bf k}_\bot^2\xi_{\bf k}]\nonumber\\
&&\times[4\xi_{\bf k}i\omega_n-4(\xi_{\bf k}^2-h^2)-\Lambda_+-\Lambda_-]\nonumber\\
&&+8\Delta^2\xi_{\bf k}(\Lambda_++\Lambda_-)\Bigg\}.
\end{eqnarray}
The above results can be simplified by using the method of Laurent expansion. We have
\begin{eqnarray}
&&X(i\omega_n,{\bf k})=2\sum_{\alpha=\pm}\frac{(i\omega_n)^2+(E_{\bf k}^\alpha)^2}{[(i\omega_n)^2-(E_{\bf k}^\alpha)^2]^2},\nonumber\\
&&Z(i\omega_n,{\bf k})=\frac{2\xi_{\bf k}}{\zeta_{\bf k}}\sum_{\alpha=\pm}\alpha\frac{(i\omega_n)^2+(E_{\bf k}^\alpha)^2}
{[(i\omega_n)^2-(E_{\bf k}^\alpha)^2]^2},\nonumber\\
&&Y(i\omega_n,{\bf k})=\sum_{\alpha=\pm}\left[\alpha\bigg(\frac{E_{\bf k}^2}{\zeta_{\bf k}}
+h^2\frac{E_{\bf k}^4+\lambda^2{\bf k}_\bot^2\Delta^2}{\zeta_{\bf k}^3}\bigg)
+\bigg(2-\frac{\lambda^2{\bf k}_\bot^2\xi_{\bf k}^2}{\zeta_{\bf k}^2}\bigg)\right]
\frac{1}{(i\omega_n)^2-(E_{\bf k}^\alpha)^2}\nonumber\\
&&\ \ \ \ \ \ \ \ \ \ \ \ \ \ \ \
+\frac{\lambda^2{\bf k}_\bot^2\xi_{\bf k}^2}{\zeta_{\bf k}^2}\sum_{\alpha=\pm}
\frac{(i\omega_n)^2+(E_{\bf k}^\alpha)^2}{[(i\omega_n)^2-(E_{\bf k}^\alpha)^2]^2}.
\end{eqnarray}
Completing the Matsubara frequency summation, we obtain
\begin{eqnarray}
n_s^\bot&=&n-\sum_{\bf k}\sum_{\alpha=\pm}\frac{{\bf k}_\bot^2}{2}
\left(1+\frac{\lambda^2\xi_{\bf k}}{\zeta_{\bf k}}\right)^2\left(\frac{1}{4T}{\rm sech}^2\frac{E_{\bf k}^\alpha}{2T}\right)\nonumber\\
&-&\lambda^2\sum_{\bf k}\sum_{\alpha=\pm}\left[\alpha\bigg(\frac{E_{\bf k}^2}{2\zeta_{\bf k}}
+h^2\frac{E_{\bf k}^4+\lambda^2{\bf k}_\bot^2\Delta^2}{2\zeta_{\bf k}^3}\bigg)
+\bigg(1-\frac{\lambda^2{\bf k}_\bot^2\xi_{\bf k}^2}{2\zeta_{\bf k}^2}\bigg)\right]
\frac{1-2f(E_{\bf k}^\alpha)}{2E_{\bf k}^\alpha},\nonumber\\
n_s^\|&=&n-\sum_{\bf k}\sum_{\alpha=\pm}{\bf k}_\|^2\left(\frac{1}{4T}{\rm sech}^2\frac{E_{\bf k}^\alpha}{2T}\right).
\end{eqnarray}
At zero temperature, we have $f(E)\rightarrow0$ and $(1/4T){\rm sech}^2(E/2T)\rightarrow\delta(E)$.

\bibliographystyle{model1-num-names}

\end{document}